\documentclass[a4paper,11pt]{article}
\pdfoutput=1
\usepackage{jheppub}
\usepackage[T1]{fontenc}
\usepackage{amsmath}
\usepackage{amssymb}
\usepackage{epsfig}
\usepackage{subfigure}
\usepackage{float}
\usepackage{verbatim} 
\usepackage{pstricks}
\usepackage{color}
\usepackage{slashed}
\usepackage{multirow}
\usepackage{pstricks}
\usepackage{color}
\usepackage{booktabs}
\usepackage{subfigure}
\usepackage{epstopdf}
\newcommand{\FDF}{(\varphi^\dagger i\!\!\overleftrightarrow{D}_\mu\varphi)}
\newcommand{\FDFI}{(\varphi^\dagger i\!\!\overleftrightarrow{D}^I_\mu\varphi)}
\newcommand{\sw}{s_{W}}
\newcommand{\cw}{c_{W}}
\newcommand{\yt}{ y_t}
\newcommand{\yb}{ y_b}

\begin{document}

\title{Dimension-six electroweak top-loop effects in Higgs production and decay}
\author[a]{Eleni Vryonidou,}
\author[b]{Cen Zhang}

\abstract{
We study the next-to-leading order electroweak corrections to Higgs processes
from dimension-six top-quark operators in the Standard Model Effective Field
Theory approach.  We consider the major production channels, including $WH$,
$ZH$, and VBF production at the LHC, and $ZH$, VBF production at future lepton
colliders, as well as the major decay channels including $H\to \gamma\gamma,
\gamma Z, Wl\nu,Zll,b\bar b,\mu\mu,\tau\tau$.  The results show that within the
current constraints, top-quark operators can shift the signal strength of the
loop-induced processes, i.e.~$H\to \gamma\gamma,\gamma Z$, by factors of
$\sim\mathcal{O}(1)-\mathcal{O}(10)$, and that of the tree-level processes,
i.e.~all remaining production and decay channels, by $\sim5-10\%$ at the LHC,
and up to $\sim15\%$ at future lepton colliders.  This implies that essentially
all Higgs channels have started to become sensitive to top-quark couplings, and
in particular, Higgs observables at high luminosity LHC as well as future
lepton colliders, even below the $t\bar t$ threshold, will improve our
knowledge of top-quark couplings.  We derive the sensitivities of Higgs
measurements to top-quark operators at the high luminosity LHC, using
projections for both inclusive and differential measurements.  We conclude that
treating the dimension-six top-quark sector and the Higgs/electroweak sector
separately may not continue to be a good strategy.  A global analysis combining
Higgs and top-quark measurements is desirable, and our calculation and
implementation provide an automatic and realistic simulation tool for this
purpose.
} 

\affiliation[a]{CERN, Theoretical Physics Department, Geneva 23 CH-1211, Switzerland}
\affiliation[b]{ Institute of High Energy Physics, and School of Physical
	Sciences, University of Chinese Academy of Sciences P.O.Box 918-4,
	Beijing, P.R.China
}


\maketitle
\section{Introduction}

Deviations in the top-quark and Higgs couplings are often studied within the
Standard Model Effective Field Theory (SMEFT) approach \cite{Weinberg:1978kz,
Buchmuller:1985jz,Leung:1984ni}.  Even though the SMEFT is a global approach
where different measurements are supposed to be combined and studied together,
in practice, several main sectors are often considered separately.
Dimension-six (dim-6) operators in the top-quark sector are analysed with top
measurements
\cite{Buckley:2015nca,Buckley:2015lku,Cirigliano:2016nyn,Rosello:2015sck,
deBlas:2015aea,Alioli:2017ces,Bernardo:2014vha,Tonero:2014jea,Cao:2015doa,
Jung:2014kxa,Zhang:2012cd,Greiner:2011tt}, while those in the Higgs sector are
analysed with Higgs and triple gauge-boson coupling (TGC) measurements
\cite{Corbett:2012ja,Butter:2016cvz,Englert:2015hrx,Falkowski:2015jaa,
Falkowski:2014tna,Ellis:2014dva,Ellis:2014jta,Ellis:2018gqa}, the rationale
behind being that the interplay between the two sectors is negligible with the
accuracy of the current measurements.

This assumption may not continue to be true as the LHC experiments
improve on higher integrated luminosities and higher energies.  Once the
precision reaches the point where loop corrections become relevant, the top- and
the Higgs-sectors should not be viewed separately.  In particular, top-quark
operators could play a role in Higgs measurements through electroweak (EW) top
and bottom loops and have non-trivial impacts.
The goal of this paper is to
answer the following two questions: 1) will this happen at the LHC, and/or
future lepton colliders? and 2) can Higgs measurements help to constrain
top-quark couplings?

\begin{figure}[th]
	\begin{center}
		\includegraphics{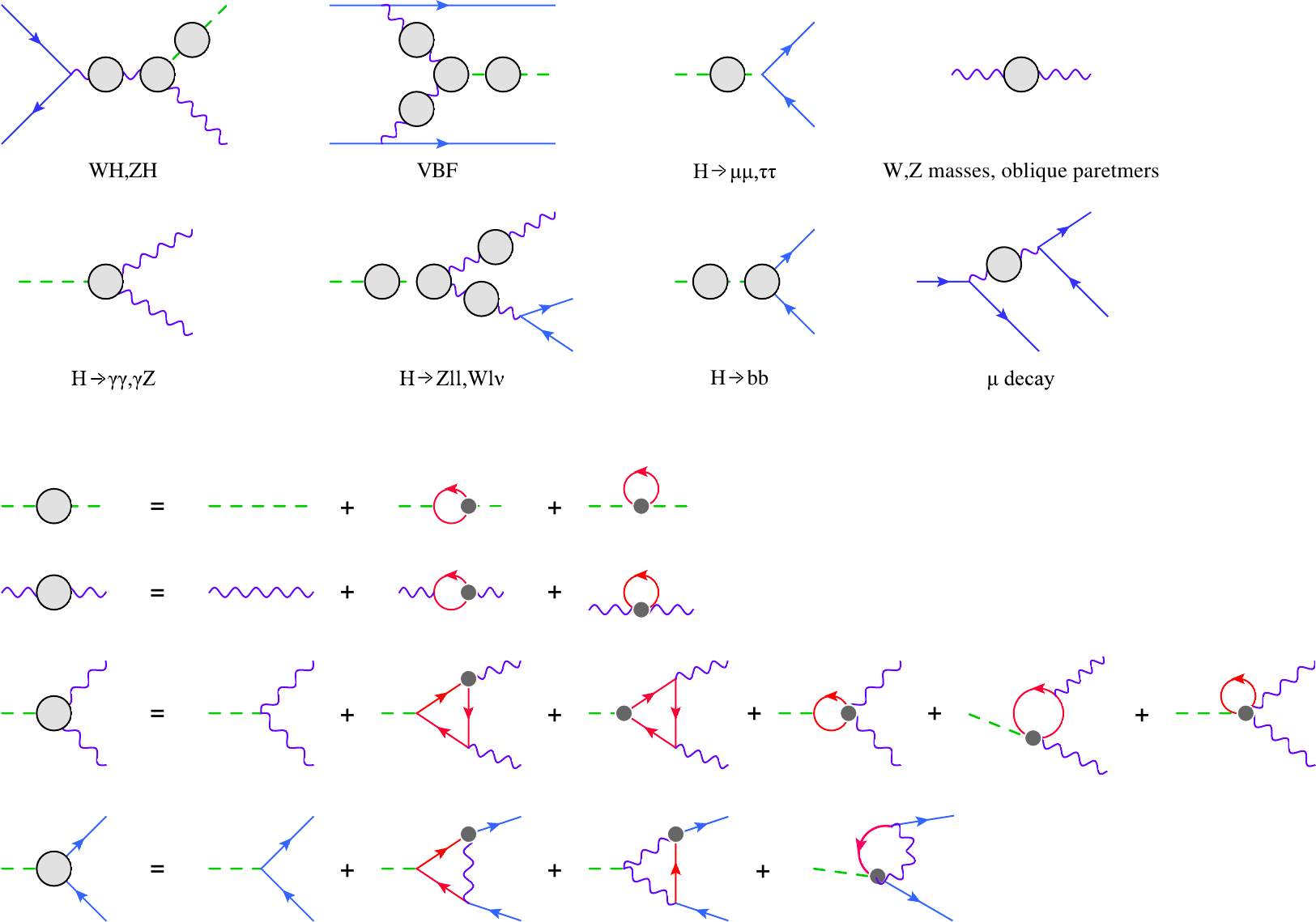}
	\end{center}
	\caption{Electroweak contributions from two-fermion top-quark operators
		to the production and decay of the Higgs boson, the oblique
		parameters, and the SM input parameters which we will take as
		$M_W$, $M_Z$ and $G_F$ (from muon decay).  Here blue fermion
		lines represent light fermions, and red fermion lines represent
		the top quark.  Large grey blobs denote collectively the SM
		contributions and dim-6 EW corrections from top-quark loops, as
	illustrated below the processes.  Dark small blobs represent the dim-6
operator insertions.  Diagrams that can be obtained by crossing legs or
reversing fermion flows are not shown.}
	\label{fig:diagram}
\end{figure}

To answer these two questions, in this work we compute the EW loop-induced
contributions from dim-6 top-quark two-fermion operators in the following main
Higgs production and decay processes:
\begin{flalign}
	&\mbox{production at LHC:}\ \mbox{VBF},\ WH,\ ZH
	\nonumber\\
	&\mbox{production at lepton collider:}\ ZH,\ \mbox{VBF}
	\nonumber\\
	&\mbox{decay:}\ H\to\gamma\gamma,\ \gamma Z,\ Wl\nu,\ 
	Zll,\ b\bar b,\ \mu^+\mu^-,\ \tau^+\tau^-.
	\nonumber
\end{flalign}
All relevant contributions are shown in Figure~\ref{fig:diagram}. 

Loop corrections in the SMEFT with dim-6 operators have been studied in the
literature.  The loop contributions in $gg\to H$ and $H\to gg$ have been
presented in \cite{Degrande:2012gr,Deutschmann:2017qum}. 
Top loop induced $gg\to ZH$, $gg\to HH$ and $gg\to Hj$ have also been
considered in \cite{Bylund:2016phk,Englert:2016hvy,Maltoni:2016yxb}.
Some of the decay processes, including $H\to WW^*,ZZ^*,\gamma\gamma$ and
$Z\gamma$, have been studied in
Refs.~\cite{Hartmann:2015oia,Hartmann:2015aia,Ghezzi:2015vva,Dawson:2018pyl}.
All other results in this work, in particular the next-to-leading order (NLO)
EW corrections for the production channels, are new and are relevant for future
Higgs and top studies at the LHC.  Note that loop corrections to $H\to b\bar b$
and $\tau^+\tau^-$ from four-fermion operators have been computed in
\cite{Gauld:2015lmb}, while the NLO
QCD corrections to $WH$, $ZH$, VBF and $t\bar tH$ production processes are also
known \cite{Mimasu:2015nqa,Degrande:2016dqg,Maltoni:2016yxb}, but they are not
relevant for the purpose of this work.  We do not consider $t\bar tH$
production because its leading contribution comes from the tree level top
Yukawa operator.  Our approach is based on the MadGraph5\_aMC@NLO (MG5\_aMC)
framework \cite{Alwall:2014hca}.  It is part of the ongoing efforts of
automating NLO EFT simulations for colliders
\cite{Zhang:2016snc,Degrande:2014tta,Franzosi:2015osa,Zhang:2016omx,
Bylund:2016phk,Maltoni:2016yxb,Degrande:2016dqg,Degrande:2018fog},
and is a first step towards including NLO EW corrections.

Formally, the top loop contributions in Higgs measurements are part of the NLO
EW corrections to the dim-6 operators. One might think that the effects must be
small relative to the tree level contributions from other EW and Higgs
operators.  However, they could be important because the top quark operators
enter for the first time at the one-loop level, and in this sense these are
leading order (LO) contributions.  Therefore it is important to know their
sizes.  An interesting and similar example in the Higgs sector is that one can
set bounds on the Higgs self coupling $\lambda_3$, by using the
$\lambda_3$-dependent EW corrections, which enter the single Higgs boson cross
section starting at the one-loop level
\cite{McCullough:2013rea,Gorbahn:2016uoy,Degrassi:2016wml,
Bizon:2016wgr,DiVita:2017eyz}.  The problem we consider in this work is in
analogy.
The one-loop contribution of some top operator, $O_t$, relative to
the tree level ones from another, say, Higgs operator $O_H$, are proportional
to the ratio of their coefficients, i.e.~$\mathcal{O}(\alpha_{EW}C_t/C_H)$
instead of just $\mathcal{O}(\alpha_{EW})$. Given that the current constraints
on the Higgs operators are in general much stronger than those on the top
operators, it is likely that the $C_t/C_H$ factor enhances the top-loop induced
contributions, so that they are of more physical relevance.  This is one of the
reasons why we would like to focus on the top-quark operators at one loop
instead of the regular NLO EW corrections to Higgs and EW operators, as the
latter are naively of order $\mathcal{O}(\alpha_{EW})$, and therefore less
important.  Another reason is that there are processes that are loop-induced in
the SM, such as $gg\to H$ and $H\to \gamma\gamma$, and for them the top-loop
induced dim-6 contributions are not small.

The above arguments also apply to other non-top operators, which could
	enter Higgs processes at one-loop. While their effects could be potentially
	interesting, in this work we are mostly interested in top and Higgs
	physics, and so we start with the interplay of these two classes of
	operators. This is a first step justified by the
	fact that the top quark has the strongest coupling to Higgs. If non-top
	operators contribute to Higgs processes at one loop, it is more likely that these
	effects are better constrained by processes without a Higgs, to
	such a level that their loop contributions are not important in
	Higgs measurements.  This is not the case for top-quark operators due to
the large Yukawa coupling of the top quark.

The collider sensitivity to the loop contributions from effective operators 
may depend on our assumptions.
Let us briefly comment on this. Once including one-loop top operator contributions,
the cross section (or decay width) of the Higgs boson takes the following form:
\begin{flalign}
	\sigma=C_H(\mu_{EFT})\sigma_\mathrm{tree}+C_t\frac{\alpha_{EW}}{\pi}\left(
	\log\frac{Q^2}{\mu_{EFT}^2}\sigma_\mathrm{log} + \sigma_\mathrm{fin}\right)
	\label{eq:logfin}
\end{flalign}
where $C_{H}(\mu_{EFT})$ is the coefficient of some Higgs operator, $O_H$,
that enters the process at the tree level and $\mu_{EFT}$ is the scale at which it
is defined.  $C_t$ is the coefficient of some top-quark operator $O_t$ which mixes into
$O_H$.  $Q^2$ is the energy of the process.  A measurement of $\sigma$ will give
us information about a linear combination of $C_H$ and $C_t$.  
Even though one cannot immediately infer the constraint on $C_t$,
this piece of information itself is already useful, as it can be combined with
other measurements, and eventually the degeneracy will be lifted.
It is however important to have all measured quantities in such a fit expressed
in terms of operator coefficients defined at a common scale $\mu_{EFT}$, as
one can clearly see from Eq.~(\ref{eq:logfin}), the linear combination of $C_H$
and $C_t$ depends on the scale $\mu_{EFT}$.

In this work, as a first step, we will focus on studying the collider
sensitivity to the loop effects.  We do not perform any global fit,
and we will ignore the $C_H$ coefficient, except for two Higgs operators which
we will use as examples to demonstrate that distinguishing between tree-level
and loop-level contributions is in principle possible only by using Higgs
data. The consequence of neglecting $C_H$ is that the experimental sensitivity
on $C_t$ depends on the scale $\mu_{EFT}$ at which $C_H$ is set to 0.  We
consider two options:
\begin{itemize}
	\item 
Take $\mu_{EFT}=\Lambda$.  The underlying assumption is that new physics
effects at high scale are mainly captured by top-quark operators.  The large
scale $\mu_{EFT}$ in this case can be considered as a proxy of renormalisation
group (RG) running and mixing effects to the scale of measurement.
The contributions from $C_t$ will be relatively large due to the logarithmic
terms, leading to relatively tighter limits.  This however does not mean that
the finite terms are not important, see for example discussions in
\cite{Hartmann:2015oia,Zhang:2016omx,Maltoni:2016yxb}.  The disadvantage of
this approach is that resulting limits rely on strong assumptions, and in
particular it is difficult to combine these limits 
with other analyses, as the assumptions at scale $\Lambda$ can be different.
\item Take $\mu_{EFT}^2=Q^2$, where $Q$ is the scale of the measurement.
This is a bottom-up point of view, where the coefficient $C_H$ is assumed to have
already evolved down to scale $Q^2$ to absorb the $\sigma_\mathrm{log}$
contributions.  The resulting sensitivity will become weaker, but it is a fair
estimate of the expected sensitivity in a global analysis.  As we will discuss
later, this is because the finite terms are crucial for discriminating the
top-loop induced effects from the tree-level contributions of other Higgs
operators, which cannot be avoided in a real bottom-up global SMEFT fit.
\end{itemize}
In this work we will present results for both options. For the first
we will take $\mu_{EFT}=\Lambda=1$ TeV and for the second $\mu_{EFT}=m_H=125$ GeV.  
Besides, we will also show that, even with Higgs operators
included, by combining observables at different energies or using differential
measurements, it is possible to lift the degeneracy between top and Higgs
operators.  This is because the finite term $\sigma_{fin}/\sigma_{tree}$ is process
dependent, unlike the logarithmic term $\sigma_{log}/\sigma_{tree}$.

While we are mostly interested in LHC physics, for completeness
we will also discuss the same effects at possible
future lepton colliders.  An $e^+e^-$ collider is an ideal machine for
determining possible deviations in the Higgs sector.  Several proposals of such
Higgs factories have been made, including the Circular Electron Positron
Collider (CEPC) in China \cite{CEPC-SPPCStudyGroup:2015csa}, the Future
Circular Collider with $e^+e^-$ (FCC-ee) at CERN \cite{Gomez-Ceballos:2013zzn},
and the International Linear Collider (ILC) in Japan \cite{Baer:2013cma}.  The
Compact Linear Collider (CLIC) at CERN \cite{CLIC:2016zwp} could also run at
higher center-of-mass energies.  The precision on Higgs signal strengths at
these machines could reach $\mathcal{O}(1\%)-\mathcal{O}(0.1\%)$ level, and
therefore one has to carefully investigate possible loop contributions from
deviations in the top-quark sector.  We will show that our results imply 
that future lepton colliders will be sensitive to top-quark couplings even
below the $t\bar t$ threshold.

This paper is organised as follows.  In Section~\ref{sec:op} we discuss the
relevant effective operators in this study.  In Section~\ref{sec:calc} we
briefly outline our calculation strategy, implementation and validation, and in
particular discuss the renormalisation scheme.  We present our major numerical
results in Section~\ref{sec:num}.  Section~\ref{sec:imp} is devoted to a
discussion of the physics implications of our results, including impacts at the
LHC, at future lepton colliders, the potential sensitivities at high-luminosity
LHC, possible improvements by using differential distributions, and the
possibilities to discriminate between tree-level and loop-level operator
contributions.  In Section~\ref{sec:conc} we conclude.

\section{Operators}
\label{sec:op}
We consider the effective Lagrangian at dim-6
\begin{flalign}
	\mathcal{L}_{EFT}=\mathcal{L}_{SM}+\sum_i\frac{C_i}{\Lambda^2}O_i+\dots
\end{flalign}
where we consider CP-even operators only.\footnote{CP-odd top-quark operators could enter at one loop
but they will interfere with the SM contribution only in CP-odd observables,
and will not affect the main production and decay rates of the Higgs.}
Two classes of operators are relevant in this work.  The first is the set of
two-fermion top-quark operators that could enter Higgs measurements via loop
effects, including \cite{Grzadkowski:2010es}
\begin{align}\nonumber
	&O_{t\varphi}
	=\bar{Q} t\tilde\varphi\: (\varphi^{\dagger}\varphi)+h.c.
	,
	&O^{(1)}_{\varphi Q} =\FDF (\bar{Q}\gamma^\mu Q)
	,\\ \nonumber
	&O^{(3)}_{\varphi Q}
	=\FDFI (\bar{Q}\gamma^\mu\tau^I Q)
	,
	&O_{\varphi t}
	=\FDF (\bar{t}\gamma^\mu t)
	,\\ \nonumber
	&O_{\varphi tb} =(\tilde\varphi^\dagger iD_\mu\varphi)
	(\bar{t}\gamma^\mu b)+h.c.
	,
	&O_{tW} =(\bar{Q}\sigma^{\mu\nu}\tau^It)\:\tilde{\varphi}W_{\mu\nu}^I+h.c.
	,\\ 
	&O_{tB}
	=(\bar{Q}\sigma^{\mu\nu} t)\:\tilde{\varphi}B_{\mu\nu}+h.c.
	,
	\label{ops:1}
\end{align}
and we define
\begin{align}
	&O^{(+)}_{\varphi Q}\equiv
	\frac{1}{2}\left(  O^{(1)}_{\varphi Q}+O^{(3)}_{\varphi Q}\right)
	&O^{(-)}_{\varphi Q}\equiv
	\frac{1}{2}\left(  O^{(1)}_{\varphi Q}-O^{(3)}_{\varphi Q}\right),
\end{align}
so that the operators $O^{(+)}_{\varphi Q}$ and $O^{(-)}_{\varphi Q}$
modify the $Zbb$ and $Ztt$ couplings respectively.  
The coefficient of $O^{(+)}_{\varphi Q}$ is constrained by the LEP experiment
\cite{Patrignani:2016xqp}, but we include it for a complete study of the 
loop-induced sensitivity.  Four-fermion operators could have similar loop-effects
in Higgs measurements, but we will leave them for future study.
The chromo-dipole operator,
\begin{equation}
	O_{tG}=\left(\bar Q\sigma^{\mu\nu}T^At\right)\tilde\varphi G^A_{\mu\nu}\,
\end{equation}
enters $ggH$ at one loop.  Since its contribution has been studied up to
two loops \cite{Deutschmann:2017qum}, we will not include it in this study.
When presenting the physics impact in Section~\ref{sec:imp}, the $ggH$ loop
will be included as it is the dominant production channel, but we will only
consider the contribution from $O_{t\varphi}$, which is a simple rescaling factor.

The above operators are the main objects of this study.  To correctly take into
account their impact on Higgs measurements, RG
running and mixing effects
\cite{Jenkins:2013zja,Jenkins:2013wua,Alonso:2013hga}
need to be considered.  We thus introduce the second set of operators that enter
the same processes at the tree level and will
provide the corresponding counter terms:
\begin{align}
	&O_{\varphi WB}
	=\varphi^\dagger\tau^I\varphi W^I_{\mu\nu}B^{\mu\nu}
	,
	&O_{\varphi W}
	=\varphi^\dagger \varphi W^I_{\mu\nu}W^{I\mu\nu}
	,\nonumber\\
	&O_{\varphi B}
	=\varphi^\dagger \varphi B_{\mu\nu}B^{\mu\nu}
	,
	&O_{\varphi \square}
	=\left( \varphi^\dagger\varphi \right)\square
	\left( \varphi^\dagger\varphi \right)
	,\nonumber\\
	&O_{\varphi D}
	=\left( \varphi^\dagger D^\mu\varphi \right)^*
	\left( \varphi^\dagger D_\mu \varphi \right)
	,
	&O_{W}
	=iD^\mu\varphi^\dagger\tau^ID^\nu\varphi W^I_{\mu\nu}
	,\nonumber\\
	&O_{B}
	=iD^\mu\varphi^\dagger D^\nu\varphi B_{\mu\nu}
	,
	&O_{b\varphi}=(\varphi^\dagger\varphi)\bar Qb\varphi
	,\nonumber\\
	&O_{\mu\varphi}=(\varphi^\dagger\varphi)\bar l_2e_2\varphi,
	&O_{\tau\varphi}=(\varphi^\dagger\varphi)\bar l_3e_3\varphi,
	\label{ops:2}
\end{align}
where the subscripts 2,3 for the lepton doublet $l$ and singlet $e$ are flavour
indices.  The above operators are sufficient to provide all mixing counter terms
needed in this study to guarantee physically meaningful results at the loop
level.  Note that in the Warsaw basis \cite{Grzadkowski:2010es}, top quark
operators could mix into light-fermion operators, in particular the ones
that involve EW gauge bosons.  
This is slightly inconvenient for a study of the loop-induced Higgs couplings,
as some of the counter terms manifest as light-fermion interactions.
Fortunately, these effects turn out to be universal, in the sense that they can
be captured by dim-6 operators which involve SM bosons only, up to suitable
field redefinitions \cite{Wells:2015uba}.
For this reason, instead of introducing these light-fermion operators in our
basis, we follow \cite{Hagiwara:1993ck} and include  $O_{W}$ and $O_{B}$ with a
slightly different convention.  This means we replace the following two
combinations of Warsaw basis operators
\begin{flalign}
	&O^{(3)}_{\varphi q}+O^{(3)}_{\varphi l},
	\\
	&\frac{1}{6}O^{(1)}_{\varphi q}-\frac{1}{2}O^{(1)}_{\varphi l}
	+\frac{2}{3}O_{\varphi u}-\frac{1}{3}O_{\varphi d}-O_{\varphi e},
\end{flalign}
by $O_{W}$ and $O_{B}$, using the equations of motion.  The counter terms provided
by them are equivalent, but the physical interpretation is more clear.
Note that these two operators project out the flat directions in the
precision EW tests \cite{Grojean:2006nn,Brivio:2017bnu}, and so in this basis it
is clear that the precision constraints only apply to two operators,
i.e.~$O_{\varphi WB}$ and $O_{\varphi D}$, which is convenient for our 
analysis.  Note also that in this basis the mixing pattern of the EW operators
becomes different than in the Warsaw basis.

Throughout the paper we will refer to the first class (Eq.~\ref{ops:1}) as top operators, and the
second (Eq.~\ref{ops:2}) as Higgs operators.

\section{Calculation and renormalisation}
\label{sec:calc}
In this section we briefly describe our calculation for the EW corrections in
Higgs processes and precision EW observables from top-quark operators.
All relevant Feynman diagrams are shown in Figure~\ref{fig:diagram}.

At dim-6 at one loop, of all the operators we study,
the $gg\to H$ production and $H\to gg$ decay channels
receive a contribution only from $O_{t\varphi}$, which is easy to include with a
rescaling factor.  All other processes, except for $H\to bb$, share the same
kinds of contributions from top-loop operators, shown as the large grey circles
in Figure~\ref{fig:diagram}.  Thus a very efficient way to obtain results is to
implement all these contributions in MG5\_aMC \cite{Alwall:2014hca}, with the
help of FeynRules \cite{Alloul:2013bka}, and use the reweighting functionality
\cite{Mattelaer:2016gcx} in MG5\_aMC to compute the dim-6 loop contributions. The
reweighting is particularly simple in this case because there are no real
corrections.  For the loop computation,
we work in the Feynman gauge.  Gauge fixing is done following
	\cite{Dedes:2017zog}, in a way similar to the SM, that cancels the
Goldstone-gauge boson mixing and leads to SM-like propagators.
In addition we need to provide the corresponding
electroweak UV and R2 counter terms.  The R2 counter terms need to be provided
only for the $HH$, $VV$ two-point functions and $HVV$ three-point functions.
These are computed by using FeynArts \cite{Hahn:2000kx} interfaced with
FeynCalc \cite{Mertig:1990an,Shtabovenko:2016sxi}.  For terms involving
$\gamma^5$ we follow the scheme of
\cite{Kreimer:1989ke,Korner:1991sx,Kreimer:1993bh}, and have checked that our
results for the SM pieces agree with Ref.~\cite{Shao:2011tg}. The UV counter
terms come from the renormalisation of the theory, which we will describe in
the following subsections.  In particular, the UV counter terms needed for our
purpose are $HH$, $VV$, $HVV$, $ffV$, and $ffH$.

Finally, $h\to bb$ has a unique contribution from $W-t$ loops, not shared with
other channels. We compute it by using FeynArts
and FormCalc \cite{Hahn:1998yk}.  The renormalisation is similar to the other
channels. For the contribution from the $O_{\varphi tb}$, our result for the
finite part agrees with that of Ref.~\cite{Alioli:2017ces}.
We have also repeated our calculation in the $R_\xi$ gauge, and checked that
the results are $\xi$ independent.

The Yukawa operators, $O_{t\varphi}$, $O_{b\varphi}$, $O_{\mu\varphi}$
and $O_{\tau\varphi}$ can change the quark and lepton masses already at the
tree level.  In the on-shell mass scheme these effects should be canceled by
redefining the SM Yukawa terms, which is equivalent to making the following
replacement:
\begin{flalign}
	\varphi^\dagger\varphi\to \varphi^\dagger\varphi-\frac{v^2}{2}
\end{flalign}
in their definitions, where $v$ is the Higgs vev, i.e.~they only represent
deviations from the SM Yukawa terms. In our calculation we will use the
definition after this replacement, and do not consider the dim-6 shift
to fermion masses.

Throughout the calculation, we assume that the CKM matrix is identity.
	We are interested in the main decay channels of the Higgs boson, where
	quark flavor changing effect does not play a role.  Moreover, in
	top-quark loops, any flavor changing effects are suppressed by two
powers of the off-diagonal component of the CKM matrix. For these reasons we
believe that an identity CKM matrix is a good approximation for our purpose.

We have validated our implementation by computing the $H\to 4l$ and $H\to 2l2\nu$
decay processes and comparing with FormCalc. The implementation provides a
simulation tool for all processes shown in Figure~\ref{fig:diagram}, allowing
us to generate events associated with weights corresponding to dim-6 top-loop
contributions, apart from $H\to b\bar b$.  Both total rates as well as
differential distributions can be efficiently obtained from the weighted
events.  Note that this implementation can also be used to compute other
non-Higgs processes involving dim-6 top loops, such as $Z$-pole processes as
well as Drell-Yan at the LHC, provided that no additional counter terms are
needed.

\subsection{Dim-6 renormalisation}
\label{sec:dim6ren}
The dim-6 operator coefficients can be renormalised using the $\overline{MS}$
scheme
\begin{flalign}
	&C_i\Rightarrow Z_{ij}C_j =C_i+ \delta Z_{ij}C_j,
	\\
	&\delta Z_{ij}
	=\frac{\alpha}{2\pi}\Delta(\mu_{EFT})\frac{1}{\epsilon}\gamma_{ij}
	\label{eq:Z}
\end{flalign}
with $\delta Z_{ij}$ the anomalous dimension matrix, which has been obtained in
Refs.~\cite{Jenkins:2013zja,Jenkins:2013wua,Alonso:2013hga} in the Warsaw
basis.  Under our operator basis, the relevant terms in matrix $\gamma_{ij}$
are:
\begin{equation}
	\begin{array}{|c|lllllll|}
		\hline
		& O_j=O_{\varphi t} 
		& O_{\varphi Q}^{(+)}
		& O_{\varphi Q}^{(-)}
		& O_{\varphi tb}
		& O_{tW}
		& O_{tB}
		& O_{t\varphi}
		\\\hline
		O_i=O_{\varphi WB}
		& \frac{1}{3\sw\cw}
		& \frac{1}{3\sw\cw}
		& -\frac{1}{6\sw\cw}
		& 0
		& -\frac{5\yt}{2e\cw }
		& -\frac{3\yt}{2e\sw }
		& 0
		\\[5pt]
		O_{\varphi D}
		& -6\frac{\yt^2}{e^2}
		& 3\frac{\yt^2-\yb^2}{e^2}
		& 3\frac{\yt^2-\yb^2}{e^2}
		& -6\frac{\yt\yb}{e^2}
		&0&0&0
		\\[5pt]
		O_{\varphi\square}
		& -\frac{3}{2}\frac{\yt^2}{e^2}
		& -\frac{3\yt^2+6\yb^2}{2e^2}
		& \frac{6\yt^2+3\yb^2}{2e^2}
		& 3\frac{\yt\yb}{e^2}
		&0&0&0
		\\[5pt]
		O_{\varphi W}
		& 0 & \frac{1}{4\sw^2} & -\frac{1}{4\sw^2}
		& 0 & \frac{3\yt}{2e\sw} &0 &0
		\\[5pt]
		O_{\varphi B}
		& \frac{1}{3\cw^2}
		& \frac{1}{12\cw^2}
		& \frac{1}{12\cw^2}
		&0 &0
		&\frac{5\yt}{2e\cw} &0
		\\[5pt]
		O_{W}
		&0 &\frac{1}{e\sw} & -\frac{1}{e\sw}
		&0 &0 &0 &0
		\\[5pt]
		O_{B}
		&\frac{4}{3e\cw}
		&\frac{1}{3e\cw}
		&\frac{1}{3e\cw} &0 &0 &0 &0
		\\[5pt]
		O_{b\varphi}
		&0&-\frac{\yb}{2\cw^2}
		& \yb\frac{-4\lambda+3\yt^2+7\yb^2}{4e^2}
		&\frac{3\yt}{4\sw^2}
		&\frac{\yt\yb}{2e\sw}&0&\frac{3\yt\yb}{4e^2}
		\\
		&&+\yb\frac{8\lambda-3\yt^2-5\yb^2}{4e^2}
		&&-\yt\frac{2\lambda+\yt^2-6\yb^2}{2e^2}
		&&&
		\\[5pt]
		O_{\mu\varphi}
		&0&-\frac{3y_\mu(\yt^2+\yb^2)}{2e^2}
		&\frac{3y_\mu(\yt^2+\yb^2)}{2e^2}
		&\frac{3\yt\yb y_\mu}{e^2}
		&0&0&\frac{3\yt y_\mu}{2e^2}
		\\[5pt]
		O_{\tau\varphi}
		&0&-\frac{3y_\tau(\yt^2+\yb^2)}{2e^2}
		&\frac{3y_\tau(\yt^2+\yb^2)}{2e^2}
		&\frac{3\yt\yb y_\tau}{e^2}
		&0&0&\frac{3\yt y_\tau}{2e^2}
		\\\hline
	\end{array}
		\nonumber
\end{equation}
with
\begin{flalign}
	\Delta(x)\equiv\Gamma(1+\epsilon)\left( \frac{4\pi \mu^2}{x^2} \right)^\epsilon\ ,
\end{flalign}
and the operator coefficients are defined at the scale $\mu_{EFT}$.
Here the Yukawa couplings are defined by
	$y_{t,b}\equiv \sqrt{2}m_{t,b}/v$.

In this work, we slightly modify the $\overline{MS}$ scheme, by introducing the
oblique parameters $S$ and $T$ as renormalisation conditions. 
This is done by modifying Eq.~(\ref{eq:Z}):
\begin{flalign}
	\delta Z_{ij}
	=\frac{\alpha}{2\pi}\Delta(\mu_{EFT})\left(
	\frac{1}{\epsilon}+\Delta_{ij}\right)\gamma_{ij}
\end{flalign}
for $O_i=O_{\varphi WB},O_{\varphi D}$, and fixing $\Delta_{ij}$ by
requiring that the dim-6 contribution to $S$ and $T$ up to one loop is exactly the
measured value.  This can be done because we are using a basis
which, apart from the top-quark operators that
are not relevant in precision EW tests, includes only oblique operators. Therefore $S$ and $T$ 
can be defined as the outcome of a global fit for the EW sector, under the
oblique assumption, 
as described in \cite{Patrignani:2016xqp}.  At the tree level, they
correspond exactly to the coefficients of $O_{\varphi WB}$ and $O_{\varphi D}$.

The main reason for doing this, is that we are interested in the top-quark loop
effects in the directions that do not lead to severe inconsistency with
precision EW observables.  This implies that $C_{\varphi WB}$ and $C_{\varphi
D}$ should always take the values that minimise the inconsistency between
top-loop effects and precision EW measurements.  A complete global fit for the
EW sector is required to fully address this problem \cite{Zhang:2012cd}, but in
this work we take a simplified approach.  We assume BSM effects are dominated
by operators in Eqs.~(\ref{ops:1})-(\ref{ops:2}), i.e.~they are oblique, and so
the oblique parameters summarise the main constraints from precision EW
observables.  The $S$ and $T$ parameters can be used to fix the values of
$C_{\varphi WB}$ and $C_{\varphi D}$, by simply setting $S=T=0$.  This approach
implicitly assumes that gauge boson self-energy corrections from top-quark
loops are approximately linear functions of $q^2$, which is not strictly true,
but is enough for a sensitivity study.

In the $\overline{MS}$ scheme, $S=T=0$ does not imply that we can simply drop
$C_{\varphi WB}(\mu_{EFT})$ and $C_{\varphi D}(\mu_{EFT})$ in our analysis.
Instead, due to the top-quark operators contributing to $S$ and $T$ at the loop
level, this approximation implies that we need to set the coefficients of
$C_{\varphi WB}(\mu_{EFT})$ and $C_{\varphi D}(\mu_{EFT})$ to values that
exactly cancel the contributions from top-quark operators at one loop.  These
coefficients will then give other contributions in other Higgs processes.  A
more convenient way to take these additional contributions into account, is
simply to use $S$ and $T$ as renormalisation conditions, so that the
renormalised values for $C_{\varphi WB}$ and $C_{\varphi D}$ correspond to the
physical values of $S$ and $T$, and so with our approximation they can be 
excluded from the analysis.  The physical results are always independent of
renormalisation scheme at the order of this calculation.  The numerical
results of $S$ and $T$ parameters in terms of top operator coefficients are
given in Section~\ref{sec:num}.

\subsection{SM renormalisation}
The SM parameters are renormalised in the on-shell scheme, following
\cite{Denner:1991kt}.  In particular, the following parameters are split into
renormalised quantities and renormalisation constants:
\begin{flalign}
	&e_0=(1+\delta Z_e)e,
	\\
	&M_{W,0}^2=M_W^2+\delta M_W^2
	\\
	&M_{Z,0}^2=M_Z^2+\delta M_Z^2
	\\
	&M_{H,0}^2=M_H^2+\delta M_H^2.
\end{flalign}
Wave function renormalisation is defined as follows:
\begin{flalign}
	W_0^{\pm}&=\left( 1+\frac{1}{2}\delta Z_W \right)W^\pm
	\\
	Z_0&=\left( 1+\frac{1}{2}\delta Z_{ZZ} \right)Z+\frac{1}{2}\delta Z_{ZA}A
	\\
	A_0&=\frac{1}{2}\delta Z_{AZ}Z +\left( 1+\frac{1}{2}\delta Z_{AA} \right)A
	\\
	H_0&=\left( 1+\frac{1}{2}\delta Z_H \right)H.
\end{flalign}
The renormalisation of the tadpole does not show up in any
counter term relevant in this calculation, so we do not consider it here.

To determine the dim-6 contributions in the renormalisation constants, we
compute the two-point functions of the gauge bosons and the Higgs boson, from
the top-quark operators.  They can be written as
\begin{flalign}
	\bar\Sigma_{VV}^{(6)}(q^2)&=\Sigma_{VV}^{(6)}(q^2)+\Sigma_{VV}^{CT}(q^2)
	\nonumber\\
	&=\sum_{i} C_i\Sigma_{VV}^{i,loop}(q^2)
	+\sum_{i,j} \delta Z_{ji}C_i\Sigma_{VV}^{j,tree}(q^2)
\end{flalign}
where in the last line, index $i$ runs over all top-quark operators, and $j$
runs over all Higgs operators that are needed to provide counter terms.
$\Sigma_{VV}^{i/j,loop/tree}(q^2)$ are the corresponding operator contributions
to the $VV$ transverse two-point function.  The Higgs-boson self energy
$\bar\Sigma_{HH}$ is defined similarly. The dim-6 contributions in the
renormalisation constants (denoted by the superscript $(6)$) are determined by
these two-point functions, namely
\begin{equation}
	\begin{array}{ll}
	\delta M_W^{2(6)}=\Re\bar\Sigma_{WW}^{(6)}(M_W^2),\qquad
	&
	\delta M_Z^{2(6)}=\Re\bar\Sigma_{ZZ}^{(6)}(M_Z^2),
	\\
	\delta M_H^{2(6)}=\Re\bar\Sigma_{HH}^{(6)}(M_H^2),\qquad
	&\delta Z_W^{(6)}=-\Re\left.\frac{\partial\bar\Sigma_{WW}^{(6)}(k^2)}
	{\partial k^2}\right|_{k^2=M_W^2},
	\\
	\delta Z_{ZZ}^{(6)}=-\Re\left.\frac{\partial\bar\Sigma_{ZZ}^{(6)}(k^2)}
	{\partial k^2}\right|_{k^2=M_Z^2},\qquad
	&
	\delta Z_{AA}^{(6)}=-\left.\frac{\partial\bar\Sigma_{AA}^{(6)}(k^2)}
	{\partial k^2}\right|_{k^2=0},
	\\
	\delta Z_{AZ}^{(6)}=-2\Re\frac{\bar\Sigma_{AZ}^{(6)}(M_Z^2)}{M_Z^2},\qquad
	&
	\delta Z_{ZA}^{(6)}=2\frac{\bar\Sigma_{AZ}^{(6)}(0)}{M_Z^2},
	\\
	\delta Z_{H}^{(6)}=-\Re\left.\frac{\partial\bar\Sigma_{HH}^{(6)}(k^2)}
	{\partial k^2}\right|_{k^2=M_H^2},\qquad
	&
	\delta Z_{e}^{(6)}=
	\frac{1}{2}\left.\frac{\partial\bar\Sigma_{AA}^{(6)}(k^2)}{\partial
	k^2}\right|_{k^2=0}-\frac{\sw}{\cw}\frac{\bar\Sigma^{(6)}_{AZ}(0)}{M_Z^2}.
\end{array}
\end{equation}
In particular, the renormalized $Z$ and $A$ two-point functions
	become diagonal if the external lines are on their mass shell.
The last equation defines the electric charge at zero momentum transfer.  For
the operators we have included, there are no large logarithmic terms arising
from the $b$-quark mass, and so there is no need to define the running electric charge to
resum the logs, as we only need the dim-6 EW contributions from the top loop.
Note that the explicit expressions for $\Sigma_{VV}^{(6)}(q^2)$ have been
given in Ref.~\cite{Zhang:2012cd}, with an overall minus sign due to different conventions. In $H\to b\bar b$, the $b$-quark mass is renormalized
	in a similar way.  The expressions for renormalisation constants
	can be found in Ref.~\cite{Denner:1991kt}, and therefore we do not
repeat them here.

The above renormalisation constants would be sufficient to determine all the
relevant counter terms, if we used $\alpha$, $M_Z$ and $M_W$ as the input
parameters.  Conventionally, EW corrections are often computed with
$\alpha$, $M_Z$ and $G_F$ as input parameters.  In our case, however, it is
more convenient to use $M_W$, $M_Z$ and $G_F$ instead. This is because $M_W$
enters the final state phase space, and we do not want its mass to depend on
the dim-6 coefficients, because it would be particularly inconvenient for
using the reweighting technique.  To switch to the $M_W$, $M_Z$ and $G_F$ scheme,
we use
\begin{equation}
	M_W^2\left( 1-\frac{M_W^2}{M_Z^2} \right)=
	\frac{\pi\alpha}{\sqrt{2}G_F}\left( 1+\Delta r \right),
\end{equation}
where the $\Delta r$ contribution from the top-quark operators is
\begin{flalign}
	\Delta_r^{(6)}=\left.\frac{\partial\bar\Sigma_{AA}^{(6)}(k^2)}
	{\partial k^2}\right|_{k^2=0}
	-\frac{c_W^2}{s_W^2}\left( 
	\frac{\bar\Sigma_{ZZ}^{(6)}(M_Z^2)}{M_Z^2}
	-\frac{\bar\Sigma_{WW}^{(6)}(M_W^2)}{M_W^2}
	\right)+\frac{\bar\Sigma^{(6)}_{WW}(0)-\bar\Sigma_{WW}^{(6)}(M_W^2)}{M_W^2}.
\end{flalign}
One can simply define
\begin{flalign}
	\bar\alpha=\alpha(1+\Delta r)
\end{flalign}
so that the tree level relation between $M_W$, $M_Z$, $G_F$, and $\alpha$ holds.
To switch to the new scheme, one just needs to modify the renormalisation
constant for $\alpha$:
\begin{equation}
	\delta \bar \alpha^{(6)}=\delta \alpha^{(6)}
	+(\alpha-\bar \alpha)_{dim-6}.
\end{equation}

\subsection{Counter terms}
The counter term Feynman rules are determined by the renormalisation constants
and the RG mixing matrix.  For example, consider the $HZZ$ vertex, whose
tree level Feynman rule is
\begin{equation}
	\Gamma_{HZZ}^{\mu\nu}=i \frac{eM_W}{\sw\cw^2}g^{\mu\nu}
	+\sum_{i}C_i\Gamma_{HZZ,i}^{\mu\nu},
\end{equation}
where the subscript $i$ covers all Higgs operators that have a contribution,
including $O_{\varphi WB}$, $O_{\varphi W}$, $O_{\varphi B}$, $O_{\varphi D}$,
$O_{W}$, and $O_{B}$.  The corresponding dim-6 counter term has two parts.  The
first is the dim-6 contribution in the SM renormalisation, given by
\begin{flalign}
i \frac{eM_W}{\sw\cw^2}g^{\mu\nu}
\left[ \delta Z_e^{(6)}
	+\frac{\delta M_W^{2(6)} }{2M_W^2}
	-\frac{\delta \sw^{(6)}}{\sw}
	-\frac{2\delta \cw^{(6)}}{\cw}
	+\delta Z_{ZZ}^{(6)}+\frac{1}{2}\delta Z_{H}^{(6)}
\right]
\end{flalign}
where
\begin{flalign}
	\delta \cw&=\frac{\cw}{2}\left( \frac{\delta M_W^2}{M_W^2}
	-\frac{\delta M_Z^2}{M_Z^2}\right)
	\\
	\delta \sw&=-\frac{\cw}{\sw}\delta \cw
\end{flalign}
while the second is from the RG mixing between dim-6 coefficients, given by
\begin{equation}
	\sum_{i,j}C_j\delta Z_{ij}\Gamma_{HZZ,i}^{\mu\nu}
\end{equation}
where $i$ runs over the Higgs operators and $j$ runs over the top-quark operators.

\section{Numerical results}
\label{sec:num}
We are now ready to present the numerical results of our computation.  We use the
following input parameters:
\begin{flalign}
	&M_Z=91.1876\ \mbox{GeV},\quad M_W=80.385\ \mbox{GeV},
	\\
	&G_F=1.16638\times 10^{-5}\ \mbox{GeV}^{-2},\quad M_H=125\ \mbox{GeV},
	\\
	&M_t=172.5\ \mbox{GeV},\quad M_b=4.7\ \mbox{GeV}.
\end{flalign}
As we have explained in the introduction, all channels are computed for two
cases: $\mu_{EFT}=M_H=125$ GeV and $\mu_{EFT}=\Lambda=1000$ GeV.  Results will
be displayed as ratios w.r.t to LO SM predictions:
\begin{flalign}
	\mu_i\equiv
	\frac{\sigma_i}{\sigma^\mathrm{SM}}
	\ \mbox{or} \ 
	\frac{\Gamma_i}{\Gamma^\mathrm{SM}}
	\equiv 1+\sum_j C_j\left( \frac{1\ \mbox{TeV}^2}{\Lambda^2} \right)
	\mu_{ij}
	\label{eq:mu}
\end{flalign}
where $i$ denotes channel, $j$ runs over all contributing operators.
$\mu_{ij}$ is the relative deviation in channel $i$ from operator $j$ if
$C_j/\Lambda^2=1/\mbox{TeV}^2$.  $\mu_{ij}$ for all decay processes are given
in percentage, in Table~\ref{tab:res1}, and for all production processes at the
LHC at 13 TeV and future lepton colliders are given in
Tables~\ref{tab:res2}-\ref{tab:res5}.  For lepton colliders we consider 250 and
350 GeV centre-of-mass energies, to cover possible scenarios at CEPC, FCC-ee,
and ILC, which are planned to collect data at both 240$\sim$250 GeV and 350 GeV
\cite{cepc,fcc,Fujii:2015jha,Barklow:2015tja}.  We
present results for two beam polarisation configurations $P(e^+,e^-)=(\pm 30\%,
\mp 80\%)$.  In the 350 GeV case, we only consider the $O_{t\varphi}$
contributions, as the sensitivity to other operators is unlikely to compete with
the direct production modes \cite{Durieux:2017gxd,Durieux:2018tev}.
In all processes we set the renormalisation and factorisation scale to $M_H$.
For $VH$ production no cuts are applied, whilst for VBF we apply the following jet 
cuts at the LHC: 
\begin{equation}
p_T^j>20\, \textrm{GeV}, |y_j|<5, |y_{j1}-y_{j2}|>3, M_{jj}>130\, \textrm{GeV}. 
\end{equation}No cuts are applied for the lepton collider results. 
In both cases the Higgs and vector bosons are not decayed.  
We note here that for $ZH$  production at the LHC we 
consider only the $q\bar{q}$ contribution. The impact of top operators on the
gluon fusion contribution, $gg\to ZH$ have already been considered in
\cite{Bylund:2016phk,Englert:2016hvy}. 

Note that these results are computed with our modified
$\overline{MS}$ scheme.  For completeness, in Appendix~\ref{sec:app1} we also
present results computed with the standard $\overline{MS}$ scheme. 

\begin{table}
\begin{tabular}{|ll|lllllll|}
		\hline
		channel & $\mu_\mathrm{EFT}$ [GeV]
		& $O_{\varphi t}$
		& $O_{\varphi Q}^{(+)}$
		& $O_{\varphi Q}^{(-)}$
		& $O_{\varphi tb}$
		& $O_{tW}$
		& $O_{tB}$
		& $O_{t\varphi}$
		\\\hline
		$H\to bb$ & 125
		&-0.15 & -0.06 & 0.24 & -1.13 & -0.28 & 0 & -0.18
		\\
		$H\to bb$ & 1000
		&0.79 & 0.54 & -1.25 & -8.16 & 0.34 & 0 & 0.29
		\\\hline
		$H\to \mu\mu,\tau\tau$ & 125
		&-0.15 & 0.001 & 0.15 & 0 & 0 & 0 & -0.27
		\\
		$H\to \mu\mu,\tau\tau$ & 1000
		&0.79 & 0.002 & -0.79 &0 & 0 & 0 & 0.68
		\\\hline
		$H\to \gamma\gamma$ & 125
		&-3.37&5.86&2.64&0&-56.4&-117.9&3.45
		\\
		$H\to \gamma\gamma$ & 1000
		&6.95& 16.2& -2.52& 0 & 14.0& 101.3& 3.45
		\\\hline
		$H\to Z\gamma$ & 125
		& 0.51 & 2.20 & 2.74 &0&-39.5& 14.0 & 0.72
		\\
		$H\to Z\gamma$ & 1000
		&4.35 & 6.04 &0.83 &0&33.9&-51.6&0.72
		\\\hline
		$H\to Zll$ & 125
		&-0.54& -0.10 & 0.56& -0.00& 0.19& -0.06& 0.08
		\\
		$H\to Zll$ & 1000
		&0.33& 0.74& -1.25& -0.06& 0.05& 0.33& 0.08
		\\\hline
		$H\to Wl\nu$ & 125
		&-0.15& -0.24& 0.38& 0.00& -0.13& 0& -0.03
		\\
		$H\to Wl\nu$ & 1000
		&0.79& 0.63& -1.42& -0.05& 0.33& 0& -0.03
		\\\hline
\end{tabular}
\caption{Percentage deviation $\mu_{ij}$ for decay channel $i$ and operator $j$. }
\label{tab:res1}
\end{table}

\begin{table}
\begin{tabular}{|ll|lllllll|}
		\hline
		channel & $\mu_\mathrm{EFT}$ [GeV]
		& $O_{\varphi t}$
		& $O_{\varphi Q}^{(+)}$
		& $O_{\varphi Q}^{(-)}$
		& $O_{\varphi tb}$
		& $O_{tW}$
		& $O_{tB}$
		& $O_{t\varphi}$
		\\\hline
		$p p \to ZH$ & 125
                 & -0.30 & 0.21 & 0.21 & -0.00 & 1.00 & -0.06 & -0.02
		\\
		$p p \to ZH$ & 1000
                 & 0.57 & 0.11 & -0.66 & -0.06 & -2.75 & -0.44 & -0.02
		\\\hline
		$p p \to WH$ & 125
                 & -0.15 & -0.04 & 0.19 & 0.00 & 0.43 & 0 & -0.21
		\\
		$p p \to WH$ & 1000
                 & 0.79 & -0.27 & -0.52 & -0.05 & -4.08 & 0 & -0.21 
		\\\hline
		$p p \to H j j$ & 125
                 & -0.26 & -0.24 & 0.51 & 0.01 & 0.05 & -0.00 & 0.03
		\\
		$ p p \to H j j$ & 1000
                 &   0.68 & 0.94 & -1.61 & -0.04 & 0.28 & -0.00 & 0.03
		\\\hline
\end{tabular}
\caption{Percentage deviation $\mu_{ij}$ for production channel $i$ and
operator $j$ at LHC 13 TeV. }
\label{tab:res2}
\end{table}

\begin{table}
\begin{tabular}{|ll|lllllll|}
		\hline
		channel & $\mu_\mathrm{EFT}$ [GeV]
		& $O_{\varphi t}$
		& $O_{\varphi Q}^{(+)}$
		& $O_{\varphi Q}^{(-)}$
		& $O_{\varphi tb}$
		& $O_{tW}$
		& $O_{tB}$
		& $O_{t\varphi}$
		\\\hline
		$e^+ e^- \to ZH$ & 125
                 & -0.40 &  -0.21  & 0.22  & -0.00	& 1.82 & -0.25 & 0.01
		\\
		$e^+ e^- \to ZH$ & 1000
                 & 0.78 & -0.10 & -0.71 & -0.05 & -2.71 & 0.62 & 0.01
		\\\hline
		$e^+ e^- \to H \nu \nu $ & 125
		& -0.15 & -0.26 & 0.41 & 0.01 & -0.08 & 0 & -0.01
		\\
		$e^+ e^- \to H \nu \nu$ & 1000
		& 0.79 & 0.76 & -1.55 & -0.04 & 0.13 & 0 & -0.01
		\\\hline
		$e^+ e^- \to H e^+ e^-$ & 125
		& -0.51 & -0.27 & 0.56 & 0.00 & 0.72 & 0.79 & 0.08
		\\
		$ e^+ e^- \to H e^+ e^-$ & 1000
		& 0.28 & 0.76 & -1.50 & -0.05 & 0.77 & -0.71 & 0.08 
		\\\hline
\end{tabular}
\caption{Percentage deviation $\mu_{ij}$ for production channel $i$ and operator $j$ for 250 GeV and $P(e^+,e^-)=(30\%,-80\%)$. }
\label{tab:res3}
\end{table}

\begin{table}
\begin{tabular}{|ll|lllllll|}
		\hline
		channel & $\mu_\mathrm{EFT}$ [GeV]
		& $O_{\varphi t}$
		& $O_{\varphi Q}^{(+)}$
		& $O_{\varphi Q}^{(-)}$
		& $O_{\varphi tb}$
		& $O_{tW}$
		& $O_{tB}$
		& $O_{t\varphi}$
		\\\hline
		$e^+ e^- \to ZH$ & 125
                 & -0.44 & 0.36 & 0.55 & -0.01 & -0.62 & 0.17 & 0.06
		\\
		$e^+ e^- \to ZH$ & 1000
                 & 0.00 & 1.14 & -1.42 & -0.06 & -1.35 & -2.35 & 0.06
		\\\hline
		$e^+ e^- \to H \nu \nu $ & 125
		& -0.15 & -0.26 & 0.41 & 0.01 & -0.01 & 0 & -0.01
		\\
		$e^+ e^- \to H \nu \nu$ & 1000
		& 0.79 & 0.76 & -1.55 & -0.04 & 0.01  & 0 & -0.01
		\\\hline
		$e^+ e^- \to H e^+ e^-$ & 125
		&-0.62 & 0.14 & 0.66 & -0.01 & 0.32 & 1.40 & 0.05
		\\
		$ e^+ e^- \to H e^+ e^-$ & 1000
		& 0.30 & 0.95& -1.08 & -0.06 & -0.60 & -0.85  & 0.05
		\\\hline
\end{tabular}
\caption{Percentage deviation $\mu_{ij}$ for production channel $i$ and operator $j$ for 250 GeV and $P(e^+,e^-)=(-30\%,80\%)$. }
\label{tab:res4}
\end{table}

\begin{table}
\centering
\begin{tabular}{|l|ll|}

		\hline
		channel $O_{t\varphi}$ & P(30\%,-80\%) & P(-30\%,80\%)
		\\\hline
		$e^+ e^- \to ZH$ &
                   -0.15  & 0.01
		\\\hline
		$e^+ e^- \to H \nu \nu $ 
		& -0.01 & -0.01
		\\\hline
		$e^+ e^- \to H e^+ e^-$ 
		& 0.10 & 0.08
		\\\hline
\end{tabular}
\caption{Percentage deviation $\mu_{ij}$ for production channel $i$ and operator $O_{t\varphi}$ for 350 GeV and $P(e^+,e^-)=(\pm 30\%,\mp 80\%)$.  The results are identical at any $\mu_\mathrm{EFT}$. }
\label{tab:res5}
\end{table}

From Tables~\ref{tab:res1}-\ref{tab:res5} we see that loop-induced decay
channels $H\to \gamma\gamma$ and $H\to Z\gamma$ can show large deviations
from the SM.  For all other processes, the relative deviation caused by a top
operator with $C_j/\Lambda^2=1/\mbox{TeV}^2$ is around the percent level.  One
also observes that $\mu_{EFT}=1000$ GeV results are in general larger
than the $\mu_{EFT}=125$ GeV ones, due to the logarithmic terms, but $\mu_{EFT}=125$ GeV
deviations are not negligible at all.  There are also entries where the $\mu_{EFT}=125$ GeV
result is larger.

Another observation is that these results depend on the scheme of input
parameters used in the calculation.  For example, if we used the $\alpha$,
$M_Z$ and $G_F$ scheme, the operator $O_{tB}$ would give a contribution to
$H\to Wl\nu$ and $pp\to HW$, because its contribution to the $ZZ$ two-point
function could affect the $Z$ mass, which in turn enters $g_w$ and $M_W$.  But
these contributions vanish if we use the $M_W$, $M_Z$ and $G_F$ input scheme,
because the renormalisation condition for $G_F$ will fix the renormalisation
constant of $g_w$ so that it only depends on the $WW$ two-point function, to
which $O_{tB}$ does not contribute. The same applies to $O_{\varphi t}$,
i.e.~$O_{\varphi t}$ gives no contribution in $H\to Wl\nu$ and $pp\to HW$
if the standard $\overline{MS}$ scheme is used.  However it does give a
contribution to the $T$ parameter, and because we use $T$ as one of the
renormalisation conditions to fix $C_{\varphi D}$, a non-zero contribution
arises due to $O_{\varphi D}$ modifying the Higgs wave-function.  In
general, most results we have obtained so far will depend on the input scheme
used in the calculation. This is also why we want to include
constraints from precision EW tests, so that the relation between $\alpha$,
$M_W$, $M_Z$ and $G_F$ is not significantly modified.  In Table~\ref{tab:res6}
we present results for $S$, $T$ and $U$ in the original $\overline{MS}$ scheme.
They are parameterised by 
\begin{flalign}
	S \equiv \sum_j C_j\left( \frac{1\ \mbox{TeV}^2}{\Lambda^2} \right)
	S_j\,,
\end{flalign}
and similarly for $T$, $U$. Note that $U$ is finite and is scheme independent~\cite{
Greiner:2011tt}.
\begin{table}
	\centering
\begin{tabular}{|ll|lllllll|}
		\hline
		 & $\mu_\mathrm{EFT}$ [GeV]
		& $O_{\varphi t}$
		& $O_{\varphi Q}^{(+)}$
		& $O_{\varphi Q}^{(-)}$
		& $O_{\varphi tb}$
		& $O_{tW}$
		& $O_{tB}$
		& $O_{t\varphi}$
		\\\hline
		S & 125
		&-0.017 & 0.029 & 0.013 & 0 & 0.084 & 0.095 & 0
		\\
		S & 1000
		& 0.035 & 0.081 & -0.013 & 0 & -0.504 & -0.565 & 0
		\\\hline
		T & 125
		&-0.186 & 0.022 & 0.165 &  0.003 & 0 & 0 &0
		\\
		T & 1000
		& 1.016 & -0.579 & -0.436 & 0.036 & 0 & 0 &0
		\\\hline
		U & 125
		&0.010 & -0.048 & -0.016 & 0.001 & 0.090 & 0 & 0
		\\
		U & 1000
		&0.010 & -0.048 & -0.016 & 0.001 & 0.090 & 0 & 0
		\\\hline
\end{tabular}
\caption{Deviation in $S$, $T$ and $U$ parameters due to top operators.}
\label{tab:res6}
\end{table}

The precision EW tests contain more information than $S$, $T$ and $U$.  This
information is lost in the $STU$ formalism simply because the two-point
functions of gauge bosons are approximated as linear functions of the momentum
squared.  A complete analysis for precision EW data without using these oblique
parameters can be performed to obtain better constraints, see
Ref.~\cite{Zhang:2012cd} for more details.  Since a global fit is not the goal
of this study, in this paper we will only use the oblique parameters to simplify
the analysis of the precision EW part, but one should keep in mind that in
principle more information can be obtained.

We briefly demonstrate that physical results are scheme-independent.
Consider $H\to \gamma\gamma$ with $\mu_{EFT}=125$ GeV as an example. Taking the
standard $\overline{MS}$ results in Table~\ref{tab:res11} in
Appendix~\ref{sec:app1}, we have the following dim-6 contributions to the
width:
\begin{flalign}
	\Gamma_{\gamma\gamma}^{(6)}
	=\Gamma_{\gamma\gamma}^{\mathrm{SM}}
	(1+25.74C_{\varphi WB}-0.733 C_{tW}-1.367 C_{tB}+0.0345 C_{t\varphi})
	\label{eq:1}
\end{flalign}
where we have used that $O_{\varphi WB}$ gives the following
$H\to\gamma\gamma$ amplitude:
\begin{equation}
	4\frac{C_{\varphi WB}\sw\cw}{\Lambda^2}\left( 
	\frac{M_H^2}{2}\epsilon_1\cdot \epsilon_2
	-(p_1\cdot\epsilon_2) (p_2\cdot \epsilon_1)\right).
\end{equation}
Following Table~\ref{tab:res6}, the $S$ parameter is given by
\begin{equation}
	S=-12.9C_{\varphi WB}
	-0.017 C_{\varphi t} + 0.029 C_{\varphi Q}^{(+)}
	+ 0.013 C_{\varphi Q}^{(-)} + 0.084 C_{tW} + 0.095 C_{tB}\,
	\label{eq:2}
\end{equation}
where the coefficient of $C_{\varphi WB}$ is given by
\begin{equation}
	\alpha_{EW} S=\frac{C_{\varphi WB}4v^2\sw\cw}{\Lambda^2}\ .
\end{equation}
Combining Eqs.~(\ref{eq:1}) and (\ref{eq:2}) and eliminating $C_{\varphi WB}$,
we find the expression of $\Gamma^{(6)}_{\gamma\gamma}$ for a fixed $S$ parameter,
which agrees with results in Table~\ref{tab:res1}.  One can easily check the
same relation holds for $\mu_{EFT}=1000$ GeV and for $H\to \gamma Z$.

Finally, we note here that our results in Tables \ref{tab:res1}-\ref{tab:res6}
provide the $\mathcal{O}(1/\Lambda^2)$ deviation from the SM predictions from
the top operators.  Our computational setup allows us to also obtain the
$\mathcal{O}(1/\Lambda^4)$ contribution coming from squaring the 1-loop
contributions. As an example in Table \ref{tab:res7} we show the percentage
deviation from the SM induced at $\mathcal{O}(1/\Lambda^4)$ by the top
operators for $C_i=1$ and $\Lambda=1$ TeV for $ZH$ production at the LHC.  We
find that these high-order contributions are suppressed by about two orders of
magnitude compared to the $\mathcal{O}(1/\Lambda^2)$ contributions in Table
\ref{tab:res2}. Given the current limits on the operator coefficients, we can
safely ignore these contributions. A similar behaviour is expected also for the
other processes studied in this work, except for loop-induced processes in the
SM, i.e.~$gg\to H$, $H\to \gamma\gamma,\gamma Z,gg$.  For these processes, the
SM contributions are suppressed, so dim-6 squared terms can be potentially
important.  Their quadratic dependence on the operator coefficients 
will be kept in Section~\ref{seq:i1} and
\ref{seq:i2}, where we only use current limits to constrain top-quark
operators.

\begin{table}
\begin{tabular}{|ll|lllllll|}
		\hline
		channel & $\mu_\mathrm{EFT}$ [GeV]
		& $O_{\varphi t}$
		& $O_{\varphi Q}^{(+)}$
		& $O_{\varphi Q}^{(-)}$
		& $O_{\varphi tb}$
		& $O_{tW}$
		& $O_{tB}$
		& $O_{t\varphi}$
		\\\hline
		$p p \to ZH$ & 125
                 & 0.0014 & 0.0078 & 0.0041 & $4.0\times 10^{-8}$ & 0.0085 & 0.0012 & $3.7 \times 10^{-5}$
		\\
		$p p \to ZH$ & 1000
                 & 0.0022 & 0.0062 & 0.0047 & $7.5 \times  10^{-6}$ & 0.037 & 0.0054 & $3.7 \times 10^{-5}$
		\\\hline
\end{tabular}
\caption{Percentage deviation at $\mathcal{O}(1/\Lambda^4)$ for $ZH$ production at LHC 13 TeV. }
\label{tab:res7}
\end{table}
 
\section{Physics implications}
\label{sec:imp}
In this section we discuss the impact of our numerical results.

\subsection{Impact on Higgs measurements at the LHC}
\label{seq:i1}
The first consequence of these loop contributions in Higgs measurements is that
they can shift the measured signal strengths at the LHC.  The current
constraints on top-quark operator coefficients are not very stringent compared
to Higgs operator constraints, and so within the allowed range, the
loop-induced contributions could potentially affect the signal strengths of all
Higgs channels.  Without knowing exactly the coefficients of top-quark
operators, these contributions can only be dealt with as theoretical
uncertainties, and should be considered in all Higgs measurements that are
analysed using a SMEFT approach.  These theoretical uncertainties cannot be
avoided in a bottom-up view of the SMEFT, as in general there is no strong
motivation to overlook a certain type of operators \cite{Jiang:2016czg}, in
particular at the energy scale of the measurements, where the RG effects will
take place and mix different types of operators.

\begin{table}
\begin{center}
\begin{tabular}{|llll|}
	\hline
	Operator & Top Fitter  &  RHCC &$\sigma_{t\bar{t}H}$ \cite{Maltoni:2016yxb} 
	\\\hline
	$C_{\varphi tb}$ & &   [-5.28,5.28] &  
	\\
	$C_{\varphi Q}^{(3)}$ & [-2.59,1.50] &  & 
	\\
	$C_{\varphi Q}^{(1)}$ & [-3.10,3.10] &&
	\\
	$C_{\varphi t}$ & [-9.78,8.18] &&
	\\
	$C_{tW}$ & [-2.49,2.49] & &
	\\
	$C_{tB}$ & [-7.09,4.68] &&
	\\
	$C_{t\varphi}$ &  &   &[-6.5,1.3]
	\\\hline
\end{tabular}
\end{center}
\caption{Individual limits on operator coefficients, from
the Top Fitter Collaboration \cite{Buckley:2015lku},
right-handed charged currents (RHCC) (tree-level only) \cite{Alioli:2017ces},
and $t\bar tH$ cross section \cite{Maltoni:2016yxb}.  $\Lambda=1$ TeV is assumed.
\label{tab:constraints}
}
\end{table}

To estimate the size of possible contributions from top operators, we consider
the current constraints on top-quark operator coefficients from direct
measurements. These constraints originate from processes where these operators
enter already at the tree level. First, the TopFitter collaboration performed a
global fit (excluding $O_{\varphi tb}$) at LO using both Tevatron and LHC data
for top production and decay \cite{Buckley:2015lku}.  Individual limits are
given for each operator, by setting other operator coefficients to zero.  In
addition, the $O_{\varphi tb}$ operator gives rise to right handed $Wtb$
coupling, which is constrained at tree-level by single top and top decay
measurements, and indirectly at loop-level by $B$ meson decay and $h\to b\bar
b$ \cite{Alioli:2017ces}.  Here we only use the direct limits. For the Yukawa
operator $\mathcal{O}_{t\varphi}$, we follow the approach in
Ref.~\cite{Maltoni:2016yxb}, and update the analysis with the recent 13 TeV
measurements in
Refs.~\cite{CMS:2017lgc,CMS:2017vru,Aaboud:2017jvq,Aaboud:2017rss}.  Again, we
do not use the loop-induced $gg\to H$ process.  In Table~\ref{tab:constraints}
we list these constraints. 

Using all the individual direct constraints and neglecting correlations, we can
approximately identify the allowed region in the 7-dimensional
parameter space at 95\% CL, by reconstructing a $\chi^2$.  We shift all the intervals so
that they are centred at zero. Within the range allowed by the
constraints, using the results presented in Table~\ref{tab:res1} and
Table~\ref{tab:res2}, we find the maximum and minimum percentage deviations in
signal strengths in all Higgs channels, taking
$\mu_{EFT}=125$ GeV, i.e.~the scale of the measurements.  These deviations are
given in Table~\ref{tab:errors}. For example, the interval shown in the 4th row
and 3rd column means that in $pp\to ZH,\ H\to bb$, the signal strength can be
shifted by $-7\%$ to $+6\%$ by top-quark operators within the current constraints.

\begin{table}[h]
	\centering
	\small
\begin{flalign}
\begin{array}{c|ccccccc}
\text{} & \gamma \gamma  & \text{$\gamma $Z} & \text{bb} & \text{WW}^*
   & \text{ZZ}^* & \tau \tau  & \mu \mu  \\\hline
 \text{gg} & \text{(-100$\%$,1980$\%$)} & \text{(-88$\%$,200$\%$)} &
   \text{(-40$\%$,48$\%$)} & \text{(-40$\%$,47$\%$)} & \text{(-40$\%$,46$\%$)} &
   \text{(-40$\%$,48$\%$)} & \text{(-40$\%$,48$\%$)} \\
 \text{VBF} & \text{(-100$\%$,1880$\%$)} & \text{(-88$\%$,170$\%$)} &
   \text{(-6.1$\%$,5.3$\%$)} & \text{(-6.8$\%$,6.7$\%$)} & \text{(-8.8$\%$,9.2$\%$)} &
   \text{(-6.2$\%$,5.9$\%$)} & \text{(-6.2$\%$,5.9$\%$)} \\
 \text{WH} & \text{(-100$\%$,1880$\%$)} & \text{(-88$\%$,170$\%$)} &
   \text{(-5.5$\%$,4.2$\%$)} & \text{(-6.1$\%$,5.6$\%$)} & \text{(-7.8$\%$,7.9$\%$)} &
   \text{(-5.8$\%$,5.1$\%$)} & \text{(-5.8$\%$,5.1$\%$)} \\
 \text{ZH} & \text{(-100$\%$,1880$\%$)} & \text{(-87$\%$,170$\%$)} &
   \text{(-6.5$\%$,5.9$\%$)} & \text{(-7.1$\%$,7.1$\%$)} & \text{(-9.4$\%$,9.9$\%$)} &
   \text{(-6.8$\%$,6.7$\%$)} & \text{(-6.8$\%$,6.7$\%$)} \\
\end{array}
\nonumber
\end{flalign}
\caption{Possible deviations in signal strengths, due to top-quark operators, in
	major Higgs production and decay channels at the LHC.  The top-quark
	operator coefficients are allowed to vary within the current
	constraints,
	described in Table~\ref{tab:constraints}.}
	\label{tab:errors}
\end{table}

We see that for the loop-induced processes, i.e.~those in the first row and the
first two columns, the deviations are large. This simply demonstrates that for
operators like $O_{tB}$, loop-induced constraints are much stronger than the
tree-level ones.  For example, the $gg\to H\to\gamma\gamma$ signal strength can
deviate by a factor of $\sim20$ and this is mainly driven by $|C_{tB}|\approx6$.  
It implies that this channel is sensitive
to $C_{tB}$ and can be used to place much stronger bounds compared with the
current ones.  The same applies to other loop-induced channels.

For the remaining tree-level Higgs channels, the impact of top-operators
through loops is in general weaker, but remains around $5\sim10\%$, and is not
negligible even for the current precision.  Although theory uncertainties of
this size may not significantly change the result of a Higgs coupling analysis,
they will become relevant from now on, and eventually, at the high luminosity
scenario, become an important component of theory uncertainties in a bottom-up
global SMEFT analysis.  

This also implies that once the precision of Higgs measurements goes beyond
$\sim10\%$, we can even hope to use Higgs measurements to place
useful constraints on top-quark operators.  In Section \ref{sec:hl} we discuss
this possibility by estimating the sensitivity in the high luminosity scenario
of LHC.

\subsection{Impact on Higgs measurements at the future lepton colliders}
\label{seq:i2}
As we have mentioned in the introduction, at lepton colliders, the estimated
precision of Higgs signal strength measurements could reach
$\mathcal{O}(1\%)-\mathcal{O}(0.1\%)$ level.  For the Higgs measurements to
make sense at this accuracy level, one has to check carefully the top-loop
induced contributions.  If the machine is planned to run at 350 GeV, it is
likely that the top-quark operator coefficients will be determined directly
through $t\bar t$ production, except for the $O_{t\varphi}$.  However, for the
CEPC case a 350 GeV run has not been officially planned, and therefore an
interesting question is by how much the top-quark operators could change the
Higgs cross sections below the $t\bar t$ threshold through loops,
given the current constraints.  In fact, in Table~\ref{tab:res3} we see that a
top operator with a coefficient of order 1/TeV$^2$ could already have visible
effects.

In Tables~\ref{tab:errorsLC}, \ref{tab:errorsLC2} and
\ref{tab:errorsLC3} we present the possible deviations at 
lepton colliders, in a way similar to Table~\ref{tab:errors} for the LHC.  
We consider two scenarios: 1) 250 GeV run only, and we allow all top
operators to vary within the current constraints. These results
are given in Table~\ref{tab:errorsLC}. 2) runs above $t\bar t$ threshold are
planned, which will fix all operator coefficients by direct production, except
for $C_{t\varphi}$, and so only $C_{t\varphi}$ is allowed to vary.
Corresponding results are shown in Tables~\ref{tab:errorsLC2} and
\ref{tab:errorsLC3}, for 250 and 350 GeV runs respectively.  All these
deviations are obtained with numerical results presented in
Table~\ref{tab:res3}.  For each process, two polarisations for $(e^+,e^-)$ are
considered.

\begin{table}[h]
	\centering
	\footnotesize
\begin{flalign}
\begin{array}{c|ccccccc}
 \text{} & \gamma \gamma  & \text{$\gamma $Z} & \text{bb} & \text{WW}^*
   & \text{ZZ}^* & \tau \tau  & \mu \mu  \\\hline
 \text{ZH(+30$\%$,-80$\%$)} & \text{(-100$\%$,1900$\%$)} &
   \text{(-87$\%$,160$\%$)} & \text{(-7.5$\%$,7.5$\%$)} &
   \text{(-8.3$\%$,8.6$\%$)} & \text{(-11$\%$,11$\%$)} &
   \text{(-8$\%$,8.3$\%$)} & \text{(-8$\%$,8.3$\%$)} \\
 \text{ZH(-30$\%$,+80$\%$)} & \text{(-100$\%$,1870$\%$)} &
   \text{(-88$\%$,180$\%$)} & \text{(-7.6$\%$,7.1$\%$)} &
   \text{(-8.1$\%$,7.9$\%$)} & \text{(-10$\%$,11$\%$)} &
   \text{(-7.6$\%$,7.3$\%$)} & \text{(-7.6$\%$,7.3$\%$)} \\
 \text{WWF(+30$\%$,-80$\%$)} & \text{(-100$\%$,1880$\%$)} &
   \text{(-88$\%$,170$\%$)} & \text{(-5.7$\%$,4.7$\%$)} &
   \text{(-6.5$\%$,6.2$\%$)} & \text{(-8.1$\%$,8.3$\%$)} &
   \text{(-5.9$\%$,5.3$\%$)} & \text{(-5.9$\%$,5.3$\%$)} \\
 \text{WWF(-30$\%$,+80$\%$)} & \text{(-100$\%$,1880$\%$)} &
   \text{(-88$\%$,170$\%$)} & \text{(-5.7$\%$,4.7$\%$)} &
   \text{(-6.5$\%$,6.2$\%$)} & \text{(-8.1$\%$,8.3$\%$)} &
   \text{(-5.9$\%$,5.3$\%$)} & \text{(-5.9$\%$,5.3$\%$)} \\
 \text{ZZF(+30$\%$,-80$\%$)} & \text{(-100$\%$,1790$\%$)} &
   \text{(-88$\%$,180$\%$)} & \text{(-11$\%$,8.6$\%$)} &
   \text{(-11$\%$,9.6$\%$)} & \text{(-13$\%$,12$\%$)} &
   \text{(-11$\%$,9$\%$)} & \text{(-11$\%$,9$\%$)} \\
 \text{ZZF(-30$\%$,+80$\%$)} & \text{(-100$\%$,1730$\%$)} &
   \text{(-88$\%$,180$\%$)} & \text{(-14$\%$,11$\%$)} &
   \text{(-14$\%$,12$\%$)} & \text{(-15$\%$,15$\%$)} &
   \text{(-14$\%$,11$\%$)} & \text{(-14$\%$,11$\%$)}
\end{array}
\nonumber
\end{flalign}
\caption{Possible deviations in signal strengths (in percent) caused by
	top-quark operators, in Higgs production (WWF for $WW$ fusion and ZZF
	for $ZZ$ fusion) and decay channels, at an $e^+e^-$ collider
	at 250 GeV centre-of-mass energy.  All
	top-quark operator coefficients are allowed to vary within the current
	constraints, described in Table~\ref{tab:constraints}.}
	\label{tab:errorsLC}
\end{table}

\begin{table}[h]
	\centering
	\footnotesize
\begin{flalign}
\begin{array}{c|ccccccc}
 \text{} & \gamma \gamma  & \text{$\gamma $Z} & \text{bb} & \text{WW}^*
   & \text{ZZ}^* & \tau \tau  & \mu \mu  \\\hline
 \text{ZH(+30$\%$,-80$\%$)} & \text{(-17$\%$,19$\%$)} &
   \text{(-7.3$\%$,7$\%$)} & \text{(-4$\%$,3.3$\%$)} &
   \text{(-4.5$\%$,4$\%$)} & \text{(-4.9$\%$,4.4$\%$)} &
   \text{(-3.6$\%$,3$\%$)} & \text{(-3.6$\%$,3$\%$)} \\
 \text{ZH(-30$\%$,+80$\%$)} & \text{(-17$\%$,19$\%$)} &
   \text{(-7.5$\%$,7.2$\%$)} & \text{(-4.1$\%$,3.5$\%$)} &
   \text{(-4.7$\%$,4.2$\%$)} & \text{(-5.1$\%$,4.6$\%$)} &
   \text{(-3.8$\%$,3.2$\%$)} & \text{(-3.8$\%$,3.2$\%$)} \\
 \text{WWF(+30$\%$,-80$\%$)} & \text{(-17$\%$,18$\%$)} &
   \text{(-7.2$\%$,7$\%$)} & \text{(-3.9$\%$,3.3$\%$)} &
   \text{(-4.4$\%$,3.9$\%$)} & \text{(-4.9$\%$,4.3$\%$)} &
   \text{(-3.5$\%$,2.9$\%$)} & \text{(-3.5$\%$,2.9$\%$)} \\
 \text{WWF(-30$\%$,+80$\%$)} & \text{(-17$\%$,18$\%$)} &
   \text{(-7.2$\%$,7$\%$)} & \text{(-3.9$\%$,3.3$\%$)} &
   \text{(-4.4$\%$,3.9$\%$)} & \text{(-4.9$\%$,4.3$\%$)} &
   \text{(-3.5$\%$,2.9$\%$)} & \text{(-3.5$\%$,2.9$\%$)} \\
 \text{ZZF(+30$\%$,-80$\%$)} & \text{(-17$\%$,19$\%$)} &
   \text{(-7.6$\%$,7.3$\%$)} & \text{(-4.2$\%$,3.6$\%$)} &
   \text{(-4.8$\%$,4.3$\%$)} & \text{(-5.2$\%$,4.7$\%$)} &
   \text{(-3.9$\%$,3.3$\%$)} & \text{(-3.9$\%$,3.3$\%$)} \\
 \text{ZZF(-30$\%$,+80$\%$)} & \text{(-17$\%$,19$\%$)} &
   \text{(-7.5$\%$,7.2$\%$)} & \text{(-4.1$\%$,3.5$\%$)} &
   \text{(-4.7$\%$,4.2$\%$)} & \text{(-5.1$\%$,4.6$\%$)} &
   \text{(-3.8$\%$,3.2$\%$)} & \text{(-3.8$\%$,3.2$\%$)} 
\end{array}
\nonumber
\end{flalign}
\caption{Possible deviations in signal strengths (in percent) caused by
	top-quark operators, in Higgs production (WWF for $WW$ fusion and ZZF
	for $ZZ$ fusion) and decay channels, at an $e^+e^-$ collider
	at 250 GeV centre-of-mass energy.  Only the coefficient
	of $O_{t\varphi}$ is allowed to vary within the current constraints,
	described in Table~\ref{tab:constraints}.} \label{tab:errorsLC2}
\end{table}

\begin{table}[h]
	\centering
	\footnotesize
\begin{flalign}
\begin{array}{c|ccccccc}
 \text{} & \gamma \gamma  & \text{$\gamma $Z} & \text{bb} & \text{WW}^*
   & \text{ZZ}^* & \tau \tau  & \mu \mu  \\\hline
 \text{ZH(+30$\%$,-80$\%$)} & \text{(-17$\%$,18$\%$)} &
   \text{(-6.7$\%$,6.4$\%$)} & \text{(-3.4$\%$,2.7$\%$)} &
   \text{(-3.9$\%$,3.3$\%$)} & \text{(-4.4$\%$,3.8$\%$)} &
   \text{(-3$\%$,2.4$\%$)} & \text{(-3$\%$,2.4$\%$)} \\
 \text{ZH(-30$\%$,+80$\%$)} & \text{(-17$\%$,19$\%$)} &
   \text{(-7.3$\%$,7.1$\%$)} & \text{(-4$\%$,3.4$\%$)} &
   \text{(-4.5$\%$,4$\%$)} & \text{(-5$\%$,4.4$\%$)} &
   \text{(-3.6$\%$,3$\%$)} & \text{(-3.6$\%$,3$\%$)} \\
 \text{WWF(+30$\%$,-80$\%$)} & \text{(-17$\%$,18$\%$)} &
   \text{(-7.3$\%$,7$\%$)} & \text{(-3.9$\%$,3.3$\%$)} &
   \text{(-4.5$\%$,3.9$\%$)} & \text{(-4.9$\%$,4.4$\%$)} &
   \text{(-3.6$\%$,2.9$\%$)} & \text{(-3.6$\%$,2.9$\%$)} \\
 \text{WWF(-30$\%$,+80$\%$)} & \text{(-17$\%$,18$\%$)} &
   \text{(-7.3$\%$,7$\%$)} & \text{(-3.9$\%$,3.3$\%$)} &
   \text{(-4.5$\%$,3.9$\%$)} & \text{(-4.9$\%$,4.4$\%$)} &
   \text{(-3.6$\%$,2.9$\%$)} & \text{(-3.6$\%$,2.9$\%$)} \\
 \text{ZZF(+30$\%$,-80$\%$)} & \text{(-17$\%$,19$\%$)} &
   \text{(-7.7$\%$,7.4$\%$)} & \text{(-4.3$\%$,3.7$\%$)} &
   \text{(-4.9$\%$,4.3$\%$)} & \text{(-5.3$\%$,4.8$\%$)} &
   \text{(-4$\%$,3.4$\%$)} & \text{(-4$\%$,3.4$\%$)} \\
 \text{ZZF(-30$\%$,+80$\%$)} & \text{(-17$\%$,19$\%$)} &
   \text{(-7.6$\%$,7.4$\%$)} & \text{(-4.3$\%$,3.7$\%$)} &
   \text{(-4.8$\%$,4.3$\%$)} & \text{(-5.2$\%$,4.7$\%$)} &
   \text{(-3.9$\%$,3.3$\%$)} & \text{(-3.9$\%$,3.3$\%$)}
\end{array}
\nonumber
\end{flalign}
\caption{Possible deviations in signal strengths (in percent) caused by
	top-quark operators, in Higgs production (WWF for $WW$ fusion and ZZF
	for $ZZ$ fusion) and decay channels, at an $e^+e^-$ collider
	at 350 GeV centre-of-mass energy.  Only the coefficient
	of $O_{t\varphi}$ is allowed to vary within the current constraints,
	described in Table~\ref{tab:constraints}.} \label{tab:errorsLC3}
\end{table}

In the first scenario,
apart from the large deviations in the loop-induced decay $\gamma\gamma$ and
$\gamma Z$, a $5-15\%$ level deviations are common in all channels.
These results suggest that the current sensitivities to top-quark couplings
can be already improved by up to an order of magnitude even with an $e^+e^-$
collider below the $t\bar t$ threshold.  In the second scenario, deviations
are smaller at $\sim5\%$, but are sufficient to further pin down the top
Yukawa coupling.  More reliable estimates of the potential constraints that can
be placed on top-quark couplings would require a global analysis including both
the Higgs and the top operators, which we will leave for future studies.

\subsection{Potential at high luminosity LHC}
\label{sec:hl}
Given that the precision of Higgs measurements will be largely improved at the
high luminosity LHC (HL-LHC), the potential $\sim5-10\%$ deviations in Higgs
measurements can be used to place constraints on the top-quark operator
coefficients.  This does not mean that we will have 7 more free parameters to fit
in the Higgs sector, because the top operators are also constrained by direct
top-quark measurements.  However, it is likely that one has to combine the two
sectors to obtain the correct exclusion limits on both top and Higgs operators.
To see if this is the case, in this section we will estimate the sensitivity of
dim-6 top-loop effects at the HL-LHC at 3000 fb$^{-1}$. 

To this end, we perform a $\chi^2$ analysis including all top operators but
fixing all the Higgs operator coefficients to zero, at $\mu_{EFT}=M_H=125$ GeV
and at $\mu_{EFT}=\Lambda=1000$ GeV.  As discussed in the introduction, the
first scale choice gives an estimate of sensitivity from a bottom-up point of
view, while the second is from a top-down point view taking into account RG
running and mixing effects.  In the $\chi^2$ analysis we assume the measured
values will be exactly the same as the SM predictions.  For the projection of
future signal strength measurements, we follow Ref.~\cite{Maltoni:2017ims},
where in Table~3 the statistical and systematic errors for $gg\to H$, VBF, $WH$
and $ZH$ production, with $ZZ^*$, $\gamma\gamma$, $WW^*$, $\tau\tau$, $\mu\mu$
decay channels are all documented.
We take one half of the theory errors, to
account for possible theory improvements in the future, and we have checked
that in any case the resulting sensitivities are affected by theory errors only
at the percent level.  QCD corrections are potentially important, but they
are likely to cancel in the signal strengths, when taking ratios w.r.t SM cross
sections, so we will not consider them.  For the $H\to bb$ channel we use
Ref.~\cite{ATL-PHYS-PUB-2014-011}, where we have assumed that two-lepton and
one-lepton channels correspond exactly to $ZH$ and $WH$ production.  Finally we
consider the $Z\gamma$ signal strength taken from \cite{ATL-PHYS-PUB-2014-006},
assuming that the production channel is dominated by $gg\to H$. 

Top-operator contributions to signal strengths can be easily computed using
results presented in Tables~\ref{tab:res1} and \ref{tab:res2}. 
We assume that the percentage deviations do not change much from 13 TeV
to 14 TeV. The modifications to the Higgs total width are taken into account.
A specific $X$-like
production channel may contain components from all five major production
mechanisms, including ggF, VBF, $WH$, $ZH$ and $ttH$.  Numerical results for
VBF, $WH$, $ZH$ are presented in Table~\ref{tab:res2}, while for the other two
channels we only need to take into account the leading effect from the top
Yukawa operator $O_{t\varphi}$, which rescales the total cross section.

As we have mentioned in Section~\ref{sec:dim6ren}, the consistency of precision
EW observables is important for the results to be scheme independent.  As
explained in Section~\ref{sec:calc}, to simplify the analysis we assume that
the $S$ and $T$ parameters are measured accurately and set them to zero. Thanks
to our renormalisation scheme, this approximation simply means that we can
exclude the operators $O_{\varphi WB}$ and $O_{\varphi D}$ from the analysis.
We however take into account the $U$ parameter which is scheme-independent up
to dim-6.  The current bound is $\pm0.1$, taken from the PDG
\cite{Patrignani:2016xqp}.

\begin{table}
	\centering
\begin{tabular}{|l|ccccccc|}
	\hline
	Operator &
	$C_{\varphi t}$& $C_{\varphi Q}^{(+)}$& $C_{\varphi Q}^{(-)}$& $C_{\varphi tb}$&
	$C_{tW}$ & $C_{tB}$ & $C_{t\varphi} $
	\\\hline
	$\mu_{EFT}=125$ GeV &
 2.5 & 1.3 & 3.2 & 9.3 & 0.2 & 0.07 & 0.9 \\
	$\mu_{EFT}=1000$ GeV &
 1.3 & 0.5 & 4.3 & 1.3 & 0.6 & 0.08 & 0.9 \\
	Current&
 9.0 & 5.1 & 5.1 & 5.3 & 2.5 & 5.9 & 3.9 
	\\\hline
\end{tabular}
\caption{Sensitivity of Higgs measurements at HL-LHC to top-quark operators,
	compared with current constraints.  Here sensitivity is defined as one
	half of the size of interval of coefficient $C_i$ at 95\% CL, assuming
	$\Lambda=1$ TeV, and all other operators coefficients are set to zero.
}	
	\label{tab:ind}
\end{table}

A total $\chi^2$ is constructed by using Higgs data plus the $U$ parameter.  We
truncate the $\chi^2$ at the quadratic order in $C$, the Wilson coefficients.
The one-sigma interval for any single parameter is given by $\Delta\chi^2=1$. The
individual sensitivities on operator coefficients are given in
Table~\ref{tab:ind}, where we compare the results of the two scenarios
$\mu_{EFT}=M_H=125$ GeV, and $\mu_{EFT}=\Lambda=1000$ GeV, and the current
constraints. Here, ``sensitivity'' is defined as one half of the size of
interval of coefficient $C_i$ at 95\% CL, assuming $\Lambda=1$ TeV, and all
other operators vanish.\footnote{
	We refrain from using words such as constraints, bounds, or limits because
	there is no data yet and because a real global fit would have to take
	into account the full set of Higgs operators in the same analysis.
The results we present here reflect the sensitivities of measurements to top
operators, but not the actual constraints.}
We can see that sensitivities on the first four
coefficients are comparable with the current direct measurements.  The last
three coefficients are even more tightly constrained.  This is mainly because they
enter $gg\to H$, $H\to \gamma\gamma$ and/or $H\to Z\gamma$, where the relative
deviations at dim-6 are large due to the loop suppression of the SM.
From Table~\ref{tab:ind} we can already conclude that Higgs measurements will
provide useful information on top-quark operators.

The individual sensitivities do not fully reflect the constraining power
of the combined analysis.  To better study the structure of the $\chi^2$,
we present the limits on each eigenstate of the quadratic terms of $\chi^2$.
This is done by writing
\begin{flalign}
	\chi^2=CMC^{T}
\end{flalign}
where 
\begin{equation}
	C=\left(\begin{array}{ccccccc}
		C_{\varphi t}& C_{\varphi Q}^{(+)}& C_{\varphi Q}^{(-)}& C_{\varphi tb}&
	C_{tW} & C_{tB} & C_{t\varphi}
\end{array}\right).
\end{equation}
By diagonalizing the matrix $M\to M^D$:
\begin{equation}
	\chi^2=CU^TM^DUC^T
\end{equation}
we find seven linear combinations of operator coefficients, given by
$UC^T$, which are statistically independent.  Then for the $i$th linear
combination, the $\chi^2$ is given by $M^D_{ii}(UC^T)_i^2$, so the one-sigma
limit on the $i$th combination will be ${M^D_{ii}}^{-\frac{1}{2}}$. 
For $\mu_{EFT}=125$ GeV, we find
\begin{flalign}
\left(
\begin{array}{ccccccc}
 -0.025 & 0.045 & 0.019 & 0.005 & -0.43 & -0.9 & -0.041 \\
 0.022 & -0.076 & -0.058 & -0.049 & 0.42 & -0.25 & 0.87 \\
 0.0012 & 0.18 & 0.068 & -0.033 & -0.77 & 0.35 & 0.49 \\
 0.26 & -0.91 & -0.26 & 0.0099 & -0.21 & 0.04 & 0.0099 \\
 0.42 & 0.27 & -0.54 & -0.68 & -0.002 & -0.0095 & -0.064 \\
 -0.48 & -0.26 & 0.41 & -0.73 & -0.0039 & 0.0076 & -0.021 \\
 0.72 & -0.00077 & 0.68 & -0.095 & 0.047 & -0.028 & -0.0092 \\
\end{array}
\right)
\nonumber\\
\times\frac{1\text{TeV}^2}{\Lambda^2}
\left( \begin{array}{ccccccc}
	C_{\varphi t}\\ C_{\varphi Q}^{(+)}\\ C_{\varphi Q}^{(-)}\\ C_{\varphi tb}\\ C_{tW} \\ C_{tB}\\ C_{t\varphi} 
\end{array} \right)
=\pm
\left( \begin{array}{c} 
	 0.0326 \\ 0.548 \\ 0.637 \\ 2.62 \\ 7.31 \\ 19.8 \\ 79.6 
\end{array}
\right)
\label{eq:eig}
\end{flalign}
We can see that all seven linear combinations can be constrained. Even though
the last one may be too weak to give meaningful constraints, we will
show that it can be significantly improved once differential distributions
are taken into account.  The first five numbers are all quite constraining.

\begin{figure}[ht]
	\centering
	\begin{minipage}{.49\linewidth}
		\includegraphics[width=1.\linewidth]{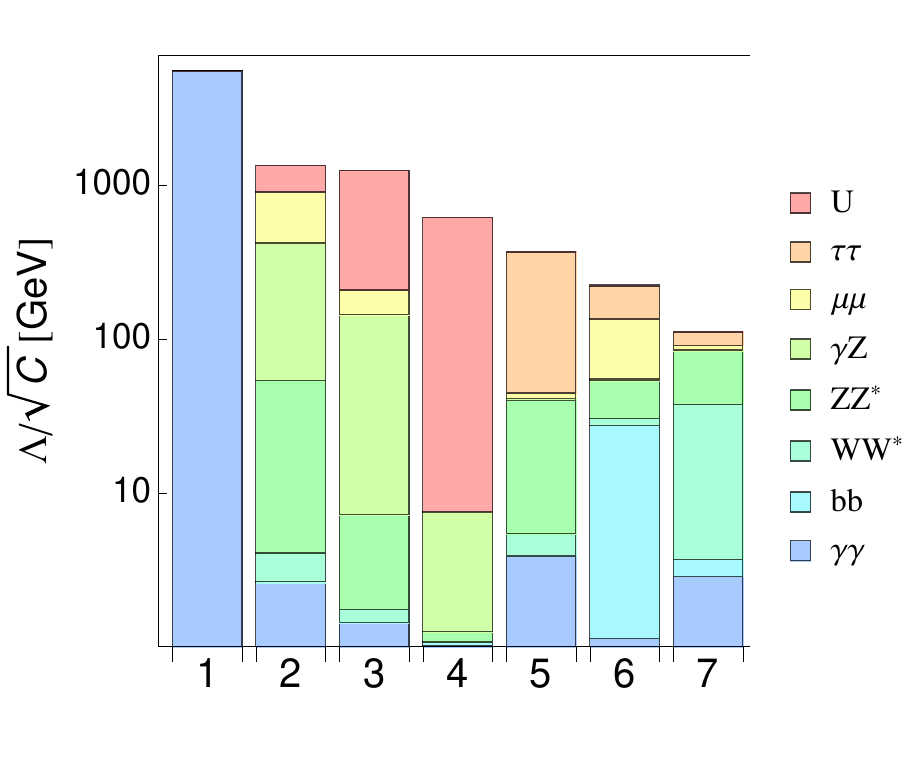}
	\end{minipage}
	\begin{minipage}{.49\linewidth}
		\includegraphics[width=1.\linewidth]{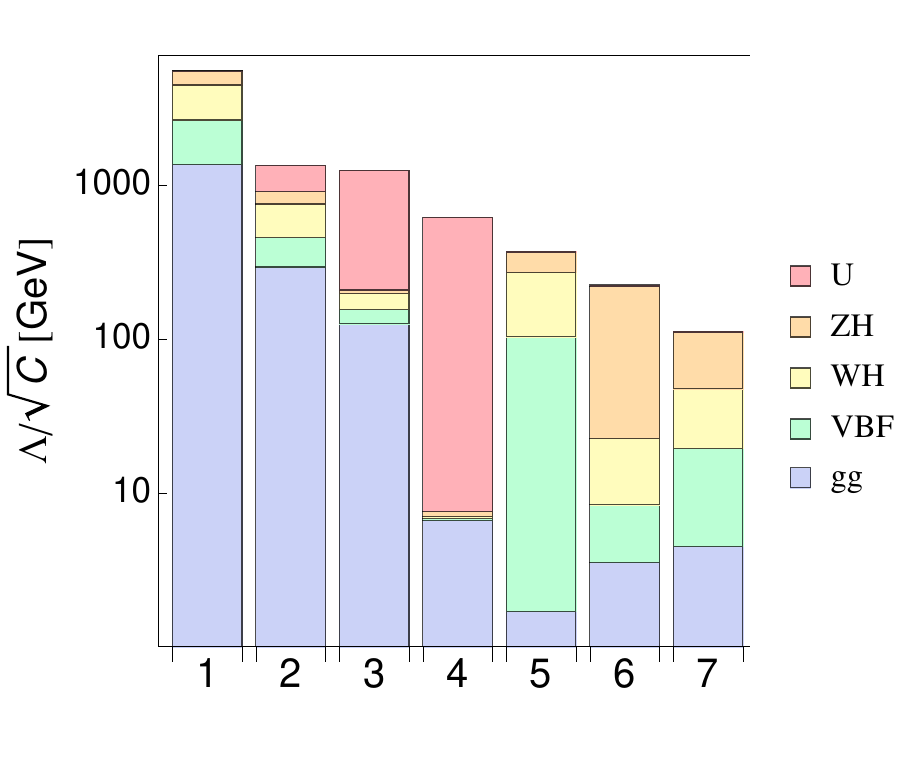}
	\end{minipage}
	\caption{Sensitivities for each eigenstate, and the relative
		contributions from each channel for $\mu_{EFT}=125$ GeV.}
	\label{fig:decay125}
\end{figure}

To see where exactly these constraints come from, for each of the eigenstates
given above, we compute the contribution from each measurement to the
eigenvalue of $\chi^2$.  Since the sensitivities
given in Eq.~(\ref{eq:eig}) are the inverse square root of these eigenvalues,
the fraction for each measurement in the eigenvalue reflects the relative
importance of that measurement in the direction that corresponds to the
eigenvectors. To have a physical intuition about what is happening, below we
give the most important operators in each of the seven eigenstates, and the
measurements that contribute the largest fractions to the corresponding
$\chi^2$:
\begin{flalign}
	\begin{array}{|c|c|c|}	
		\hline
		\text{Eigenstates} & \text{Coefficients} & \text{Channels}
		\\\hline
		1st&
		C_{tB}(81\%) &
		gg\to H \to \gamma\gamma\ (84\%)
		\\\hline
		2nd&
		C_{t\varphi}(75\%), C_{tW}(18\%) &
		gg\to H \to ZZ^*,\gamma Z,\mu\mu,WW^*\ (77\%)
		\\\hline
		3rd&
		C_{tW}(59\%),C_{t\varphi}(24\%)&
		gg\to H\to \gamma Z\ (42\%),\ U\ (25\%)
		\\\hline
		4th&
		C_{\varphi Q}^{(+)}(82\%)&
		U\ (68\%),\ gg\to H\to \gamma Z\ (28\%)
		\\\hline
		5th&
		C_{\varphi tb}(46\%),C_{\varphi Q}^{(-)}(29\%)&
		VBF\to \tau\tau,ZZ^*,\gamma\gamma\ (64\%),
		\\&&
		WH\to ZZ^*,\gamma\gamma\ (16\%)
		\\\hline
		6th&
		C_{\varphi tb}(53\%),C_{\varphi t}(23\%),C_{\varphi Q}^{(-)}(17\%)&
		ZH,WH\to b\bar b\ (59\%)
		\\&&
		gg\to H\to\mu\mu\ (17\%)
		\\\hline
		7th&
		C_{\varphi t}(52\%), C_{\varphi Q}^{(-)}(47\%)&
		ggF,VBF\to WW^* (49\%)
		\\&&
		WH,ZH\to \gamma\gamma (18\%)
		\\\hline
	\end{array}
	\nonumber
\end{flalign}
We can see, for example, the most constrained eigenvector, i.e.~the 1st one, is
mainly due to $H\to\gamma\gamma$. From Eq.~\ref{eq:eig} we see that it places a
constraint mainly on $C_{tB}$.  This is due to
its relatively large contributions to $H\to\gamma\gamma$.  Similarly, the
second one comes mostly from $gg\to H$.  The constraint is mostly on
$C_{t\varphi}$ as it enters the $ggH$ loop.  The third and the fourth are two
combinations of $U$ parameter and $gg\to H\to \gamma Z$, and from the
sensitivities we know that in the latter it is the $H\to \gamma Z$ that leads
to the bounds, on three most relevant operators.  Until this point, the most
useful information comes from processes that are loop-induced in the SM, from which
tighter constraints are expected.  The last three eigenvalues are dominated by
corrections to tree-level processes in the SM, and mainly constrain the first four
operators that give rise to vector-like $ttV$ and $tbW$ couplings.  They rely
mostly on the $VH$ and VBF production channels and $ZZ^*$, $WW^*$ and $b\bar b$
decay channels.  This information is also given in Figure~\ref{fig:decay125}
left, where the heights of the bars are the constraints of each eigenstate (in
terms of $\Lambda/\sqrt{C}$), and different colours indicate relative
contributions of different measurements to this constraint, taking into account
all decay channels and the $U$ parameter measurement. For each decay channel
all production modes are included. Similarly, the relative contributions from
all production channels (with all decay modes grouped together) and the $U$
parameter is given in Figure~\ref{fig:decay125} right.

\begin{figure}[ht]
	\centering
	\begin{minipage}{.49\linewidth}
		\includegraphics[width=1.\linewidth]{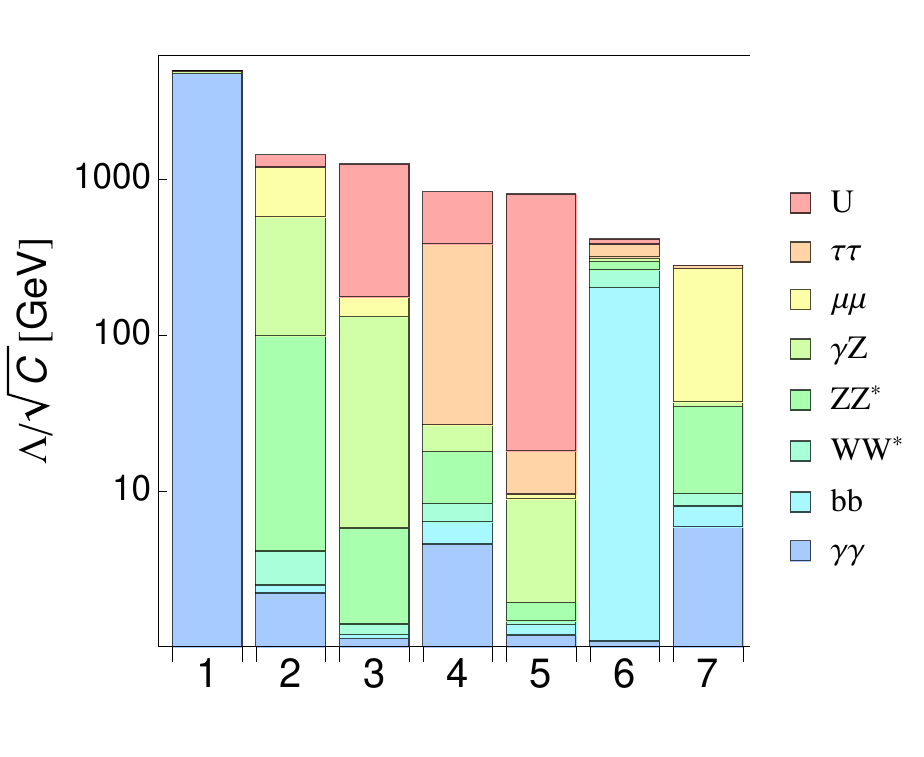}
	\end{minipage}
	\begin{minipage}{.49\linewidth}
		\includegraphics[width=1.\linewidth]{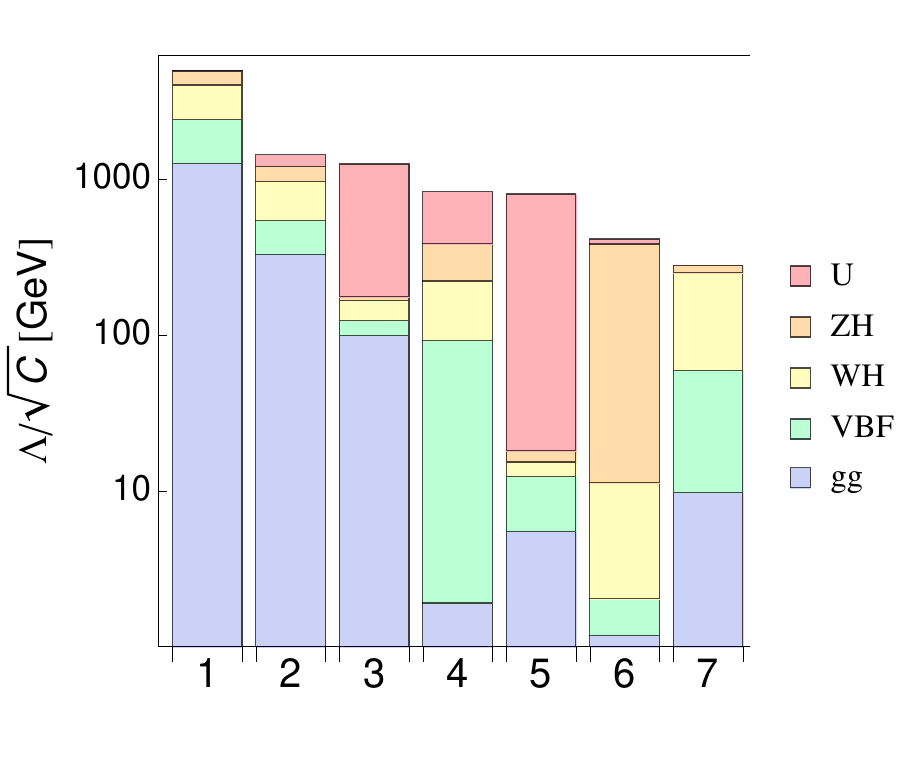}
	\end{minipage}
	\caption{Sensitivities for each eigenstates, and the relative
		contributions from each channel for $\mu_{EFT}=1000$ GeV.}
	\label{fig:decay1000}
\end{figure}

Following the same procedure for $\mu_{EFT}=1000$ GeV, we find the following eigenstates:
\begin{flalign}
\left(
\begin{array}{ccccccc}
 -0.061 & -0.15 & 0.016 & -0.045 & -0.13 & -0.98 & 0.055 \\
 -0.036 & -0.064 & 0.013 & -0.4 & -0.38 & 0.13 & 0.82 \\
 0.13 & 0.0054 & -0.018 & -0.17 & 0.9 & -0.099 & 0.36 \\
 0.11 & 0.46 & -0.091 & 0.77 & -0.048 & -0.087 & 0.4 \\
 -0.019 & -0.84 & -0.31 & 0.41 & 0.035 & 0.11 & 0.14 \\
 0.61 & 0.12 & -0.73 & -0.21 & -0.11 & -0.051 & -0.097 \\
 0.77 & -0.2 & 0.6 & 0.072 & -0.083 & 0.0012 & 0.004 
\end{array}
\right)
\nonumber\\
\times\frac{1\text{TeV}^2}{\Lambda^2}
\left( \begin{array}{ccccccc}
	C_{\varphi t}\\ C_{\varphi Q}^{(+)}\\ C_{\varphi Q}^{(-)}\\ C_{\varphi tb}\\ C_{tW} \\ C_{tB}\\ C_{t\varphi} 
\end{array} \right)
=\pm
\left( \begin{array}{c}
	 0.041 \\ 0.487 \\ 0.638 \\ 1.45 \\ 1.55 \\ 5.84 \\ 12.7
 \end{array} \right)
\label{eq:eig1000}
\end{flalign}
The sensitivities are in general better due to the log enhanced terms.
The most important channels for each eigenstate are slightly different.
We show the channel decomposition in Figure~\ref{fig:decay1000}.

\subsection{Improvements with differential distributions}
\label{sec:diff}

Higher dimensional operators typically lead to different differential
distributions than the SM, due to different Lorentz structures and sometimes an
$E^2/\Lambda^2$ enhancement.  This is expected also at the loop order, and
therefore to fully exploit this behaviour, we study differential observables for
the Higgs production processes, which can be easily simulated thanks to our
implementation.  Again, we show results both for $\mu_{EFT}=\Lambda=1$ TeV and
for $\mu_{EFT}=M_H=125$ GeV. 

\begin{figure}[ht]
	\centering
	\begin{minipage}{.49\linewidth}
		\includegraphics[width=1.\linewidth]{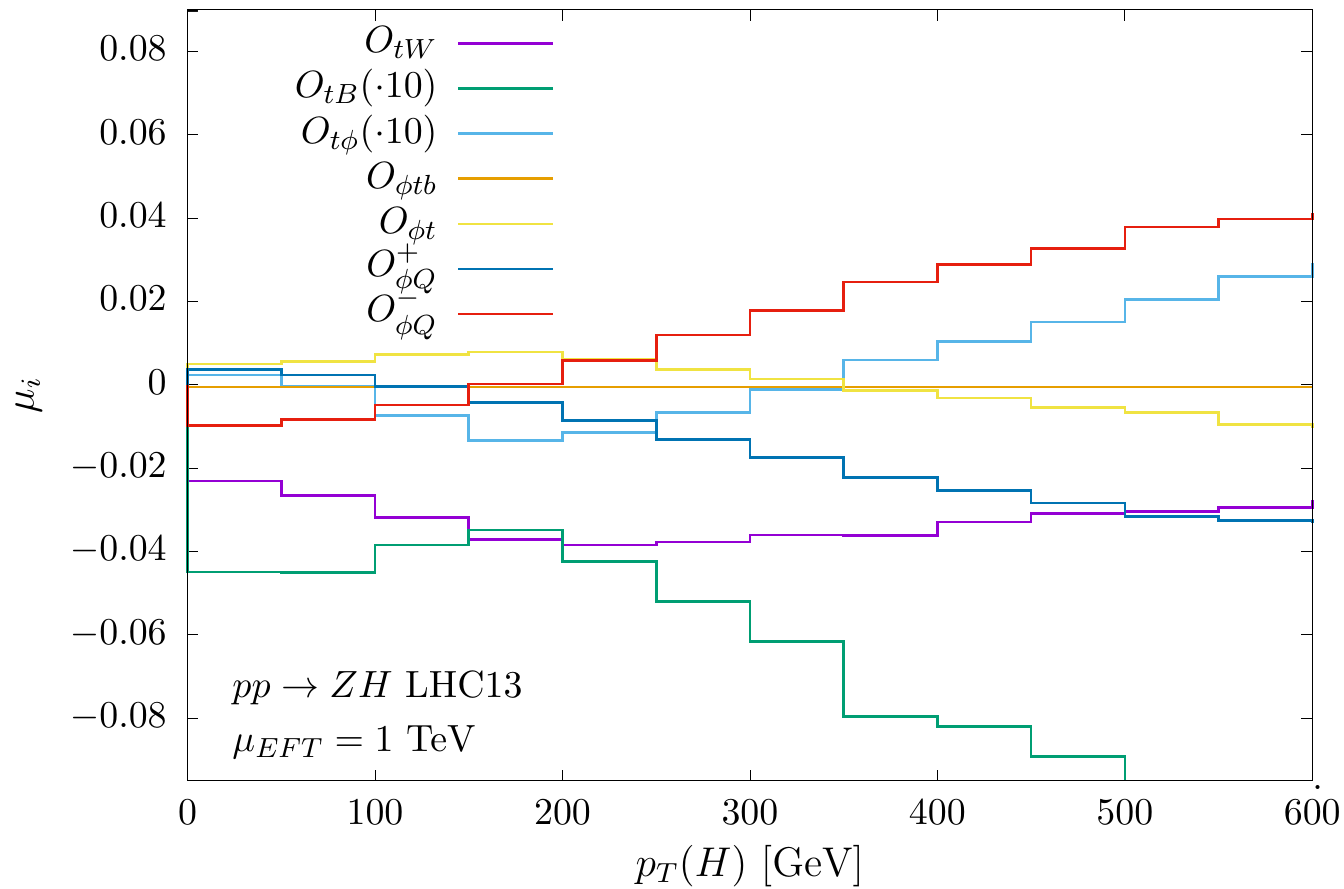}
	\end{minipage}
	\begin{minipage}{.49\linewidth}
		\includegraphics[width=1.\linewidth]{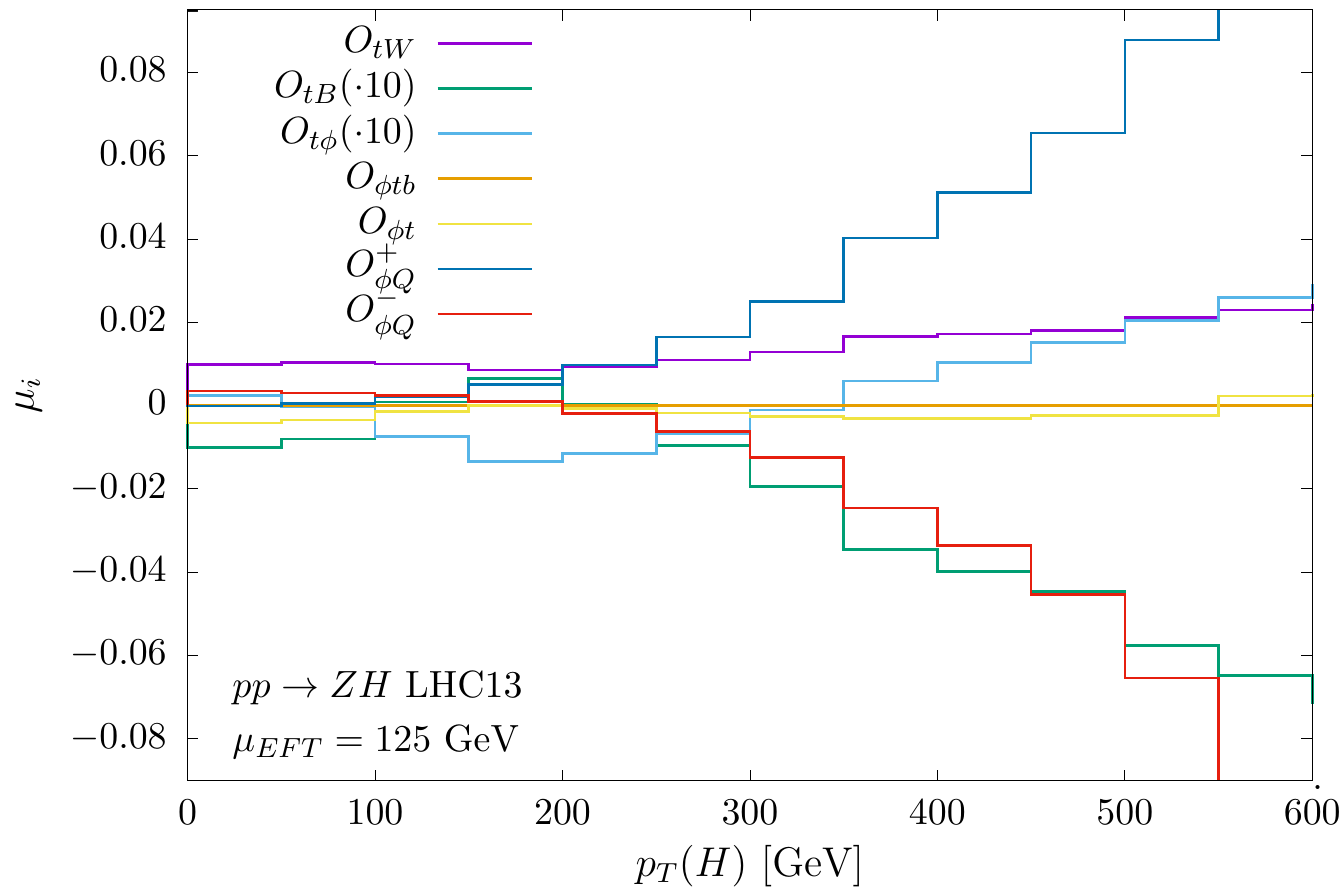}
	\end{minipage}
		\centering
	\begin{minipage}{.49\linewidth}
		\includegraphics[width=1.\linewidth]{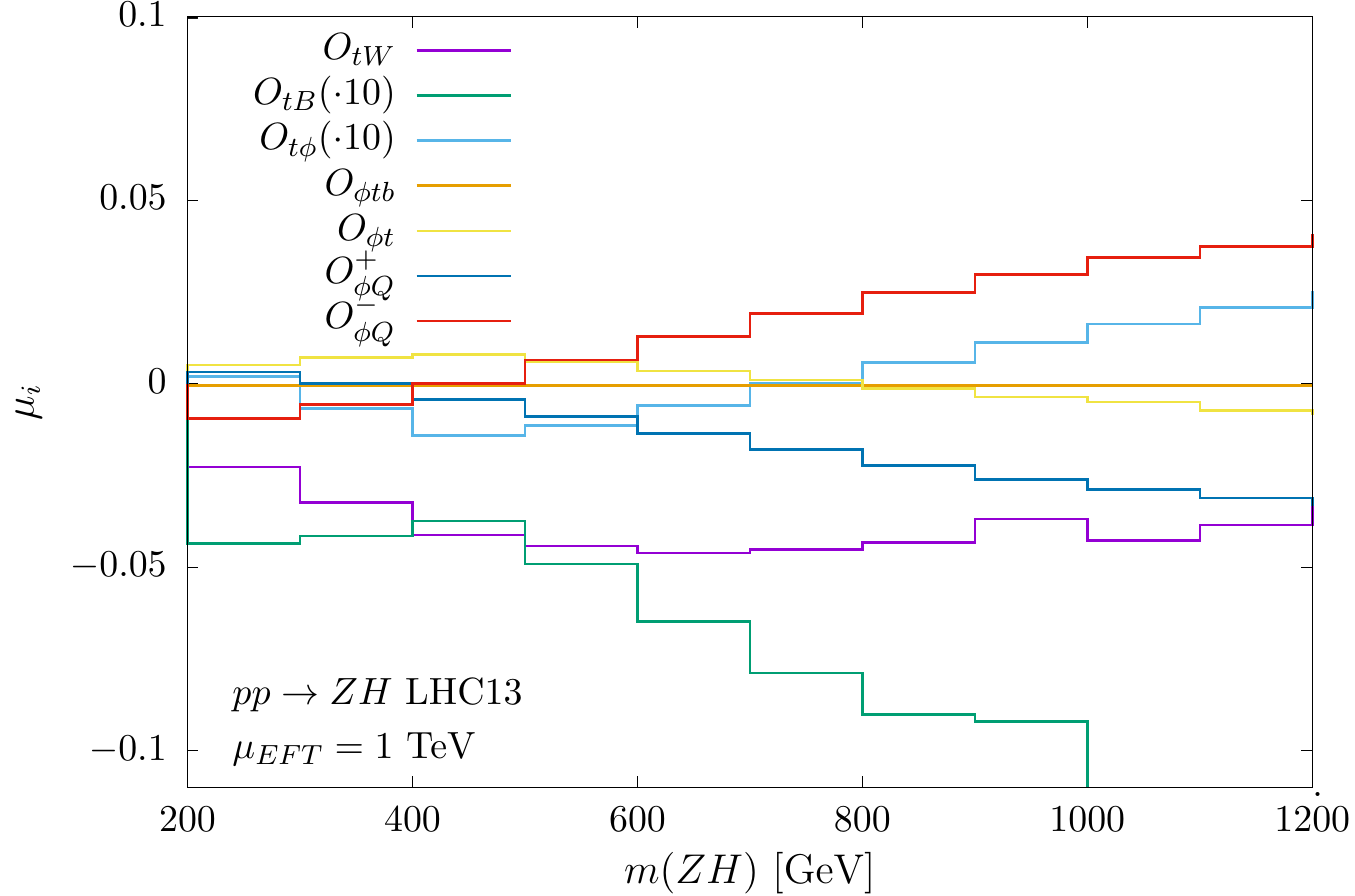}
	\end{minipage}
	\begin{minipage}{.49\linewidth}
		\includegraphics[width=1.\linewidth]{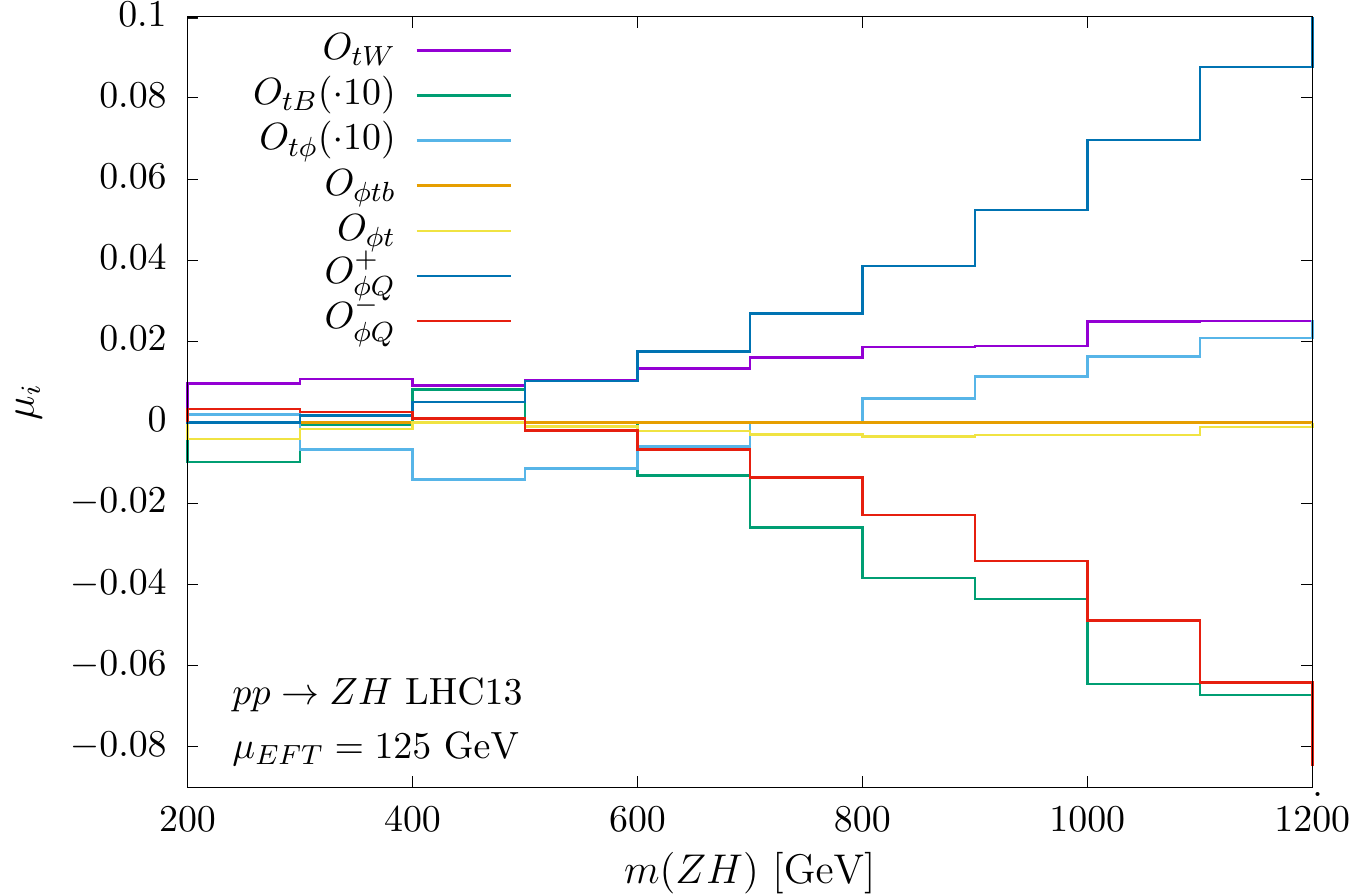}
	\end{minipage}
	\caption{Sensitivity of the Higgs transverse momentum distribution (top) and $ZH$ invariant mass distribution (bottom) in $ZH$ production for the different operators at $\mu_{EFT}=1000$ GeV (left) and $\mu_{EFT}=125$ GeV (right).}
	\label{fig:ZHdistributions}
\end{figure}

\begin{figure}[ht]
	\centering
	\begin{minipage}{.49\linewidth}
		\includegraphics[width=1.\linewidth]{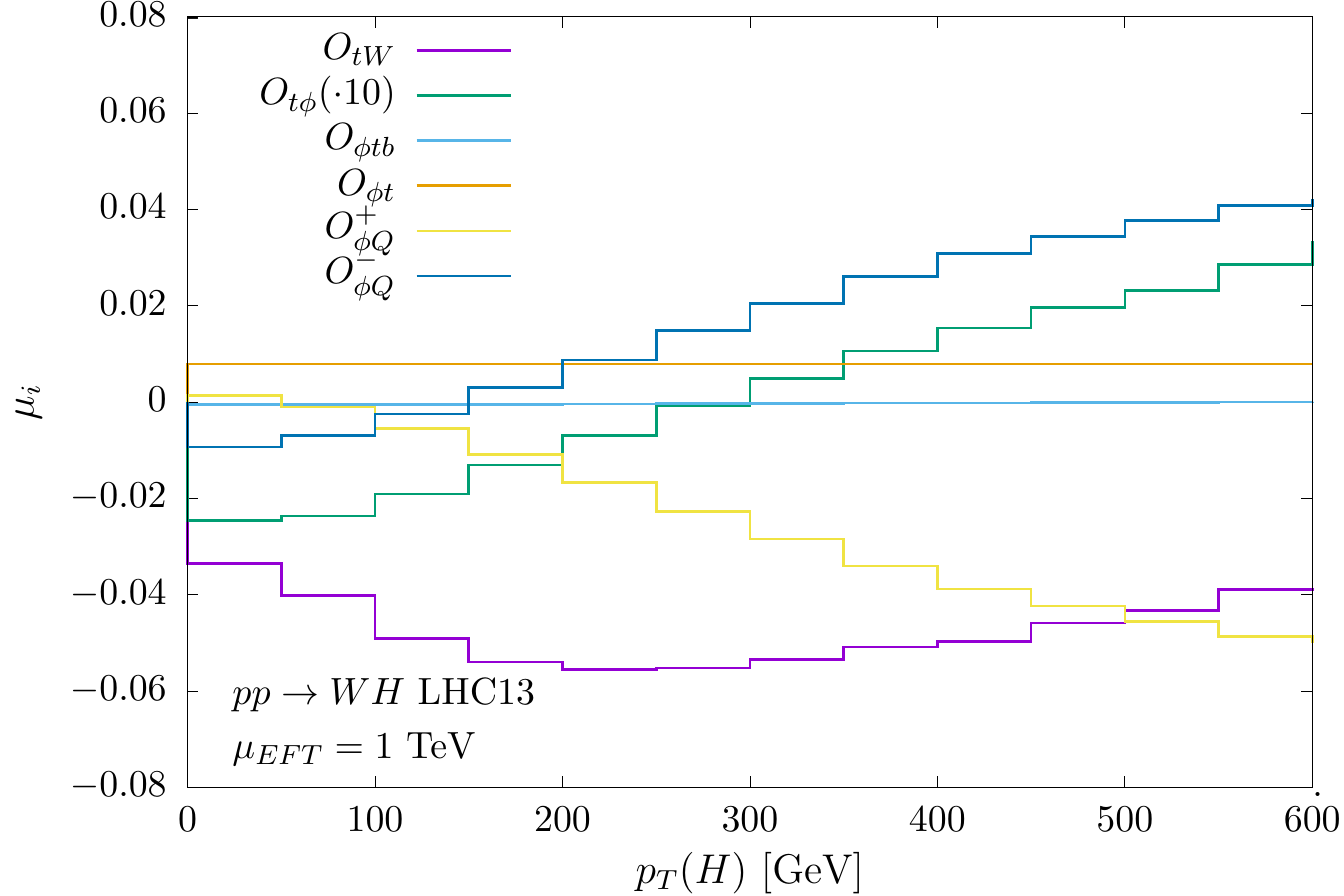}
	\end{minipage}
	\begin{minipage}{.49\linewidth}
		\includegraphics[width=1.\linewidth]{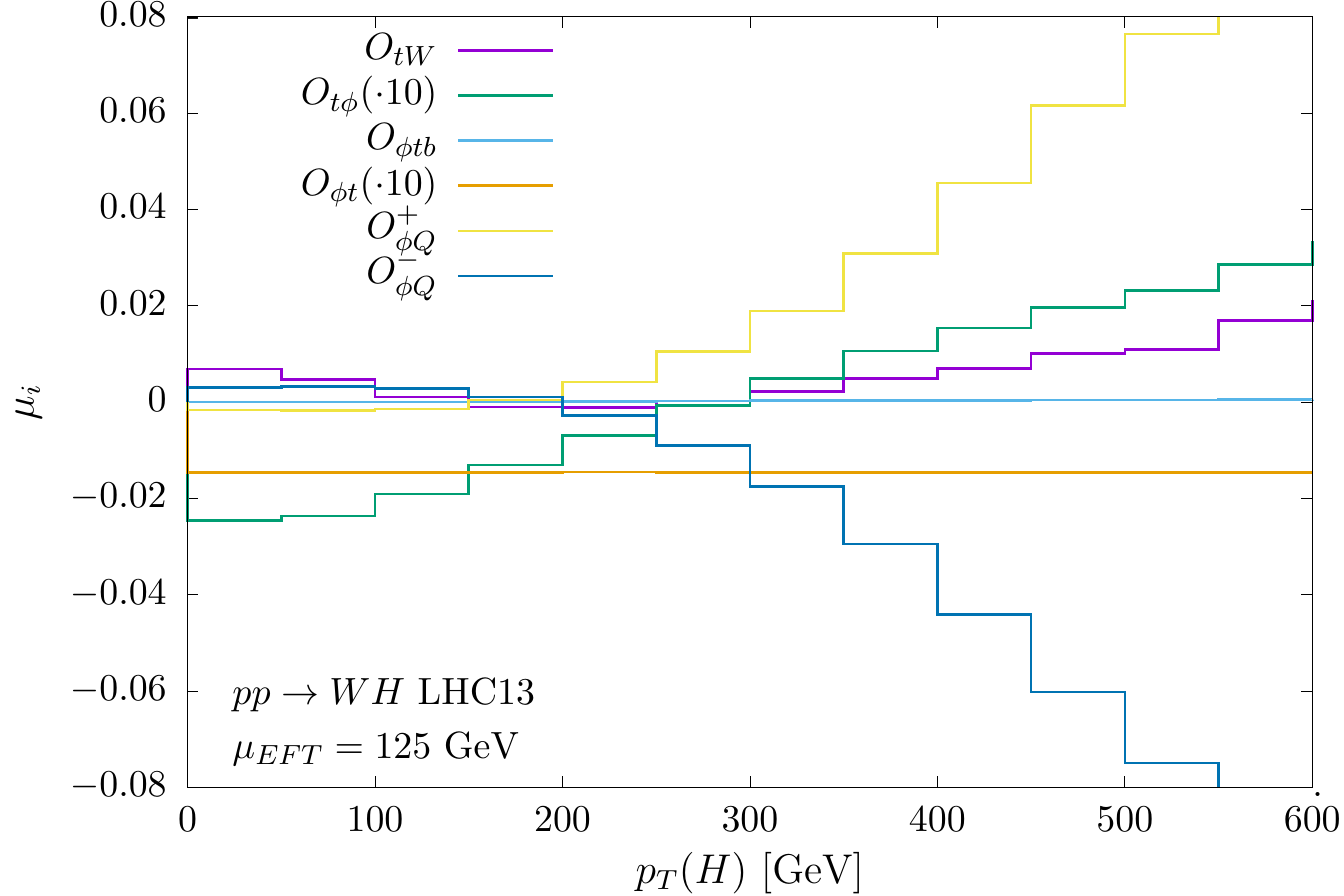}
	\end{minipage}
		\centering
	\begin{minipage}{.49\linewidth}
		\includegraphics[width=1.\linewidth]{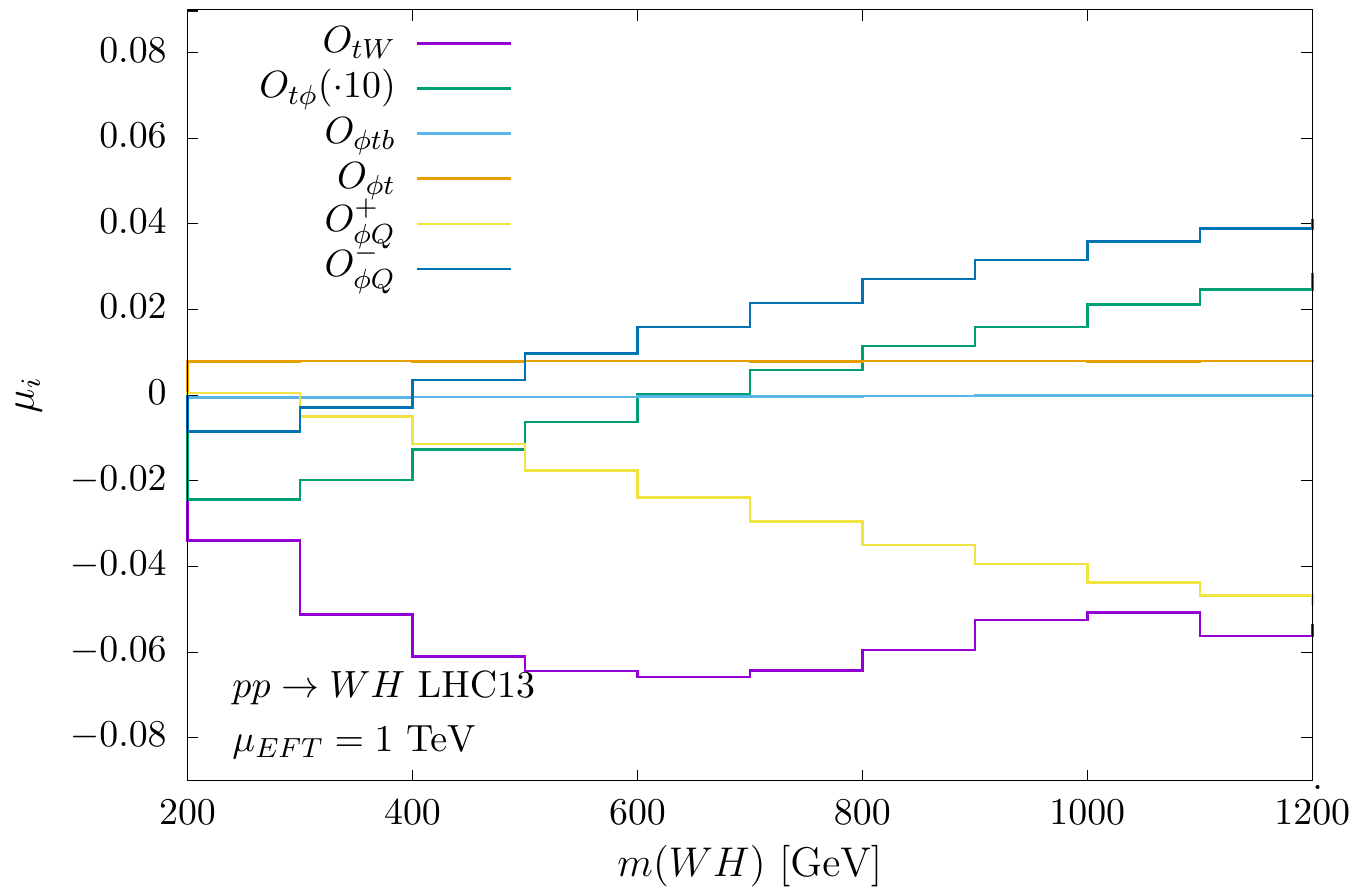}
	\end{minipage}
	\begin{minipage}{.49\linewidth}
		\includegraphics[width=1.\linewidth]{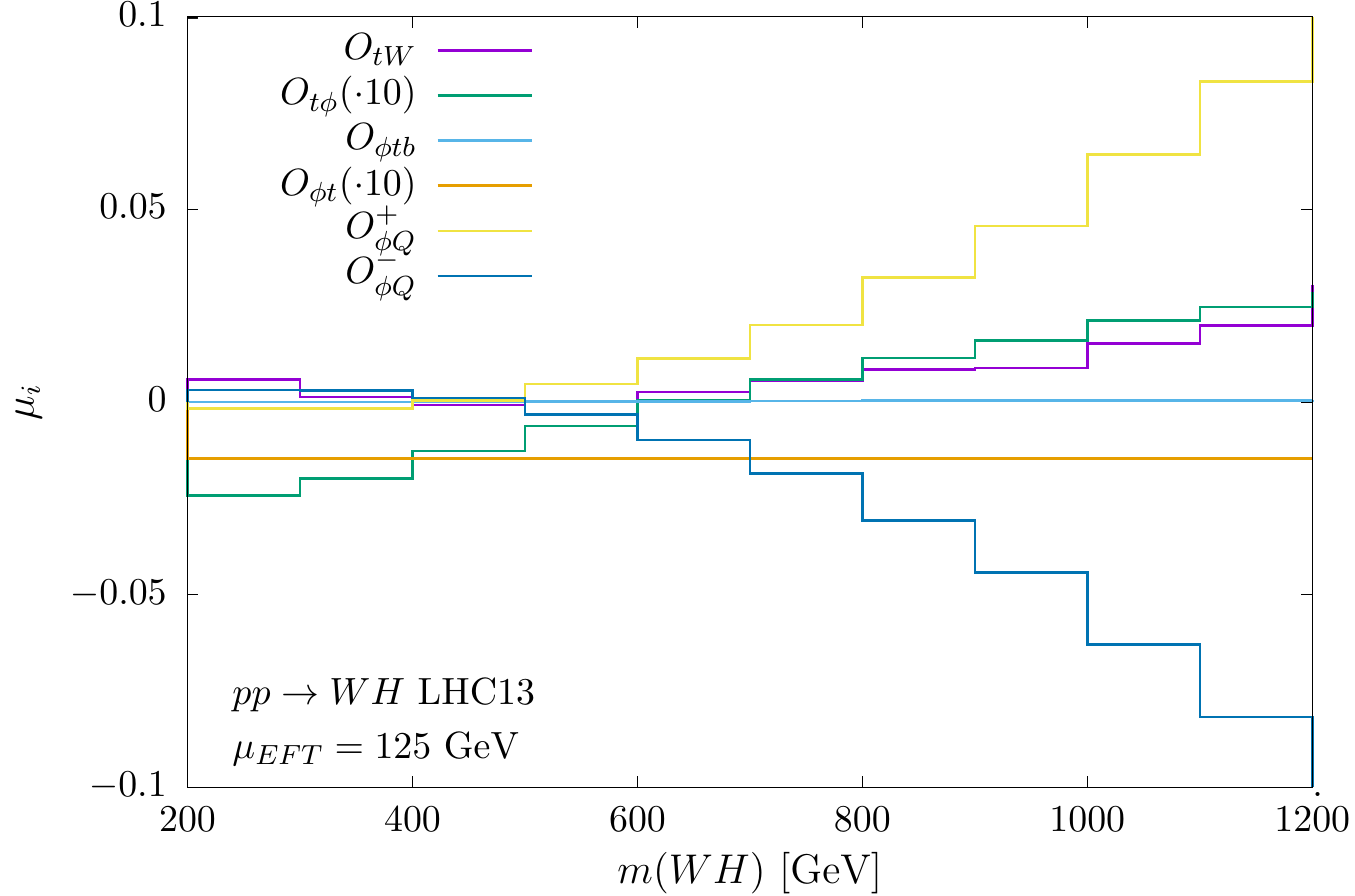}
	\end{minipage}
	\caption{Sensitivity of the Higgs transverse momentum distribution (top) and $WH$ invariant mass distribution (bottom) in $WH$ production for the different operators at $\mu_{EFT}=1000$ GeV (left) and $\mu_{EFT}=125$ GeV (right).}
	\label{fig:WHdistributions}
\end{figure}

\begin{figure}[ht]
	\centering
	\begin{minipage}{.49\linewidth}
		\includegraphics[width=1.\linewidth]{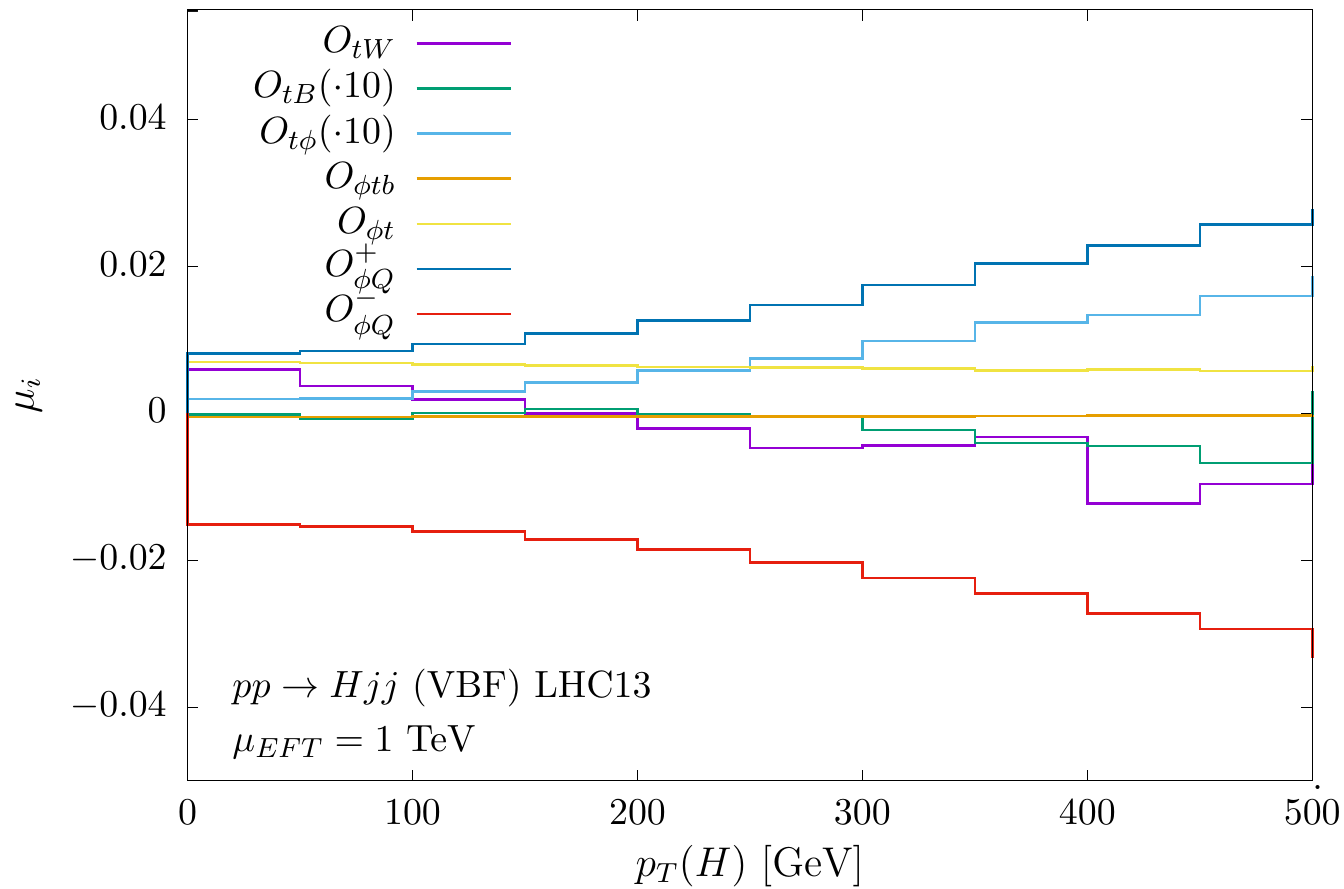}
	\end{minipage}
	\begin{minipage}{.49\linewidth}
		\includegraphics[width=1.\linewidth]{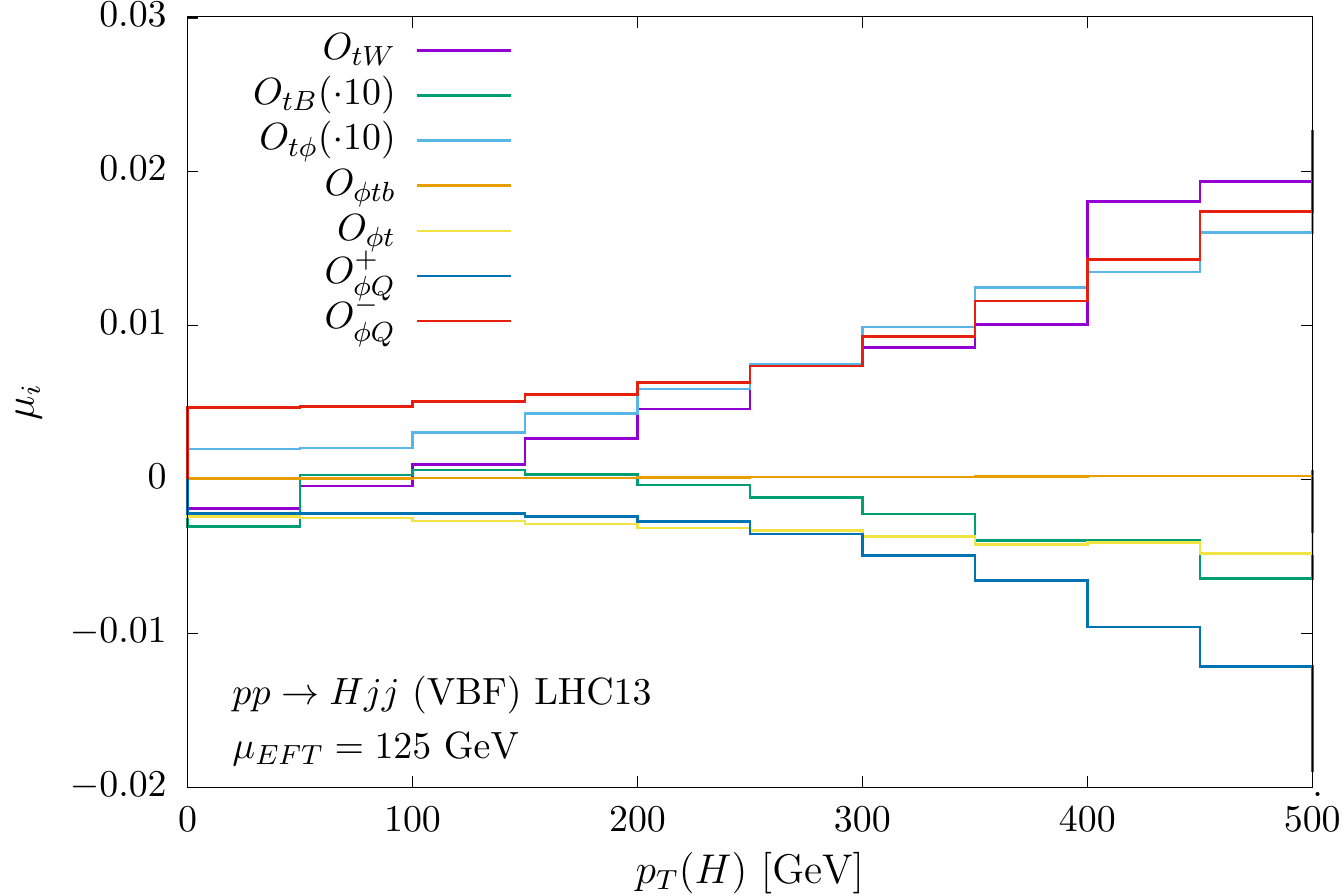}
	\end{minipage}
		\centering
	\begin{minipage}{.49\linewidth}
		\includegraphics[width=1.\linewidth]{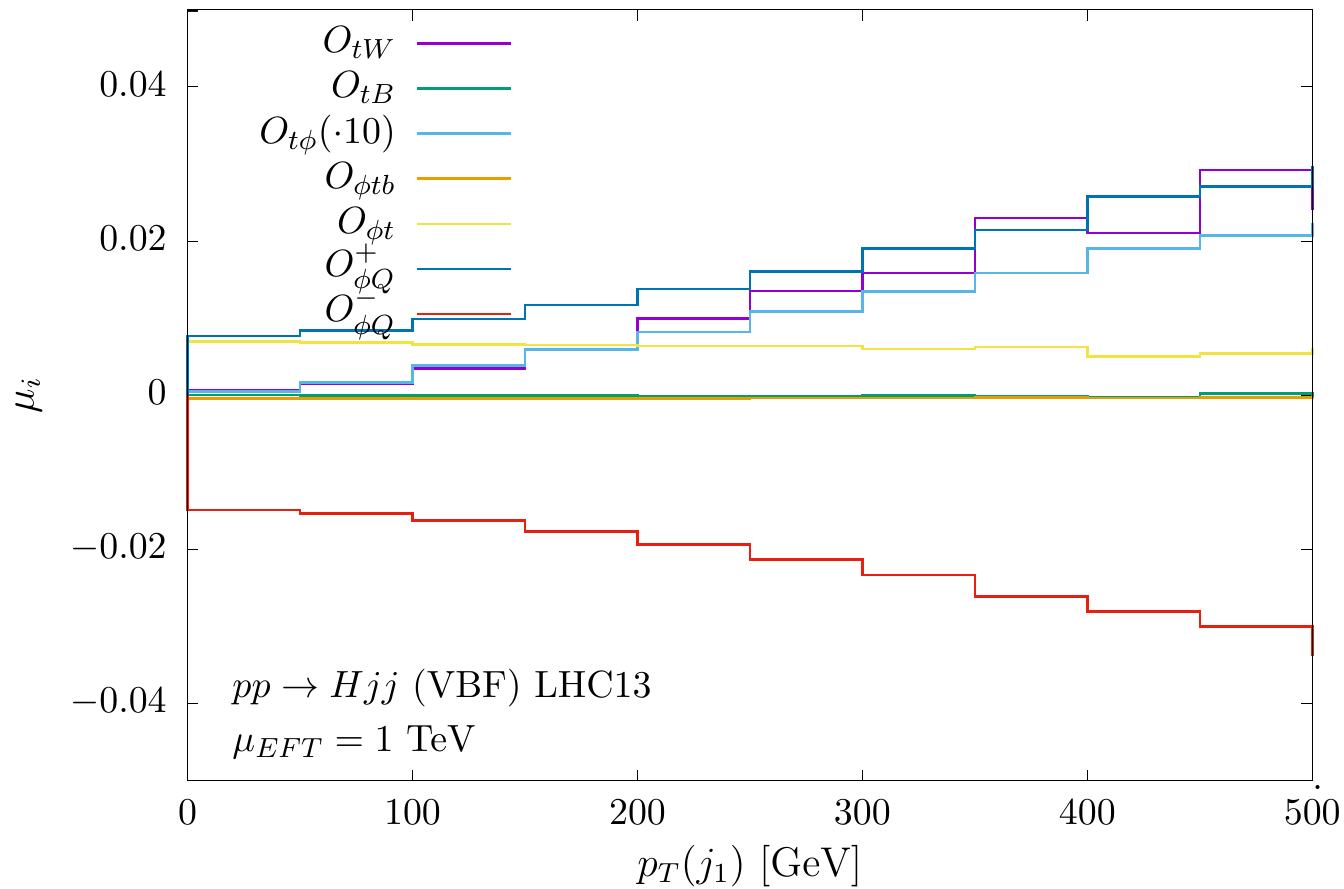}
	\end{minipage}
	\begin{minipage}{.49\linewidth}
		\includegraphics[width=1.\linewidth]{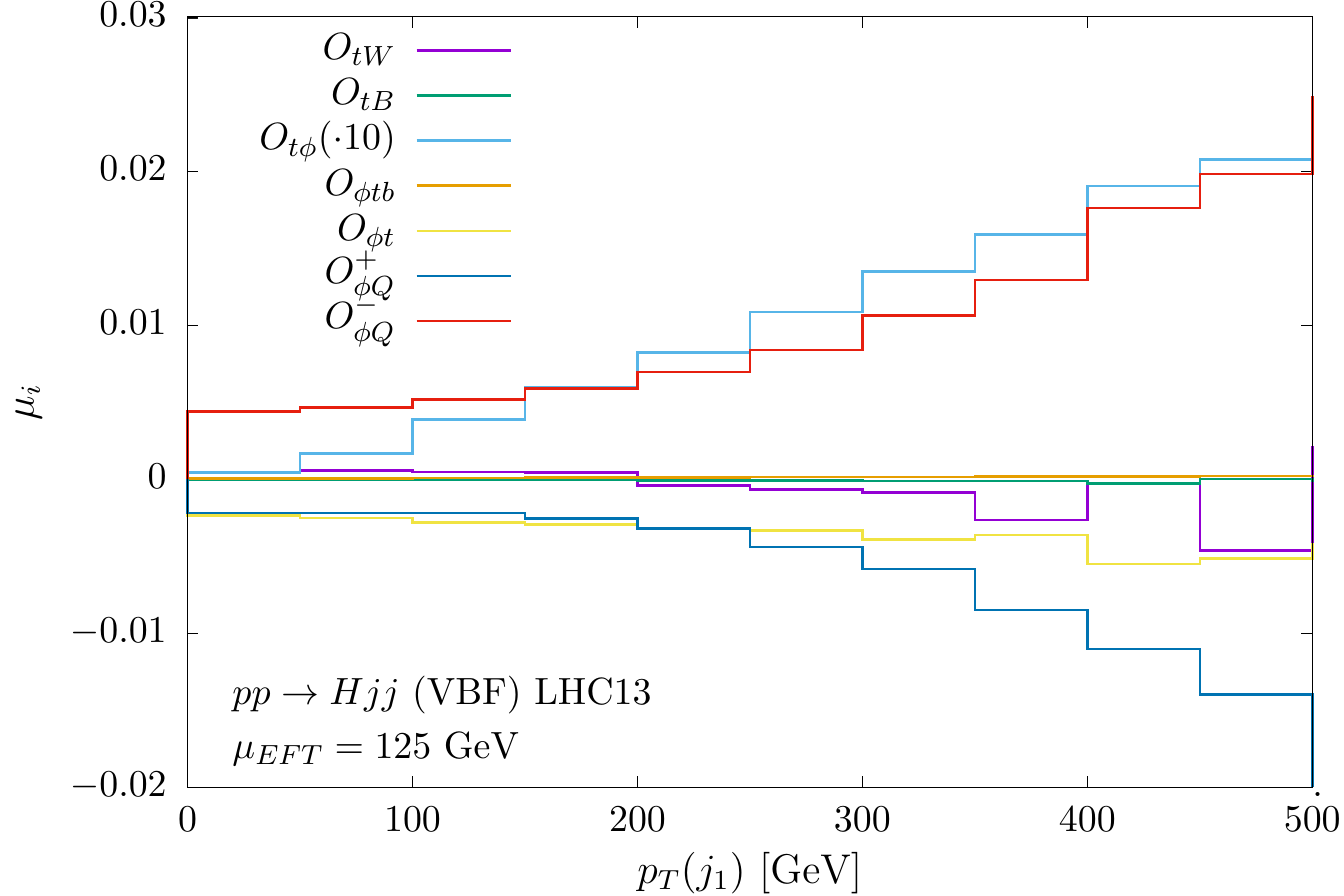}
	\end{minipage}
	\caption{Sensitivity of the Higgs (top)  and hardest jet (bottom) transverse momentum distribution in VBF Higgs production for the different operators at $\mu_{EFT}=1000$ GeV (left) and $\mu_{EFT}=125$ GeV (right). }
	\label{fig:VBFdistributions}
\end{figure}

As a representative sample of distributions, we show the transverse momentum of
the Higgs and invariant mass distribution of the $ZH$ system in $ZH$ in
Figure~\ref{fig:ZHdistributions}, whilst the corresponding distributions for $WH$
are shown in Figure~\ref{fig:WHdistributions}. For VBF we show the Higgs and
hardest jet transverse momentum distributions in
Figure~\ref{fig:VBFdistributions}.  The vertical axes are the bin-by-bin relative deviation
w.r.t the LO SM prediction, i.e.~following the definition in Eq.~(\ref{eq:mu}).

We find that in most cases, the impact of the operators increases in the tails
giving larger deviations from the SM prediction than those obtained at the
inclusive level. 
By comparing the two $\mu_{EFT}$ scales we find significant differences related
to the impact of the logarithmic terms, which are present in the predictions
for $\mu_{EFT}=1000$ GeV but absent at $\mu_{EFT}=125$. In most cases the
finite contributions which are the only ones present at $\mu_{EFT}=125$ are far
from negligible. For $VH$ production we find that the two observables we
consider, i.e. $p_T(H)$ and $m(VH)$ show similar sensitivities at $m(VH)\sim 2
p_T(H)$.  Similarly in VBF, the two $p_T$ distributions show comparable effects
but typically smaller than what is seen in $VH$ production. 
An interesting observation is that $O_{\varphi t}$ gives rise to a constant deviation in the
$WH$ production channel.  As we have explained in Section~\ref{sec:num}, this
is due to the renormalisation scheme we are using, and the contribution enters
only through Higgs wave-function renormalisation, so no different kinematics
can be generated.

Given that the top-quark operators give rise to harder $p_T(H)$ distributions,
we could use the differential information to improve the sensitivity at the HL-LHC.
To estimate the potential of differential measurements at HL-LHC, again we follow
the approach in Ref.~\cite{Maltoni:2017ims}, and assume that the number of
events in the $j$th bin of an $X$-like measurement, from production channel $i$
and decay channel $f$, is given by
\begin{flalign}
	N^{SM}_{X\text{-like},i,f,j}=r_j^XN^{SM}_{X\text{-like},i,f}
\end{flalign}
where $r_j^X$ is the ratio of the cross section of the $j$th bin with the total
cross section for process $i$. For each production-like mode we use $r_j^X$
only for the dominant production process and decay. The same assumption is made
for the background. Theoretical errors are taken to be the same as the
inclusive ones.  Systematic errors are also scaled using $r_j^X$.  Unlike 
Ref.~\cite{Maltoni:2017ims}, we consider the differential distributions in
all three channels: $WH$, $ZH$ and VBF, and use different binning. The $r_j^X$
values as well as the deviations in the signal strength, i.e.~$\mu_{ij}$ in
each $p_T(H)$ bin, for all three production channels, are given in
Tables~\ref{tab:diff1} and \ref{tab:diff2} in Appendix~\ref{sec:app1}.

The resulting individual sensitivities are not significantly different than the
inclusive analysis, as they are dominated by the loop-induced processes, such
as $gg\to H$, $H\to\gamma\gamma,\gamma Z$ where no distributions can be used.
However, by looking into the eigenvalues of the $\chi^2$ we find improvements
in particular in the last three eigenvalues, which are mainly driven by the 
top-loop corrections to SM tree-level processes. For $\mu_{EFT}=125$ GeV we have
\begin{flalign}
\left(
\begin{array}{ccccccc}
-0.025 & 0.045 & 0.019 & 0.005 & -0.43 & -0.9 & -0.041 \\
 0.024 & -0.073 & -0.058 & -0.051 & 0.4 & -0.24 & 0.88 \\
 0.00076 & 0.19 & 0.069 & -0.033 & -0.78 & 0.36 & 0.47 \\
 -0.26 & 0.91 & 0.24 & -0.0082 & 0.21 & -0.043 & -0.0097 \\
 0.39 & 0.28 & -0.64 & -0.6 & -0.0067 & -0.0076 & -0.064 \\
 0.13 & 0.2 & -0.57 & 0.78 & -0.016 & 0.0051 & 0.03 \\
 0.87 & 0.12 & 0.44 & 0.15 & 0.045 & -0.03 & -0.0043 
\end{array}
\right)
\times\frac{1\text{TeV}^2}{\Lambda^2}
\left( \begin{array}{ccccccc}
	C_{\varphi t}\\ C_{\varphi Q}^{(+)}\\ C_{\varphi Q}^{(-)}\\ C_{\varphi tb}\\ C_{tW} \\ C_{tB}\\ C_{t\varphi} 
\end{array} \right)
\nonumber\\
=\pm
\left( \begin{array}{c}
  0.0325 \\ 0.543 \\ 0.63 \\ 2.53 \\ 6.03 \\ 14.8 \\ 32.1
\end{array} \right),\ 
\mbox{compared with}
\left( \begin{array}{c}
	 0.0326 \\ 0.548 \\ 0.637 \\ 2.62 \\ 7.31 \\ 19.8 \\ 79.6 
\end{array}
\right)\mbox{from inclusive measurements.}
\end{flalign}
The improvement on the last eigenstate is about a factor of 2.5.  On average,
the Global Determinant Parameter (GDP) \cite{Durieux:2017rsg} of this fit is
improved by a factor of 0.81 compared to the inclusive results, which is not huge,
but it is very important that the weaker constraints receive larger
improvements, which means that directions that were almost flat are now lifted.  For
$\mu_{EFT}=1$ TeV we find:
\begin{flalign}
\left(
\begin{array}{ccccccc}
 -0.061 & -0.15 & 0.016 & -0.046 & -0.13 & -0.98 & 0.054 \\
 -0.034 & -0.065 & 0.022 & -0.42 & -0.35 & 0.12 & 0.83 \\
 0.13 & -0.0042 & -0.016 & -0.17 & 0.92 & -0.1 & 0.32 \\
 -0.12 & -0.24 & 0.2 & -0.83 & 0.03 & 0.058 & -0.45 \\
 -0.042 & -0.93 & -0.28 & 0.2 & 0.037 & 0.13 & 0.032 \\
 0.52 & 0.14 & -0.79 & -0.25 & -0.099 & -0.049 & -0.11 \\
 0.83 & -0.18 & 0.51 & 0.056 & -0.096 & -0.0056 & -0.0042 
\end{array}
\right)
\times\frac{1\text{TeV}^2}{\Lambda^2}
\left( \begin{array}{ccccccc}
	C_{\varphi t}\\ C_{\varphi Q}^{(+)}\\ C_{\varphi Q}^{(-)}\\ C_{\varphi tb}\\ C_{tW} \\ C_{tB}\\ C_{t\varphi} 
\end{array} \right)
\nonumber\\
=\pm
\left( \begin{array}{c} 
 0.0409 \\ 0.479 \\ 0.629 \\ 1.3 \\ 1.5 \\ 5.44 \\ 10.8 
 \end{array} \right),\ 
\mbox{compared with}
\left( \begin{array}{c}
	 0.041 \\ 0.487 \\ 0.638 \\ 1.45 \\ 1.55 \\ 5.84 \\ 12.7
\end{array} \right)
\mbox{from inclusive measurements.}
\end{flalign}
Improvements are smaller than the $\mu_{EFT}=125$ GeV case.

\begin{figure}[h]
	\begin{center}
		\includegraphics[width=1.1\linewidth]{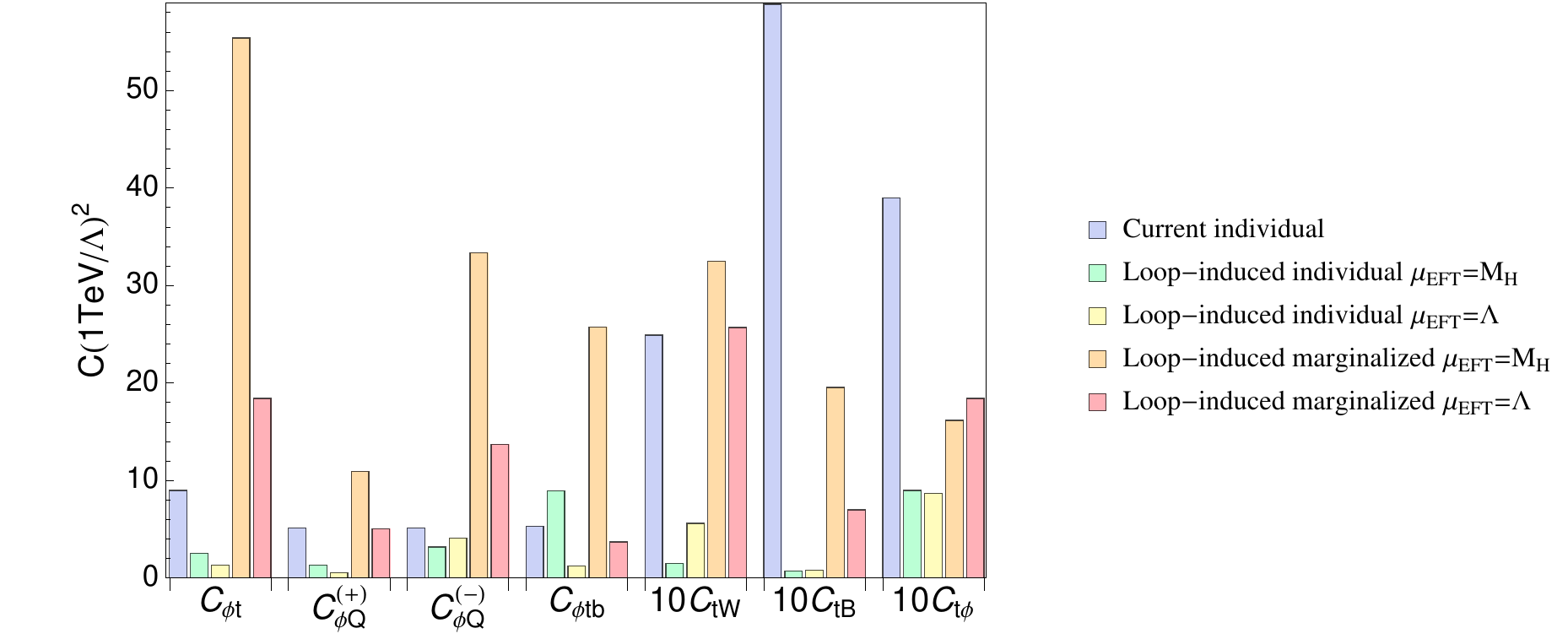}
	\end{center}
	\caption{Individual and marginalised sensitivities (i.e.~one-half of
		95\% CL interval) at HL-LHC through top loops, including
		differential measurements, compared with current individual
		limits.  $\mu_{EFT}$ is set at either $M_H=125$ GeV or
	$\Lambda=1$ TeV.}
	\label{fig:limits}
\end{figure}

In summary, after taking into account differential distributions, the one-sigma
constraints on the seven linear combinations of top operator coefficients span
the range from $\mathcal{O}(10^{-2})\mathrm{TeV}^{-2}$ to
$\mathcal{O}(10)\mathrm{TeV}^{-2}$.  This reflects the HL-LHC potential on
probing top-quark operators through EW loops.
We summarise the individual and marginalised sensitivities in
Figure~\ref{fig:limits}.  They are not stronger then the current ones from
direct top-quark measurements, but are definitely competitive.  This means that
in the near future the loop-induced effects cannot be neglected, and the only
way to correctly take them into account is to perform global SMEFT fits by
combining both the top-quark and the Higgs-boson sectors.

\subsection{Loop/tree discrimination}

Until now we have focused on the sensitivities of the top operators, and have
not considered the effects of Higgs operators at the tree level.  In a real fit,
one has to consider these contributions, and include sufficient observables so
that no blind directions remain.  In the following we briefly argue that, in
principle, we can discriminate the tree-level contribution from Higgs operators
and loop-level contribution from top operators, by only using Higgs
measurements.  This possibility relies on the finite non-logarithmic terms,
$\sigma_\mathrm{fin}$, in Eq.~(\ref{eq:logfin}).  The reason is that the
logarithmic terms are completely captured by RG effects, and are thus process-
and observable-independent.  If one considers only these effects, the discrimination
between top/Higgs operators would be impossible no matter how many observables 
are included.

As an example, consider the top-quark operator $C_{tB}$ which mixes into
$C_{\varphi B}$ and $C_{\varphi WB}$.  The latter mixing does not exist in our
scheme as we
renormalise it to a physical quantity.  Consider the measurements
$H\to\gamma\gamma, \gamma Z, WW^*,ZZ^*$ and $ZH/WH$ production.  Another
operator that enters these observables is $O_{\varphi W}$.   Assuming all these
processes are measured with 10\% precision, we can construct a $\chi^2$ and
marginalise over the $O_{\varphi W}$ operator.  Taking $\mu_{EFT}=125$ GeV,
after diagonalisation we find
\begin{flalign}
	&C_{\varphi B}+0.021C_{tB}=\pm0.0022\ (\Lambda/1\mathrm{TeV})^2\ ,
	\\
	&C_{tB}-0.021C_{\varphi B}=\pm6.7\ (\Lambda/1\mathrm{TeV})^2\ ,
\end{flalign}
so both directions are constrained and no blind direction is left.

\begin{figure}[h]
	\begin{center}
		\includegraphics[width=.5\linewidth]{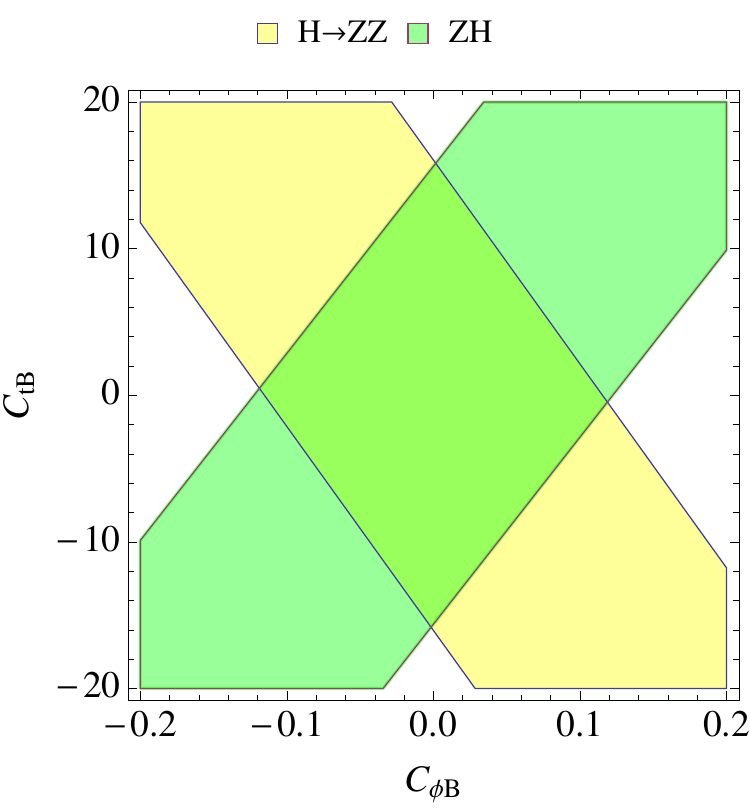}
	\end{center}
	\caption{Constraining both $C_{tB}$ (top operator, loop level) and $C_{\varphi B}$ (Higgs
operator, tree level) by combining $H\to ZZ^*$ and $ZH$ production, assuming 1\% precision on both.
$\mu_{EFT}=125$ GeV.  $\Lambda=1$ TeV.}
	\label{fig:hbtb}
\end{figure}

On the other hand, if one takes $\mu_{EFT}=1$ TeV but only includes the logarithmic terms,
the situation will be different.  In this case we find
\begin{flalign}
	&C_{\varphi B}-0.046C_{tB}=\pm0.0022\ (\Lambda/1\mathrm{TeV})^2\ ,
	\\
	&C_{tB}+0.046C_{\varphi B}=\pm\infty\ (\Lambda/1\mathrm{TeV})^2\ ,
\end{flalign}
so one combination is unconstrained.  This demonstrates that the finite
contributions in the SMEFT loop corrections are important, not only because
their sizes can be large, but more importantly, because they allow us to
discriminate the pure loop-induced effects from the tree-level contributions of
other operators, into which they could mix, by combining various measurements,
preferably at different energies, to eliminate possible unconstrained
directions in a global fit.  This is why we believe a study taking
$\mu_{EFT}=M_H=125$ GeV is a more reasonable estimate of sensitivities that can be
achieved in a bottom-up global fit, where the discriminating power between
loop- and tree-level effects is crucial for setting bounds on top operators.
In Figure~\ref{fig:hbtb} we illustrate that the $H\to ZZ^*$ and $ZH$
production are complementary in the $C_{\varphi B}-C_{tB}$ plane.

\begin{figure}[ht]
	\centering
	\begin{minipage}{.49\linewidth}
		\includegraphics[width=1\linewidth]{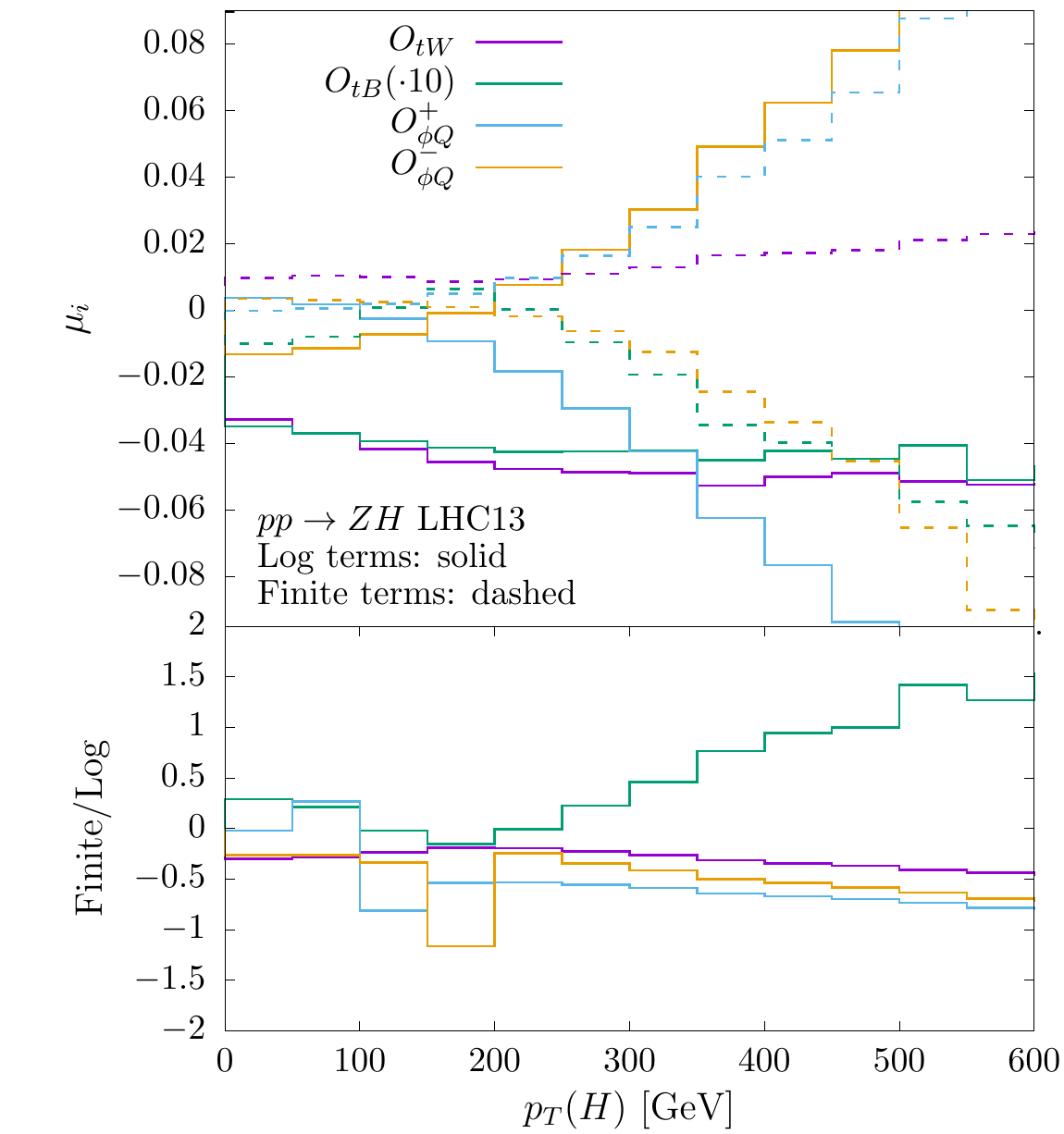}
	\end{minipage}
	\begin{minipage}{.49\linewidth}
		\includegraphics[width=1\linewidth]{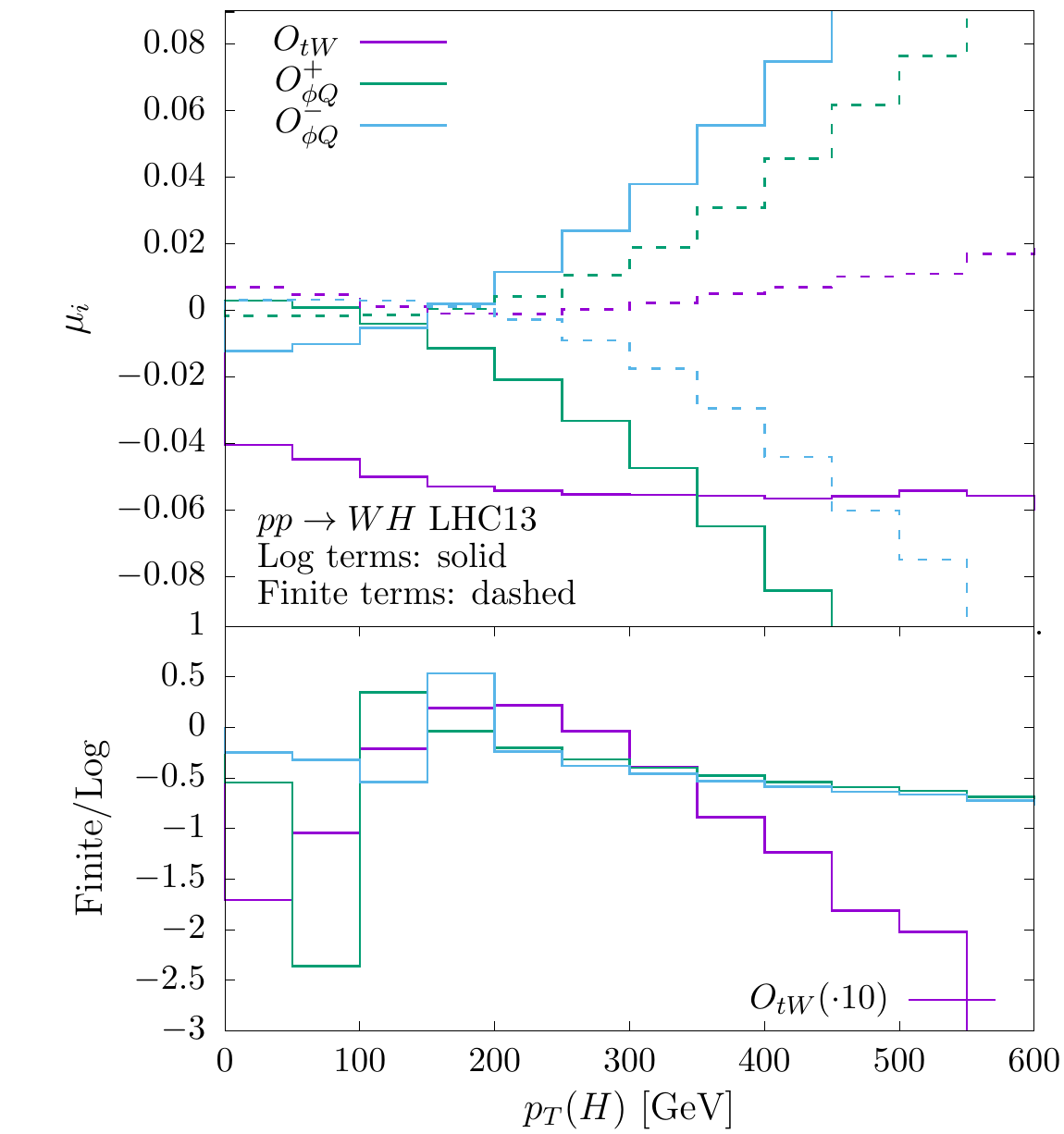}
       \end{minipage}
	\caption{Comparison of logarithmic and finite terms in the Higgs transverse momentum distribution in $ZH$ and $WH$ production for the different operators. The lower panels show the ratio of the finite over the logarithmic terms. }
	\label{fig:logfinite}
\end{figure}

Instead of combining several inclusive measurements, one could also use the
differential information in one measurement.  The differential distributions of
the logarithmic terms can not be distinguished from the tree level ones, but
they differ from the ones of the finite terms.  In Figure~\ref{fig:logfinite} we
compare the normalised distributions (over the SM) from finite and log terms
for the operators which show the most promising energy dependence in $VH$ production. These plots
demonstrate that indeed the kinematic behaviour of the finite and logarithmic
terms can be very different, and so the kinematic distributions will serve
as discriminating observables in a global fit, lifting the degeneracy between
loop and tree-level contributions.

\section{Conclusion}
\label{sec:conc}

We have computed the NLO EW corrections to Higgs production and decay processes
from dim-6 top-quark operators in the SMEFT framework. We have studied the
major production channels including VBF, $WH$ and $ZH$ at the LHC, $ZH$ and VBF
at $e^+e^-$ colliders, and the major decay channels including $H\to
\gamma\gamma,\gamma Z,Zll,Wl\nu,bb,\mu\mu,\tau\tau$.  These results are part of
the ongoing efforts of automating SMEFT simulations for colliders at NLO, and
is a first step towards including EW corrections.

These results allow us to study whether the Higgs measurements at the LHC
and future colliders are
sensitive to possible deviations in the top-quark sector. We find that, within
the current direct constraints on the top-quark sector, the top-quark operators
can shift the signal strength of the loop-induced Higgs processes, i.e.~$gg\to
H$ and $H\to \gamma\gamma,\gamma Z$ by factors
$\sim\mathcal{O}(1)-\mathcal{O}(10)$, and that of the tree-level processes,
i.e.~all remaining production and decay channels, by $\sim5-10\%$ through loop
corrections at the LHC and up to 15\% at future lepton colliders.  This implies
that with the current precision we can already learn about top-quark couplings
by using EW loops in Higgs measurements, while in the future, at the
high-luminosity scenario, all Higgs measurements can be sensitive to top-quark
couplings.  It also implies that at an $e^+e^-$ collider, even measurements
below the $t\bar t$ threshold can be sensitive to deviations in top-quark
couplings.  As a result, in a global fit for Higgs couplings based on the SMEFT
approach, theoretical uncertainties due to unknown top-quark operators are not
negligible and should be taken into account.

These results also allow us to quantitatively derive the experimental
sensitivities to top-quark operators at the HL-LHC, by only using the Higgs
observables.   Using projected inclusive measurements, we are able to constrain
all directions in the 7-dimensional parameter space spanned by all seven
operator coefficients.  We have also studied the change of differential
distributions from top-quark operators, and have found that the
signal/background ratio is enhanced at the tail of the distributions. By
considering the $p_T(H)$ distributions in VBF, $WH$ and $ZH$ production
processes, we have significantly improved the sensitivities on the most weakly
constrained eigenvectors.  The resulting one-sigma range on the seven
eigenvectors of Wilson coefficients span the range from
$\mathcal{O}(10^{-2})\mathrm{TeV}^{-2}$ to $\mathcal{O}(10)\mathrm{TeV}^{-2}$.

Finally, we have briefly discussed possible ways to discriminate the loop-level
top-quark operator contributions from the tree-level Higgs operator
contributions, which is necessary in a global fit.  We have demonstrated that
this can be done by combining several observables or looking into the
differential distributions, and that the crucial information is supplied by the
finite (i.e.~non-logarithmic) terms in the full NLO EW corrections.  Therefore,
even though they cannot be obtained using the RG matrix, it is very
important to compute these finite terms for an actual SMEFT fit.

In conclusion, as the experimental precision on Higgs measurements continues to
improve, the NLO EW corrections from dim-6 top-quark operators via top-loop
effects have started to become relevant.  For this reason, treating the dim-6
top-quark sector and the Higgs/EW sector separately may not continue to be a
good strategy.  A global SMEFT analysis taking into account both sectors by
combining Higgs and top-quark measurements is desirable.  Our calculation is a
first step towards this direction, and our implementation in the MG5\_aMC
framework provides an automatic and realistic simulation tool for this purpose.

\section*{Acknowledgements}
We thank F.~Maltoni and X.~Zhao for useful discussions and comments on the manuscript,
and O.~Mattelaer for technical assistance with the reweighting module in
MG5\_aMC.  CZ thanks M.~Trott for a discussion about the renormalisation scheme.
CZ is supported by IHEP under Contract No.~Y7515540U1. EV is supported by a
Marie Sk\l{}odowska-Curie Individual Fellowship of the European Commission's
Horizon 2020 Programme under contract number 704187 HEFTinLOOPS.

\appendix

\section{More numerical results}
\label{sec:app1}
In Tables~\ref{tab:res11} and \ref{tab:res12} we present results similar to
Tables~\ref{tab:res1} and \ref{tab:res2} but with the standard $\overline{MS}$
scheme.  In Tables~\ref{tab:diff1} and \ref{tab:diff2} we show the $r_j^X$
values for the distributions we consider in Section~\ref{sec:diff}, as well as
the deviations in the signal strength, i.e.~$\mu_{ij}$ in each $p_T(H)$
bin, for all three production channels.

\begin{table}
	\centering
\begin{tabular}{|ll|lllllll|}
		\hline
		channel & $\mu_\mathrm{EFT}$ [GeV]
		& $O_{\varphi t}$
		& $O_{\varphi Q}^{(+)}$
		& $O_{\varphi Q}^{(-)}$
		& $O_{\varphi tb}$
		& $O_{tW}$
		& $O_{tB}$
		& $O_{t\varphi}$
		\\\hline
		$H\to bb$ & 125
		&0 & -0.07 & 0.11 & -1.13 & -0.28 & 0 & -0.18
		\\
		$H\to bb$ & 1000
		&0 & -0.99 & -0.91 & -8.18 & 0.34 & 0 & 0.29
		\\\hline
		$H\to ll$ & 125
		&0 & -0.02 & 0.02 & -0.00& 0 & 0 & -0.27
		\\
		$H\to ll$ & 1000
		&0 & 0.45 & -0.45 & -0.03 & 0 & 0 & 0.68
		\\\hline
		$H\to \gamma\gamma$ & 125
		&0 & 0 & 0 & 0 & 
		-73.3 & -136.8 & 3.45
		\\
		$H\to \gamma\gamma$ & 1000
		&0 & 0 & 0 & 0 & 
		114.6 & 214.0 & 3.45
		\\\hline
		$H\to Z\gamma$ & 125
		& 1.77 & 0.03 & 1.77 & 0 &
		-45.8 & 6.97 & 0.72
		\\
		$H\to Z\gamma$ & 1000
		& 1.77 & 0.03 & 1.77 & 0 &
		71.3 & -9.69 & 0.72
		\\\hline
		$H\to Zll$ & 125
		&-0.65& -0.07& 0.65& -0.00 & 0.22& -0.02& 0.08
		\\
		$H\to Zll$ & 1000
		&0.88& 0.47& -1.49& -0.04& -0.16& 0.11& 0.08
		\\\hline
		$H\to Wl\nu$ & 125
		&0& -0.25& 0.25& 0.00& -0.13& 0& -0.03
		\\
		$H\to Wl\nu$ & 1000
		&0& 1.08& -1.08& -0.08& 0.33& 0& -0.03
		\\\hline
\end{tabular}
\caption{Percentage deviation $\mu_{ij}$ for decay channel $i$ and operator $j$,
in the $\overline{MS}$ scheme. }
\label{tab:res11}
\end{table}

\begin{table}
	\centering
\begin{tabular}{|ll|lllllll|}
		\hline
		channel & $\mu_\mathrm{EFT}$ [GeV]
		& $O_{\varphi t}$
		& $O_{\varphi Q}^{(+)}$
		& $O_{\varphi Q}^{(-)}$
		& $O_{\varphi tb}$
		& $O_{tW}$
		& $O_{tB}$
		& $O_{t\varphi}$
		\\\hline
		$p p \to ZH$ & 125
		& -0.44 & 0.16 & 0.33 & 0.00 & 0.80  & -0.29 & -0.02
		\\
		$p p \to ZH$ & 1000
		& 1.45 & -0.64 & -1.05  &  -0.02 & -1.53 & 0.93 & -0.02
		\\\hline
		$p p \to WH$ & 125
		& 0 & -0.06 & 0.06 & -0.00 & 0.43 & 0 & -0.21
		\\
		$p p \to WH$ & 1000
		&0  & 0.18  & -0.18  & -0.08 & -4.09 &  0 &  -0.21
		\\\hline
		$p p \to H j j$ & 125
		& -0.21 & -0.25 & 0.46 & 0.01 & 0.04 & -0.02 & 0.03
		\\
		$ p p \to H j j$ & 1000
		&  0.36 & 1.10 & -1.48 & -0.06 & 0.36 & 0.09 & 0.03
		\\\hline
\end{tabular}
\caption{Percentage deviation $\mu_{ij}$ for production channel $i$ and operator $j$, in the $\overline{MS}$ scheme. }
\label{tab:res12}
\end{table}

\begin{table}
	\centering
\begin{tabular}{|l|l|l|lllllll|}
		\hline
		Bin [GeV] & Channel & $r$ value
		& $O_{\varphi t}$
		& $O_{\varphi Q}^{(+)}$
		& $O_{\varphi Q}^{(-)}$
		& $O_{\varphi tb}$
		& $O_{tW}$
		& $O_{tB}$
		& $O_{t\varphi}$
		\\\hline
 \text{0-50} & \text{VBF} & 0.22 & -0.24 & -0.22 & 0.47 & 0.01 & -0.19 & -0.03 & 0.02 \\
 \text{} & \text{WH} & 0.35 & -0.15 & -0.16 & 0.31 & 0.00 & 0.69 & 0.00 & -0.25 \\
 \text{} & \text{ZH} & 0.34 & -0.42 & -0.01 & 0.35 & 0.00 & 0.98 & -0.10 & 0.02 \\
 \hline
 \text{50-100} & \text{VBF} & 0.37 & -0.25 & -0.22 & 0.47 & 0.01 & -0.04 & 0.00 & 0.02 \\
 \text{} & \text{WH} & 0.38 & -0.15 & -0.18 & 0.32 & 0.00 & 0.47 & 0.00 & -0.24 \\
 \text{} & \text{ZH} & 0.38 & -0.35 & 0.05 & 0.30 & 0.00 & 1.00 & -0.08 & 0.00 \\
 \hline
 \text{100-150} & \text{VBF} & 0.23 & -0.27 & -0.22 & 0.50 & 0.01 & 0.09 & 0.01 & 0.03 \\
 \text{} & \text{WH} & 0.16 & -0.15 & -0.14 & 0.29 & 0.00 & 0.11 & 0.00 & -0.19 \\
 \text{} & \text{ZH} & 0.17 & -0.13 & 0.20 & 0.25 & 0.00 & 0.99 & 0.01 & -0.07 \\
 \hline
 \text{150-200} & \text{VBF} & 0.1 & -0.29 & -0.24 & 0.55 & 0.01 & 0.26 & 0.00 & 0.04 \\
 \text{} & \text{WH} & 0.062 & -0.15 & 0.04 & 0.10 & 0.01 & -0.10 & 0.00 & -0.13 \\
 \text{} & \text{ZH} & 0.066 & 0.00 & 0.50 & 0.09 & 0.00 & 0.85 & 0.06 & -0.13 \\
 \hline
 \text{200-250} & \text{VBF} & 0.043 & -0.32 & -0.27 & 0.63 & 0.01 & 0.46 & 0.00 & 0.06 \\
 \text{} & \text{WH} & 0.026 & -0.15 & 0.42 & -0.27 & 0.01 & -0.12 & 0.00 & -0.07 \\
 \text{} & \text{ZH} & 0.027 & -0.07 & 0.97 & -0.19 & 0.00 & 0.93 & 0.00 & -0.11 \\
 \hline
 \text{250-300} & \text{VBF} & 0.018 & -0.33 & -0.36 & 0.73 & 0.01 & 0.73 & -0.01 & 0.07 \\
 \text{} & \text{WH} & 0.012 & -0.15 & 1.10 & -0.91 & 0.02 & 0.02 & 0.00 & -0.01 \\
 \text{} & \text{ZH} & 0.012 & -0.18 & 1.60 & -0.63 & 0.00 & 1.10 & -0.10 & -0.07 \\
 \hline
 \text{300-350} & \text{VBF} & 0.0087 & -0.37 & -0.50 & 0.93 & 0.01 & 0.85 & -0.02 & 0.10 \\
 \text{} & \text{WH} & 0.0063 & -0.15 & 1.90 & -1.70 & 0.03 & 0.22 & 0.00 & 0.05 \\
 \text{} & \text{ZH} & 0.0056 & -0.26 & 2.50 & -1.30 & 0.00 & 1.30 & -0.19 & -0.01 \\
 \hline
 \text{350-400} & \text{VBF} & 0.0038 & -0.42 & -0.66 & 1.20 & 0.02 & 1.00 & -0.04 & 0.12 \\
 \text{} & \text{WH} & 0.0034 & -0.15 & 3.10 & -2.90 & 0.03 & 0.49 & 0.00 & 0.11 \\
 \text{} & \text{ZH} & 0.0033 & -0.31 & 4.00 & -2.50 & 0.00 & 1.70 & -0.35 & 0.06 \\
 \hline
 \text{400-450} & \text{VBF} & 0.002 & -0.41 & -0.96 & 1.40 & 0.02 & 1.80 & -0.04 & 0.13 \\
 \text{} & \text{WH} & 0.002 & -0.15 & 4.60 & -4.40 & 0.04 & 0.70 & 0.00 & 0.15 \\
 \text{} & \text{ZH} & 0.0017 & -0.31 & 5.10 & -3.40 & 0.00 & 1.70 & -0.40 & 0.10 \\
 \hline
 \text{450-500} & \text{VBF} & 0.00098 & -0.48 & -1.20 & 1.70 & 0.02 & 1.90 & -0.06 & 0.16 \\
 \text{} & \text{WH} & 0.0014 & -0.15 & 6.20 & -6.00 & 0.04 & 1.00 & 0.00 & 0.20 \\
 \text{} & \text{ZH} & 0.0011 & -0.24 & 6.50 & -4.50 & 0.00 & 1.80 & -0.45 & 0.15 \\
 \hline
 \text{500+} & \text{VBF} & 0.0014 & -0.58 & -2.50 & 3.00 & 0.03 & 3.00 & -0.10 & 0.21 \\
 \text{} & \text{WH} & 0.0024 & -0.15 & 14.00 & -14.00 & 0.05 & 1.90 & 0.00 & 0.32 \\
 \text{} & \text{ZH} & 0.0021 & 0.35 & 15.00 & -12.00 & 0.00 & 2.40 & -0.71 & 0.29 \\
 \hline
\end{tabular}
\caption{$r$ values and percentage deviations $\mu_{ij}$ for VBF, $WH$ and $ZH$ production
in 11 bins of the $p_T(H)$ distribution at $\mu_{EFT}=125$ GeV.}
\label{tab:diff1}
\end{table}

\begin{table}
	\centering
\begin{tabular}{|l|l|l|lllllll|}
		\hline
		Bin [GeV] & Channel & $r$ value
		& $O_{\varphi t}$
		& $O_{\varphi Q}^{(+)}$
		& $O_{\varphi Q}^{(-)}$
		& $O_{\varphi tb}$
		& $O_{tW}$
		& $O_{tB}$
		& $O_{t\varphi}$
		\\\hline
\text{0-50} & \text{VBF} & 0.22 & 0.70 & 0.82 & -1.50 & -0.05 & 0.60 & 0.00 & 0.02 \\
 \text{} & \text{WH} & 0.35 & 0.79 & 0.13 & -0.93 & -0.05 & -3.40 & 0.00 & -0.25 \\
 \text{} & \text{ZH} & 0.34 & 0.50 & 0.37 & -0.98 & -0.05 & -2.30 & -0.45 & 0.02 \\
 \hline
 \text{50-100} & \text{VBF} & 0.37 & 0.69 & 0.85 & -1.50 & -0.05 & 0.37 & -0.01 & 0.02 \\
 \text{} & \text{WH} & 0.38 & 0.79 & -0.10 & -0.69 & -0.05 & -4.00 & 0.00 & -0.24 \\
 \text{} & \text{ZH} & 0.38 & 0.56 & 0.23 & -0.84 & -0.05 & -2.70 & -0.45 & 0.00 \\
 \hline
 \text{100-150} & \text{VBF} & 0.23 & 0.67 & 0.95 & -1.60 & -0.04 & 0.20 & 0.00 & 0.03 \\
 \text{} & \text{WH} & 0.16 & 0.79 & -0.55 & -0.24 & -0.05 & -4.90 & 0.00 & -0.19 \\
 \text{} & \text{ZH} & 0.17 & 0.73 & -0.05 & -0.48 & -0.05 & -3.20 & -0.38 & -0.07 \\
 \hline
 \text{150-200} & \text{VBF} & 0.1 & 0.65 & 1.10 & -1.70 & -0.04 & 0.00 & 0.01 & 0.04 \\
 \text{} & \text{WH} & 0.062 & 0.79 & -1.10 & 0.30 & -0.04 & -5.40 & 0.00 & -0.13 \\
 \text{} & \text{ZH} & 0.066 & 0.79 & -0.43 & 0.01 & -0.05 & -3.70 & -0.35 & -0.13 \\
 \hline
 \text{200-250} & \text{VBF} & 0.043 & 0.64 & 1.30 & -1.90 & -0.04 & -0.20 & 0.00 & 0.06 \\
 \text{} & \text{WH} & 0.026 & 0.79 & -1.70 & 0.88 & -0.04 & -5.50 & 0.00 & -0.07 \\
 \text{} & \text{ZH} & 0.027 & 0.61 & -0.86 & 0.58 & -0.05 & -3.80 & -0.42 & -0.11 \\
 \hline
 \text{250-300} & \text{VBF} & 0.018 & 0.63 & 1.50 & -2.00 & -0.04 & -0.47 & 0.00 & 0.07 \\
 \text{} & \text{WH} & 0.012 & 0.79 & -2.30 & 1.50 & -0.03 & -5.50 & 0.00 & -0.01 \\
 \text{} & \text{ZH} & 0.012 & 0.37 & -1.30 & 1.20 & -0.05 & -3.80 & -0.52 & -0.07 \\
 \hline
 \text{300-350} & \text{VBF} & 0.0087 & 0.61 & 1.80 & -2.20 & -0.04 & -0.43 & -0.02 & 0.10 \\
 \text{} & \text{WH} & 0.0063 & 0.79 & -2.80 & 2.10 & -0.03 & -5.30 & 0.00 & 0.05 \\
 \text{} & \text{ZH} & 0.0056 & 0.13 & -1.70 & 1.80 & -0.05 & -3.60 & -0.62 & -0.01 \\
 \hline
 \text{350-400} & \text{VBF} & 0.0038 & 0.58 & 2.00 & -2.50 & -0.04 & -0.32 & -0.04 & 0.12 \\
 \text{} & \text{WH} & 0.0034 & 0.79 & -3.40 & 2.60 & -0.02 & -5.10 & 0.00 & 0.11 \\
 \text{} & \text{ZH} & 0.0033 & -0.14 & -2.20 & 2.50 & -0.05 & -3.60 & -0.80 & 0.06 \\
 \hline
 \text{400-450} & \text{VBF} & 0.002 & 0.60 & 2.30 & -2.70 & -0.03 & -1.20 & -0.04 & 0.13 \\
 \text{} & \text{WH} & 0.002 & 0.79 & -3.90 & 3.10 & -0.02 & -5.00 & 0.00 & 0.15 \\
 \text{} & \text{ZH} & 0.0017 & -0.32 & -2.50 & 2.90 & -0.05 & -3.30 & -0.82 & 0.10 \\
 \hline
 \text{450-500} & \text{VBF} & 0.00098 & 0.58 & 2.60 & -2.90 & -0.03 & -0.96 & -0.07 & 0.16 \\
 \text{} & \text{WH} & 0.0014 & 0.79 & -4.20 & 3.40 & -0.01 & -4.60 & 0.00 & 0.20 \\
 \text{} & \text{ZH} & 0.0011 & -0.55 & -2.80 & 3.30 & -0.05 & -3.10 & -0.89 & 0.15 \\
 \hline
 \text{500+} & \text{VBF} & 0.0014 & 0.55 & 3.20 & -3.50 & -0.03 & -1.30 & -0.08 & 0.21 \\
 \text{} & \text{WH} & 0.0024 & 0.79 & -4.80 & 4.00 & 0.00 & -3.70 & 0.00 & 0.32 \\
 \text{} & \text{ZH} & 0.0021 & -0.96 & -3.10 & 4.00 & -0.05 & -2.80 & -1.20 & 0.29 \\
 \hline
\end{tabular}
\caption{$r$ values and percentage deviations $\mu_{ij}$ for VBF, $WH$ and $ZH$ production
in 11 bins of the $p_T(H)$ distribution at $\mu_{EFT}=1$ TeV. }
\label{tab:diff2}
\end{table}

\bibliography{refs.bib}

\providecommand{\href}[2]{#2}\begingroup\raggedright\begin{thebibliography}{10}

\bibitem{Weinberg:1978kz}
S.~Weinberg, {\it {Phenomenological Lagrangians}},  {\em Physica} {\bf A96}
  (1979) 327.

\bibitem{Buchmuller:1985jz}
W.~Buchmuller and D.~Wyler, {\it {Effective Lagrangian Analysis of New
  Interactions and Flavor Conservation}},  {\em Nucl. Phys.} {\bf B268} (1986)
  621--653.

\bibitem{Leung:1984ni}
C.~N. Leung, S.~T. Love, and S.~Rao, {\it {Low-Energy Manifestations of a New
  Interaction Scale: Operator Analysis}},  {\em Z. Phys.} {\bf C31} (1986) 433.

\bibitem{Buckley:2015nca}
A.~Buckley, C.~Englert, J.~Ferrando, D.~J. Miller, L.~Moore, M.~Russell, and
  C.~D. White, {\it {Global fit of top quark effective theory to data}},  {\em
  Phys. Rev.} {\bf D92} (2015), no.~9 091501,
  [\href{http://xxx.lanl.gov/abs/1506.08845}{{\tt 1506.08845}}].

\bibitem{Buckley:2015lku}
A.~Buckley, C.~Englert, J.~Ferrando, D.~J. Miller, L.~Moore, M.~Russell, and
  C.~D. White, {\it {Constraining top quark effective theory in the LHC Run II
  era}},  {\em JHEP} {\bf 04} (2016) 015,
  [\href{http://xxx.lanl.gov/abs/1512.03360}{{\tt 1512.03360}}].

\bibitem{Cirigliano:2016nyn}
V.~Cirigliano, W.~Dekens, J.~de~Vries, and E.~Mereghetti, {\it {Constraining
  the top-Higgs sector of the Standard Model Effective Field Theory}},  {\em
  Phys. Rev.} {\bf D94} (2016), no.~3 034031,
  [\href{http://xxx.lanl.gov/abs/1605.04311}{{\tt 1605.04311}}].

\bibitem{Rosello:2015sck}
M.~P. Rosello and M.~Vos, {\it {Constraints on four-fermion interactions from
  the $t\bar{t}$ charge asymmetry at hadron colliders}},  {\em Eur. Phys. J.}
  {\bf C76} (2016), no.~4 200, [\href{http://xxx.lanl.gov/abs/1512.07542}{{\tt
  1512.07542}}].

\bibitem{deBlas:2015aea}
J.~de~Blas, M.~Chala, and J.~Santiago, {\it {Renormalization Group Constraints
  on New Top Interactions from Electroweak Precision Data}},  {\em JHEP} {\bf
  09} (2015) 189, [\href{http://xxx.lanl.gov/abs/1507.00757}{{\tt
  1507.00757}}].

\bibitem{Alioli:2017ces}
S.~Alioli, V.~Cirigliano, W.~Dekens, J.~de~Vries, and E.~Mereghetti, {\it
  {Right-handed charged currents in the era of the Large Hadron Collider}},
  {\em JHEP} {\bf 05} (2017) 086,
  [\href{http://xxx.lanl.gov/abs/1703.04751}{{\tt 1703.04751}}].

\bibitem{Bernardo:2014vha}
C.~Bernardo, N.~F. Castro, M.~C.~N. Fiolhais, H.~Gonçalves, A.~G.~C. Guerra,
  M.~Oliveira, and A.~Onofre, {\it {Studying the $Wtb$ vertex structure using
  recent LHC results}},  {\em Phys. Rev.} {\bf D90} (2014), no.~11 113007,
  [\href{http://xxx.lanl.gov/abs/1408.7063}{{\tt 1408.7063}}].

\bibitem{Tonero:2014jea}
A.~Tonero and R.~Rosenfeld, {\it {Dipole-induced anomalous top quark couplings
  at the LHC}},  {\em Phys. Rev.} {\bf D90} (2014), no.~1 017701,
  [\href{http://xxx.lanl.gov/abs/1404.2581}{{\tt 1404.2581}}].

\bibitem{Cao:2015doa}
Q.-H. Cao, B.~Yan, J.-H. Yu, and C.~Zhang, {\it {A General Analysis of Wtb
  anomalous Couplings}},  {\em Chin. Phys.} {\bf C41} (2017), no.~6 063101,
  [\href{http://xxx.lanl.gov/abs/1504.03785}{{\tt 1504.03785}}].

\bibitem{Jung:2014kxa}
S.~Jung, P.~Ko, Y.~W. Yoon, and C.~Yu, {\it {Renormalization group-induced
  phenomena of top pairs from four-quark effective operators}},  {\em JHEP}
  {\bf 08} (2014) 120, [\href{http://xxx.lanl.gov/abs/1406.4570}{{\tt
  1406.4570}}].

\bibitem{Zhang:2012cd}
C.~Zhang, N.~Greiner, and S.~Willenbrock, {\it {Constraints on Non-standard Top
  Quark Couplings}},  {\em Phys. Rev.} {\bf D86} (2012) 014024,
  [\href{http://xxx.lanl.gov/abs/1201.6670}{{\tt 1201.6670}}].

\bibitem{Greiner:2011tt}
N.~Greiner, S.~Willenbrock, and C.~Zhang, {\it {Effective Field Theory for
  Nonstandard Top Quark Couplings}},  {\em Phys. Lett.} {\bf B704} (2011)
  218--222, [\href{http://xxx.lanl.gov/abs/1104.3122}{{\tt 1104.3122}}].

\bibitem{Corbett:2012ja}
T.~Corbett, O.~J.~P. Eboli, J.~Gonzalez-Fraile, and M.~C. Gonzalez-Garcia, {\it
  {Robust Determination of the Higgs Couplings: Power to the Data}},  {\em
  Phys. Rev.} {\bf D87} (2013) 015022,
  [\href{http://xxx.lanl.gov/abs/1211.4580}{{\tt 1211.4580}}].

\bibitem{Butter:2016cvz}
A.~Butter, O.~J.~P. Eboli, J.~Gonzalez-Fraile, M.~C. Gonzalez-Garcia, T.~Plehn,
  and M.~Rauch, {\it {The Gauge-Higgs Legacy of the LHC Run I}},  {\em JHEP}
  {\bf 07} (2016) 152, [\href{http://xxx.lanl.gov/abs/1604.03105}{{\tt
  1604.03105}}].

\bibitem{Englert:2015hrx}
C.~Englert, R.~Kogler, H.~Schulz, and M.~Spannowsky, {\it {Higgs coupling
  measurements at the LHC}},  {\em Eur. Phys. J.} {\bf C76} (2016), no.~7 393,
  [\href{http://xxx.lanl.gov/abs/1511.05170}{{\tt 1511.05170}}].

\bibitem{Falkowski:2015jaa}
A.~Falkowski, M.~Gonzalez-Alonso, A.~Greljo, and D.~Marzocca, {\it {Global
  constraints on anomalous triple gauge couplings in effective field theory
  approach}},  {\em Phys. Rev. Lett.} {\bf 116} (2016), no.~1 011801,
  [\href{http://xxx.lanl.gov/abs/1508.00581}{{\tt 1508.00581}}].

\bibitem{Falkowski:2014tna}
A.~Falkowski and F.~Riva, {\it {Model-independent precision constraints on
  dimension-6 operators}},  {\em JHEP} {\bf 02} (2015) 039,
  [\href{http://xxx.lanl.gov/abs/1411.0669}{{\tt 1411.0669}}].

\bibitem{Ellis:2014dva}
J.~Ellis, V.~Sanz, and T.~You, {\it {Complete Higgs Sector Constraints on
  Dimension-6 Operators}},  {\em JHEP} {\bf 07} (2014) 036,
  [\href{http://xxx.lanl.gov/abs/1404.3667}{{\tt 1404.3667}}].

\bibitem{Ellis:2014jta}
J.~Ellis, V.~Sanz, and T.~You, {\it {The Effective Standard Model after LHC Run
  I}},  {\em JHEP} {\bf 03} (2015) 157,
  [\href{http://xxx.lanl.gov/abs/1410.7703}{{\tt 1410.7703}}].

\bibitem{Ellis:2018gqa}
J.~Ellis, C.~W. Murphy, V.~Sanz, and T.~You, {\it {Updated Global SMEFT Fit to
  Higgs, Diboson and Electroweak Data}},
  \href{http://xxx.lanl.gov/abs/1803.03252}{{\tt 1803.03252}}.

\bibitem{Degrande:2012gr}
C.~Degrande, J.~M. Gerard, C.~Grojean, F.~Maltoni, and G.~Servant, {\it
  {Probing Top-Higgs Non-Standard Interactions at the LHC}},  {\em JHEP} {\bf
  07} (2012) 036, [\href{http://xxx.lanl.gov/abs/1205.1065}{{\tt 1205.1065}}].
  [Erratum: JHEP03,032(2013)].

\bibitem{Deutschmann:2017qum}
N.~Deutschmann, C.~Duhr, F.~Maltoni, and E.~Vryonidou, {\it {Gluon-fusion Higgs
  production in the Standard Model Effective Field Theory}},  {\em JHEP} {\bf
  12} (2017) 063, [\href{http://xxx.lanl.gov/abs/1708.00460}{{\tt
  1708.00460}}]. [Erratum: JHEP02,159(2018)].

\bibitem{Bylund:2016phk}
O.~Bessidskaia~Bylund, F.~Maltoni, I.~Tsinikos, E.~Vryonidou, and C.~Zhang,
  {\it {Probing top quark neutral couplings in the Standard Model Effective
  Field Theory at NLO in QCD}},  {\em JHEP} {\bf 05} (2016) 052,
  [\href{http://xxx.lanl.gov/abs/1601.08193}{{\tt 1601.08193}}].

\bibitem{Englert:2016hvy}
C.~Englert, R.~Rosenfeld, M.~Spannowsky, and A.~Tonero, {\it {New physics and
  signal-background interference in associated $pp\to HZ$ production}},  {\em
  EPL} {\bf 114} (2016), no.~3 31001,
  [\href{http://xxx.lanl.gov/abs/1603.05304}{{\tt 1603.05304}}].

\bibitem{Maltoni:2016yxb}
F.~Maltoni, E.~Vryonidou, and C.~Zhang, {\it {Higgs production in association
  with a top-antitop pair in the Standard Model Effective Field Theory at NLO
  in QCD}},  {\em JHEP} {\bf 10} (2016) 123,
  [\href{http://xxx.lanl.gov/abs/1607.05330}{{\tt 1607.05330}}].

\bibitem{Hartmann:2015oia}
C.~Hartmann and M.~Trott, {\it {On one-loop corrections in the standard model
  effective field theory; the $\Gamma(h \rightarrow \gamma \, \gamma)$ case}},
  {\em JHEP} {\bf 07} (2015) 151,
  [\href{http://xxx.lanl.gov/abs/1505.02646}{{\tt 1505.02646}}].

\bibitem{Hartmann:2015aia}
C.~Hartmann and M.~Trott, {\it {Higgs Decay to Two Photons at One Loop in the
  Standard Model Effective Field Theory}},  {\em Phys. Rev. Lett.} {\bf 115}
  (2015), no.~19 191801, [\href{http://xxx.lanl.gov/abs/1507.03568}{{\tt
  1507.03568}}].

\bibitem{Ghezzi:2015vva}
M.~Ghezzi, R.~Gomez-Ambrosio, G.~Passarino, and S.~Uccirati, {\it {NLO Higgs
  effective field theory and $\kappa$-framework}},  {\em JHEP} {\bf 07} (2015)
  175, [\href{http://xxx.lanl.gov/abs/1505.03706}{{\tt 1505.03706}}].

\bibitem{Dawson:2018pyl}
S.~Dawson and P.~P. Giardino, {\it {Higgs Decays to $ZZ$ and $Z\gamma$ in the
  SMEFT: an NLO analysis}},  \href{http://xxx.lanl.gov/abs/1801.01136}{{\tt
  1801.01136}}.

\bibitem{Gauld:2015lmb}
R.~Gauld, B.~D. Pecjak, and D.~J. Scott, {\it {One-loop corrections to $h\to
  b\bar b$ and $h\to \tau\bar \tau$ decays in the Standard Model Dimension-6
  EFT: four-fermion operators and the large-$m_t$ limit}},  {\em JHEP} {\bf 05}
  (2016) 080, [\href{http://xxx.lanl.gov/abs/1512.02508}{{\tt 1512.02508}}].

\bibitem{Mimasu:2015nqa}
K.~Mimasu, V.~Sanz, and C.~Williams, {\it {Higher Order QCD predictions for
  Associated Higgs production with anomalous couplings to gauge bosons}},  {\em
  JHEP} {\bf 08} (2016) 039, [\href{http://xxx.lanl.gov/abs/1512.02572}{{\tt
  1512.02572}}].

\bibitem{Degrande:2016dqg}
C.~Degrande, B.~Fuks, K.~Mawatari, K.~Mimasu, and V.~Sanz, {\it {Electroweak
  Higgs boson production in the standard model effective field theory beyond
  leading order in QCD}},  {\em Eur. Phys. J.} {\bf C77} (2017), no.~4 262,
  [\href{http://xxx.lanl.gov/abs/1609.04833}{{\tt 1609.04833}}].

\bibitem{Alwall:2014hca}
J.~Alwall, R.~Frederix, S.~Frixione, V.~Hirschi, F.~Maltoni, {\em et~al.}, {\it
  {The automated computation of tree-level and next-to-leading order
  differential cross sections, and their matching to parton shower
  simulations}},  {\em JHEP} {\bf 1407} (2014) 079,
  [\href{http://xxx.lanl.gov/abs/1405.0301}{{\tt 1405.0301}}].

\bibitem{Zhang:2016snc}
C.~Zhang, {\it {Automating Predictions for Standard Model Effective Field
  Theory in MadGraph5\_aMC@NLO}},  in {\em {Proceedings, 12th International
  Symposium on Radiative Corrections (Radcor 2015) and LoopFest XIV (Radiative
  Corrections for the LHC and Future Colliders)}}, 2016.
\newblock \href{http://xxx.lanl.gov/abs/1601.03994}{{\tt 1601.03994}}.

\bibitem{Degrande:2014tta}
C.~Degrande, F.~Maltoni, J.~Wang, and C.~Zhang, {\it {Automatic computations at
  next-to-leading order in QCD for top-quark flavor-changing neutral
  processes}},  {\em Phys. Rev.} {\bf D91} (2015) 034024,
  [\href{http://xxx.lanl.gov/abs/1412.5594}{{\tt 1412.5594}}].

\bibitem{Franzosi:2015osa}
D.~Buarque~Franzosi and C.~Zhang, {\it {Probing the top-quark chromomagnetic
  dipole moment at next-to-leading order in QCD}},  {\em Phys. Rev.} {\bf D91}
  (2015), no.~11 114010, [\href{http://xxx.lanl.gov/abs/1503.08841}{{\tt
  1503.08841}}].

\bibitem{Zhang:2016omx}
C.~Zhang, {\it {Single Top Production at Next-to-Leading Order in the Standard
  Model Effective Field Theory}},  {\em Phys. Rev. Lett.} {\bf 116} (2016),
  no.~16 162002, [\href{http://xxx.lanl.gov/abs/1601.06163}{{\tt 1601.06163}}].

\bibitem{Degrande:2018fog}
C.~Degrande, F.~Maltoni, K.~Mimasu, E.~Vryonidou, and C.~Zhang, {\it
  {Single-top associated production with a $Z$ or $H$ boson at the LHC: the
  SMEFT interpretation}},  \href{http://xxx.lanl.gov/abs/1804.07773}{{\tt
  1804.07773}}.

\bibitem{McCullough:2013rea}
M.~McCullough, {\it {An Indirect Model-Dependent Probe of the Higgs
  Self-Coupling}},  {\em Phys. Rev.} {\bf D90} (2014), no.~1 015001,
  [\href{http://xxx.lanl.gov/abs/1312.3322}{{\tt 1312.3322}}]. [Erratum: Phys.
  Rev.D92,no.3,039903(2015)].

\bibitem{Gorbahn:2016uoy}
M.~Gorbahn and U.~Haisch, {\it {Indirect probes of the trilinear Higgs
  coupling: $gg \to h$ and $h \to \gamma \gamma$}},  {\em JHEP} {\bf 10} (2016)
  094, [\href{http://xxx.lanl.gov/abs/1607.03773}{{\tt 1607.03773}}].

\bibitem{Degrassi:2016wml}
G.~Degrassi, P.~P. Giardino, F.~Maltoni, and D.~Pagani, {\it {Probing the Higgs
  self coupling via single Higgs production at the LHC}},  {\em JHEP} {\bf 12}
  (2016) 080, [\href{http://xxx.lanl.gov/abs/1607.04251}{{\tt 1607.04251}}].

\bibitem{Bizon:2016wgr}
W.~Bizon, M.~Gorbahn, U.~Haisch, and G.~Zanderighi, {\it {Constraints on the
  trilinear Higgs coupling from vector boson fusion and associated Higgs
  production at the LHC}},  {\em JHEP} {\bf 07} (2017) 083,
  [\href{http://xxx.lanl.gov/abs/1610.05771}{{\tt 1610.05771}}].

\bibitem{DiVita:2017eyz}
S.~Di~Vita, C.~Grojean, G.~Panico, M.~Riembau, and T.~Vantalon, {\it {A global
  view on the Higgs self-coupling}},  {\em JHEP} {\bf 09} (2017) 069,
  [\href{http://xxx.lanl.gov/abs/1704.01953}{{\tt 1704.01953}}].

\bibitem{CEPC-SPPCStudyGroup:2015csa}
C.-S.~S. Group, {\it {CEPC-SPPC Preliminary Conceptual Design Report. 1.
  Physics and Detector}},  tech. rep., 2015.

\bibitem{Gomez-Ceballos:2013zzn}
{\bf TLEP Design Study Working Group} Collaboration, M.~Bicer {\em et~al.},
  {\it {First Look at the Physics Case of TLEP}},  {\em JHEP} {\bf 01} (2014)
  164, [\href{http://xxx.lanl.gov/abs/1308.6176}{{\tt 1308.6176}}].

\bibitem{Baer:2013cma}
H.~Baer, T.~Barklow, K.~Fujii, Y.~Gao, A.~Hoang, S.~Kanemura, J.~List, H.~E.
  Logan, A.~Nomerotski, M.~Perelstein, {\em et~al.}, {\it {The International
  Linear Collider Technical Design Report - Volume 2: Physics}},
  \href{http://xxx.lanl.gov/abs/1306.6352}{{\tt 1306.6352}}.

\bibitem{CLIC:2016zwp}
{\bf CLICdp, CLIC} Collaboration, M.~J. Boland {\em et~al.}, {\it {Updated
  baseline for a staged Compact Linear Collider}},
  \href{http://xxx.lanl.gov/abs/1608.07537}{{\tt 1608.07537}}.

\bibitem{Grzadkowski:2010es}
B.~Grzadkowski, M.~Iskrzynski, M.~Misiak, and J.~Rosiek, {\it {Dimension-Six
  Terms in the Standard Model Lagrangian}},  {\em JHEP} {\bf 10} (2010) 085,
  [\href{http://xxx.lanl.gov/abs/1008.4884}{{\tt 1008.4884}}].

\bibitem{Patrignani:2016xqp}
{\bf Particle Data Group} Collaboration, C.~Patrignani {\em et~al.}, {\it
  {Review of Particle Physics}},  {\em Chin. Phys.} {\bf C40} (2016), no.~10
  100001.

\bibitem{Jenkins:2013zja}
E.~E. Jenkins, A.~V. Manohar, and M.~Trott, {\it {Renormalization Group
  Evolution of the Standard Model Dimension Six Operators I: Formalism and
  lambda Dependence}},  {\em JHEP} {\bf 10} (2013) 087,
  [\href{http://xxx.lanl.gov/abs/1308.2627}{{\tt 1308.2627}}].

\bibitem{Jenkins:2013wua}
E.~E. Jenkins, A.~V. Manohar, and M.~Trott, {\it {Renormalization Group
  Evolution of the Standard Model Dimension Six Operators II: Yukawa
  Dependence}},  {\em JHEP} {\bf 01} (2014) 035,
  [\href{http://xxx.lanl.gov/abs/1310.4838}{{\tt 1310.4838}}].

\bibitem{Alonso:2013hga}
R.~Alonso, E.~E. Jenkins, A.~V. Manohar, and M.~Trott, {\it {Renormalization
  Group Evolution of the Standard Model Dimension Six Operators III: Gauge
  Coupling Dependence and Phenomenology}},  {\em JHEP} {\bf 04} (2014) 159,
  [\href{http://xxx.lanl.gov/abs/1312.2014}{{\tt 1312.2014}}].

\bibitem{Wells:2015uba}
J.~D. Wells and Z.~Zhang, {\it {Effective theories of universal theories}},
  {\em JHEP} {\bf 01} (2016) 123,
  [\href{http://xxx.lanl.gov/abs/1510.08462}{{\tt 1510.08462}}].

\bibitem{Hagiwara:1993ck}
K.~Hagiwara, S.~Ishihara, R.~Szalapski, and D.~Zeppenfeld, {\it {Low-energy
  effects of new interactions in the electroweak boson sector}},  {\em Phys.
  Rev.} {\bf D48} (1993) 2182--2203.

\bibitem{Grojean:2006nn}
C.~Grojean, W.~Skiba, and J.~Terning, {\it {Disguising the oblique
  parameters}},  {\em Phys. Rev.} {\bf D73} (2006) 075008,
  [\href{http://xxx.lanl.gov/abs/hep-ph/0602154}{{\tt hep-ph/0602154}}].

\bibitem{Brivio:2017bnu}
I.~Brivio and M.~Trott, {\it {Scheming in the SMEFT... and a reparameterization
  invariance!}},  {\em JHEP} {\bf 07} (2017) 148,
  [\href{http://xxx.lanl.gov/abs/1701.06424}{{\tt 1701.06424}}].

\bibitem{Alloul:2013bka}
A.~Alloul, N.~D. Christensen, C.~Degrande, C.~Duhr, and B.~Fuks, {\it
  {FeynRules 2.0 - A complete toolbox for tree-level phenomenology}},  {\em
  Comput.Phys.Commun.} {\bf 185} (2014) 2250--2300,
  [\href{http://xxx.lanl.gov/abs/1310.1921}{{\tt 1310.1921}}].

\bibitem{Mattelaer:2016gcx}
O.~Mattelaer, {\it {On the maximal use of Monte Carlo samples: re-weighting
  events at NLO accuracy}},  {\em Eur. Phys. J.} {\bf C76} (2016), no.~12 674,
  [\href{http://xxx.lanl.gov/abs/1607.00763}{{\tt 1607.00763}}].

\bibitem{Dedes:2017zog}
A.~Dedes, W.~Materkowska, M.~Paraskevas, J.~Rosiek, and K.~Suxho, {\it {Feynman
  rules for the Standard Model Effective Field Theory in R$_{\xi}$-gauges}},
  {\em JHEP} {\bf 06} (2017) 143,
  [\href{http://xxx.lanl.gov/abs/1704.03888}{{\tt 1704.03888}}].

\bibitem{Hahn:2000kx}
T.~Hahn, {\it {Generating Feynman diagrams and amplitudes with FeynArts 3}},
  {\em Comput. Phys. Commun.} {\bf 140} (2001) 418--431,
  [\href{http://xxx.lanl.gov/abs/hep-ph/0012260}{{\tt hep-ph/0012260}}].

\bibitem{Mertig:1990an}
R.~Mertig, M.~Bohm, and A.~Denner, {\it {FEYN CALC: Computer algebraic
  calculation of Feynman amplitudes}},  {\em Comput. Phys. Commun.} {\bf 64}
  (1991) 345--359.

\bibitem{Shtabovenko:2016sxi}
V.~Shtabovenko, R.~Mertig, and F.~Orellana, {\it {New Developments in FeynCalc
  9.0}},  {\em Comput. Phys. Commun.} {\bf 207} (2016) 432--444,
  [\href{http://xxx.lanl.gov/abs/1601.01167}{{\tt 1601.01167}}].

\bibitem{Kreimer:1989ke}
D.~Kreimer, {\it {The $\gamma$(5) Problem and Anomalies: A Clifford Algebra
  Approach}},  {\em Phys. Lett.} {\bf B237} (1990) 59--62.

\bibitem{Korner:1991sx}
J.~G. Korner, D.~Kreimer, and K.~Schilcher, {\it {A Practicable gamma(5) scheme
  in dimensional regularization}},  {\em Z. Phys.} {\bf C54} (1992) 503--512.

\bibitem{Kreimer:1993bh}
D.~Kreimer, {\it {The Role of gamma(5) in dimensional regularization}},
  \href{http://xxx.lanl.gov/abs/hep-ph/9401354}{{\tt hep-ph/9401354}}.

\bibitem{Shao:2011tg}
H.-S. Shao, Y.-J. Zhang, and K.-T. Chao, {\it {Feynman Rules for the Rational
  Part of the Standard Model One-loop Amplitudes in the 't Hooft-Veltman
  $\gamma_5$ Scheme}},  {\em JHEP} {\bf 09} (2011) 048,
  [\href{http://xxx.lanl.gov/abs/1106.5030}{{\tt 1106.5030}}].

\bibitem{Hahn:1998yk}
T.~Hahn and M.~Perez-Victoria, {\it {Automatized one loop calculations in
  four-dimensions and D-dimensions}},  {\em Comput. Phys. Commun.} {\bf 118}
  (1999) 153--165, [\href{http://xxx.lanl.gov/abs/hep-ph/9807565}{{\tt
  hep-ph/9807565}}].

\bibitem{Denner:1991kt}
A.~Denner, {\it {Techniques for calculation of electroweak radiative
  corrections at the one loop level and results for W physics at LEP-200}},
  {\em Fortsch. Phys.} {\bf 41} (1993) 307--420,
  [\href{http://xxx.lanl.gov/abs/0709.1075}{{\tt 0709.1075}}].

\bibitem{cepc}
M.~Ruan, {\it {Status \& Updates from CEPC Simulation - Detector
  optimization}},  in {\em Presentation at the High Energy Physics Conference
  (IAS HKUST, 24 Jan 2017),
  http://ias.ust.hk/program/shared\_doc/2017/201701hep/HEP\_20170124\_Manqi\_Ruan.pdf}.

\bibitem{fcc}
A.~Blondel, {\it Summary fcc-ee experiments},  in {\em Presentation at the FCC
  Week (Berlin, 2 Jun 2017),
  https://indico.cern.ch/event/556692/contributions/2487579/attachments/1469993/2274251/99-Blondel-FCC-ee-summary-Berlin.pdf}.

\bibitem{Fujii:2015jha}
K.~Fujii {\em et~al.}, {\it {Physics Case for the International Linear
  Collider}},  \href{http://xxx.lanl.gov/abs/1506.05992}{{\tt 1506.05992}}.

\bibitem{Barklow:2015tja}
T.~Barklow, J.~Brau, K.~Fujii, J.~Gao, J.~List, N.~Walker, and K.~Yokoya, {\it
  {ILC Operating Scenarios}},  \href{http://xxx.lanl.gov/abs/1506.07830}{{\tt
  1506.07830}}.

\bibitem{Durieux:2017gxd}
G.~Durieux, {\it {Precision constraints on the top-quark effective field theory
  at future lepton colliders}},  {\em PoS} {\bf DIS2017} (2018) 088,
  [\href{http://xxx.lanl.gov/abs/1708.09849}{{\tt 1708.09849}}].

\bibitem{Durieux:2018tev}
G.~Durieux, M.~Perelló, M.~Vos, and C.~Zhang, {\it {Global and optimal probes
  for the top-quark effective field theory at future lepton colliders}},
  \href{http://xxx.lanl.gov/abs/1807.02121}{{\tt 1807.02121}}.

\bibitem{Jiang:2016czg}
Y.~Jiang and M.~Trott, {\it {On the non-minimal character of the SMEFT}},  {\em
  Phys. Lett.} {\bf B770} (2017) 108--116,
  [\href{http://xxx.lanl.gov/abs/1612.02040}{{\tt 1612.02040}}].

\bibitem{CMS:2017lgc}
{\bf CMS} Collaboration, C.~Collaboration, {\it {Search for the associated
  production of a Higgs boson with a top quark pair in final states with a
  $\tau$ lepton at $\sqrt{s} = 13~\mathrm{TeV}$}},  Tech. Rep.
  CMS-PAS-HIG-17-003, 2017.

\bibitem{CMS:2017vru}
{\bf CMS} Collaboration, C.~Collaboration, {\it {Search for Higgs boson
  production in association with top quarks in multilepton final states at
  $\sqrt{s}=13~\mathrm{TeV}$}},  Tech. Rep. CMS-PAS-HIG-17-004, 2017.

\bibitem{Aaboud:2017jvq}
{\bf ATLAS} Collaboration, M.~Aaboud {\em et~al.}, {\it {Evidence for the
  associated production of the Higgs boson and a top quark pair with the ATLAS
  detector}},  {\em Submitted to: Phys. Rev. D} (2017)
  [\href{http://xxx.lanl.gov/abs/1712.08891}{{\tt 1712.08891}}].

\bibitem{Aaboud:2017rss}
{\bf ATLAS} Collaboration, M.~Aaboud {\em et~al.}, {\it {Search for the
  Standard Model Higgs boson produced in association with top quarks and
  decaying into a $b\bar{b}$ pair in $pp$ collisions at $\sqrt{s}$ = 13 TeV
  with the ATLAS detector}},  {\em Submitted to: Phys. Rev. D} (2017)
  [\href{http://xxx.lanl.gov/abs/1712.08895}{{\tt 1712.08895}}].

\bibitem{Maltoni:2017ims}
F.~Maltoni, D.~Pagani, A.~Shivaji, and X.~Zhao, {\it {Trilinear Higgs coupling
  determination via single-Higgs differential measurements at the LHC}},  {\em
  Eur. Phys. J.} {\bf C77} (2017), no.~12 887,
  [\href{http://xxx.lanl.gov/abs/1709.08649}{{\tt 1709.08649}}].

\bibitem{ATL-PHYS-PUB-2014-011}
{\it {Prospects for the study of the Higgs boson in the VH(bb) channel at
  HL-LHC}},  Tech. Rep. ATL-PHYS-PUB-2014-011, CERN, Geneva, Jul, 2014.

\bibitem{ATL-PHYS-PUB-2014-006}
{\it {Update of the prospects for the $H\to Z\gamma$ search at the
  High-Luminosity LHC}},  Tech. Rep. ATL-PHYS-PUB-2014-006, CERN, Geneva, May,
  2014.

\bibitem{Durieux:2017rsg}
G.~Durieux, C.~Grojean, J.~Gu, and K.~Wang, {\it {The leptonic future of the
  Higgs}},  {\em JHEP} {\bf 09} (2017) 014,
  [\href{http://xxx.lanl.gov/abs/1704.02333}{{\tt 1704.02333}}].

\end{thebibliography}\endgroup

\end{document}